\documentclass{article}
\usepackage[T1]{fontenc}
\usepackage[latin1]{inputenc}
\usepackage{graphics}
\usepackage{graphicx}
\usepackage{longtable}
\usepackage{color}
\usepackage{pslatex}
\usepackage{hyperref}
\usepackage{sverb}
\usepackage{subfigure}

\usepackage{a4p}
\usepackage{rotating}

\usepackage{graphicx}
\usepackage{atlasphysics}
\usepackage{subfigure}
\usepackage{longtable}
\usepackage{multirow}
\usepackage{textcomp}
\usepackage{authblk,colortbl}
\usepackage{units}
\usepackage{amsmath,amssymb,amsfonts} 

\usepackage{authblk}

\usepackage{rotating}

\begin{document}

\begin{titlepage}
 
\begin{flushright} 
{ \bf IFJPAN-VIII-2016-21 
} 
\end{flushright}
 
\vskip 30 mm
\begin{center}
{\bf\huge  W production at LHC: lepton angular distributions }\\
\vskip 3 mm
{\bf\huge and  reference frames for probing hard QCD }
\end{center}
\vskip 13 mm

\begin{center}
   {\bf E. Richter-Was$^{a}$ and Z. Was$^{b}$  }\\
   \vskip 3 mm
       {\em $^a$ Institute of Physics, Jagellonian University, Lojasiewicza 11, 30-348 Cracow, Poland} \\
       {\em $^b$ Institute of Nuclear Physics, PAN, Krak\'ow, ul. Radzikowskiego 152, Poland}\\ 
\end{center}
\vspace{1.1 cm}
\begin{center}
{\bf   ABSTRACT  }
\end{center}

Precision tests of the Standard Model  in the Strong and Electroweak sectors play 
an important role, among  the physics goals of  LHC experiments. 
Because of the nature of proton-proton processes, observables based on the measurement of 
the direction and energy of leptons provide the most precise signatures. 

In the present paper, we concentrate on the  angular distribution of leptons from $W \to \ell \nu$ decays
in the lepton-pair rest-frame. The vector nature of the intermediate state imposes that distributions are 
to a good precision described by spherical polynomials of at most second order. We argue, that contrary
to general belief often expressed in the literature, the full set of angular coefficients can be measured experimentally,
despite the presence in the final state of neutrino escaping detection. 
There is thus no  principle difference with respect to the phenomenology of the   
$Z/\gamma \to \ell^+ \ell^-$ Drell-Yan process.

We show also, that with the proper choice of the coordinate frames, only one coefficient in this polynomial 
decomposition remains sizable, even in the presence of one or more high $p_T$ jets. 
The necessary stochastic choice of the frames relies on  probabilities independent from any 
coupling constants. In this way, electroweak 
effects (dominated by the $V-A$ nature of $W$ couplings to fermions)
can be better separated from the ones of strong interactions.
The separation is convenient for the measurements interpretation.

\vskip 1 cm

\vspace{0.2 cm}

\vspace{0.1 cm}
\vfill
{\small
\begin{flushleft}
{   IFJPAN-VIII-2016-21
\\ August 2016
}
\end{flushleft}
}
\end{titlepage}

\clearpage
\section{Introduction}

The main purpose of the LHC experiments \cite{Aad:2008zzm,Chatrchyan:2008aa} is to search for effects 
of New Physics. This program continues after the breakthrough
discovery of the Higgs boson \cite{Aad:2012tfa,Chatrchyan:2012xdj} and measurement of its main properties~\cite{Khachatryan:2016vau}. 
In parallel to searches of  New Physics, see eg.  
\cite{Aad:2015baa,Aad:2015iea,Khachatryan:2016kdk}, a program of measurements in the domain of Electroweak (EW) and 
Strong (QCD) interactions is on-going. 
This is the keystone for establishing the Standard Model as a fundamental theory. 
It is focused around two main directions: searches (setting upper limits)
for anomalous couplings and precision measurements of the Standard Model parameters.
Precision measurements of the production and decay of intermediate $Z$ and $W$ 
bosons represent the primary group of measurements of the second domain, 
see e.g. ~\cite{Aad:2015auj,Aad:2015uau,CMS:2014jea,Khachatryan:2016yte,Aad:2016izn}.
The study of the differential cross-sections of $W$ production and decays is essential for understanding open questions related 
to the electroweak physics, like the origin of the electroweak symmetry breaking or the source of the CP violation.

Since the discovery of the $W$ boson, its hadronic production both in $p \bar p$ and $pp$ collisions, mass and the width,
have been measured to great precision \cite{PDG2014}. To complete physical information on the production process,
measurements pursued the boson's differential distribution.
The measurements rely on
outgoing leptons of the $W \to \ell \nu$ decays in the $W$-boson rest frame. Because EW interaction of the decay vertex
are known with much better precision than the QCD interaction of the production, the measurement predominantly tests dynamics of 
QCD inprinted in the angular distributions of outgoing leptons.

In principle, the same standard formalism of the Drell-Yan production $Z \to \ell \ell$ \cite{DrellYan70} can be applied in case 
of $W \to \ell \nu$ production~\cite{Mirkes:1992hu, Mirkes:1994eb}.
The angular dependence of the differential cross-section 
can be written again as
\begin{equation}
 \label{Eq:master1}
  \frac{ d\sigma}{dp_T^2 dY d\Omega^*} = 
     \Sigma_{ \alpha=1}^{9} g_{ \alpha}( \theta,  \phi)
     \frac{3}{ 16 \pi} \frac{d \sigma ^{\alpha }}{ dp_T^2 dY}, %
\end{equation}
where the $ g_{ \alpha }( \theta,  \phi)$ represent harmonic polynomials of the second order,
multiplied by normalisation constants and by $d \sigma ^{\alpha }$ which denote helicity cross-sections, corresponding
to nine helicity configurations of $W$ matrix elements.
The angle $\theta$ and $\phi$ in $d\Omega^* = d \cos\theta d\phi$ are the polar and azimuthal decay angles 
of the lepton in  the $W$ rest-frame. The $p_T$, $Y$ denote  transverse momenta and rapidity 
of the intermediate  $W$ boson in the laboratory frame. The z-axis of the $W$ rest frame can be chosen along the
$W$ momentum of the laboratory frame (the helicity frame), or constructed from the directions of 
the two beams (the Collins-Soper frame~\cite{Collins:1977iv}). 

We rewrite  Eq.~(\ref{Eq:master1}) explicitly, defining polynomials and corresponding 
coefficients%
\begin{eqnarray}
 \label{Eq:master2}
  \frac{ d\sigma}{dp_T^2 dY d \cos\theta d\phi} & = & \frac{3}{ 16 \pi} \frac{d \sigma ^{ U+ L }}{ dp_T^2 dY}  \\
  && [  (1 + \cos^2\theta)+ 1/2\ A_0 (1 - 3 \cos^2\theta) +  A_1 \sin( 2 \theta)\cos\phi 
  + 1/2\ A_2 \sin^2\theta \cos( 2 \phi)   \nonumber\\
  &&+ A_3 \sin \theta \cos \phi +   A_4 \cos \theta + A_5 \sin^2 \theta \sin( 2 \phi) + A_6 \sin(2 \theta) \sin \phi 
  + A_7 \sin \theta\ \sin \phi ] \nonumber 
\end{eqnarray}
where $d \sigma ^{ U+ L }$ denotes the unpolarised differential cross-section 
(a convention used in several papers of the 80's). In case of $W$ boson, $(\theta, \phi)$ define the orientation 
of the charged lepton from $W \to \ell \nu$ . 
The coefficients $A_i(p_T, Y)$ are related to ratios of corresponding cross-sections for  
intermediate state  helicity configurations. 
The full set of $A_i$ coefficients has been explicitly calculated for $p \bar p \to W (\to \ell \nu)+1j $
at QCD NLO in \cite{Mirkes:1992hu, Mirkes:1994eb}.

The first term at Born level (no jets): $ (1 + \cos^2\theta)$ results from spin 1 of the intermediate boson.  
The dynamics of the production process is  hidden in the angular 
coefficients $A_i (p_T, Y)$. This allows to treat the problem in a model independent manner. 
In particular, as we will see, all the hadronic physics is described 
implicitly by the angular coefficients and it decouples from the well understood leptonic and intermediate boson physics. 
Let us stress, that the actual choice of the orientation of coordinate frames represents an important topic; 
we will return to it later.

The understanding of how QCD corrections affect lepton angular distributions is important in the measurement 
of the $W$ mass $(m_W)$, independently of whether leptonic transverse momentum or transverse mass $m_T^{W}$ of the $W$ are used. 
In fact, the first measurements of the angular coefficients explored this relation in the 
opposite way. Assuming the  mass of the $W$ boson measured by LEP, from the fit to transverse mass distribution 
of the lepton-neutrino system $m_T^W$, information on the angular 
orientation of the outgoing leptons was extracted.

The cross-section has been parametrised \cite{Cerrito2002} using only the polar-angle (i.e. integrating over azimuthal angle) as 
\begin{equation}
\frac{d \sigma}{d \cos \theta} \sim (1 + \alpha_1 \cos \theta + \alpha_2 \cos^2 \theta)
\end{equation}
\noindent
with the following relations between coefficients;
\begin{equation}
\alpha_1 = \frac{ 2 A_4} {2 + A_0}; \ \ \ \ \ \ \alpha_2 = \frac{2 - 3 A_0}{2 + A_0}.
\end{equation}
It has been estimated that 1\% uncertainty on $\alpha_2$ corresponds to a shift
of the measured $m_W$ in $p \bar p$ collision, determined by fitting the transverse mass distribution,
of approximately 10~MeV.  The $\alpha_1$ measures the forward-backward leptonic decay asymmetry. 

The measurements of  $\alpha_2$ at 1.8 TeV $p\bar p$ collisions have been conducted 
by D0 and CDF experiments and published in \cite{Acosta:2003jm, Abbott:2000aj}. It was based on the data collected 
in 1994-1995 by Fermilab's Tevatron Run Ia. The fit to $m_T^W$ was performed in  several ranges of the 
$W$ boson transvers momentum.  The measurements confirmed SM expectations, that $\alpha_2$ decreases 
with increasing $W$ boson transvers momentum, which corresponds to increase of the longitudinal component of the
$W$ boson polarisation. The ratio of longitudinally to  transversely polarised $W$ bosons in the Collins-Soper $W$ rest frame 
increases with the $W$ transverse momentum at a rate of approximately 15\% per 10 GeV. 

With more data collected during Fermilab Tevatron Run Ib, the measurement of the $W$ angular coefficients
was performed using a different technique; through direct measurement of the azimuthal angle of the charged lepton 
in the Collins-Soper rest-frame  of the $W$ boson \cite{Acosta:2005dn}. The strategy of this novel measurement
was documented in a separate paper \cite{Strologas:2005xs}. Because of the two-fold ambiguity on determining
the sign of $\cos \theta$ (due to neutrino momenta escaping detection) which was not resolved,
only the measurement of the coefficients $A_2$ and $A_3$ was performed and  angular coefficients were 
measured as function of the transverse momentum of the $W$ boson. 
The measurement was performed specifically for the $W^-$ bosons; angular coefficients of the $W^+$ were obtained by CP transformation 
of Eq.~(\ref{Eq:master2}).

The pure $V-A$ interactions of $W^{\pm}$ without QCD effects, lead for $p \bar p$ collisions 
to $\alpha_2 = 1.0$ and $\alpha_1 = 2.0$, thus to pure transversely polarised $W$ boson. 
This assumes that the $W$ boson is produced with no transverse momenta, 
and sea-quarks and gluon contributions to the structure functions can be neglected. Such simple parton-model could guide 
intuition for the $p \bar p$ collisions at Tevatron, but had to be revisited for the  $pp$ collisions at LHC. 

The dominance of quark-gluon initial states, along with the V-A nature of the coupling of the W boson to fermions
implies that at the LHC, $W$ bosons with high transverse momenta are expected to exhibit 
a different polarisation as the production mechanism is different at low $p_T^W$ and high $p_T^W$ \cite{Berger:2009ep,Bern:2011ie}.
 $W$ bosons produced with low $p_T^{W}$, and therefore moving generally along 
beam axis, exhibit a left-handed  polarisation \cite{Ellis:1991qj}. 
This is because the $W$-boson couples, in the dominant production diagram, to the left-handed component of valence quarks, and to the 
right-handed one of the sea anti-quarks. At high $p_T^{W}$, the situation becomes more complex due to contributions of higher-order processes. 
Of special interest, to quantify the validity of the QCD predictions, becomes the behavior of polarisation fractions as function of $p_T^{W}$. 
It was recently pointed out in \cite{Frederix:2014cba}, that events with high  $p_T^W$ can tests the absorptive part of the 
scattering amplitudes and hence offer a non-trivial test of perturbative QCD at one and higher-loop levels. 
In all $p_T^{W}$ ranges, the production at LHC displays therefore new characteristics: asymmetries in charge 
and momentum for $W$ bosons and their decay leptons.

The LHC experiments pursued  measurement techniques different than Tevatron. With 7 TeV data of $pp$ collisions, the helicity frame
and  not the Collins-Soper frame was used. The interest was not to measure 
$A_i$ coefficients directly but rather the helicity fractions, $f_0, f_L, f_R$. The helicity state of the $W$ boson becomes a mixture 
of the left and right handed states, whose proportions are respectively described with fractions $f_L$ and $f_R$. 
The $f_0$ denotes the fraction of longitudinally polarised $W$ bosons, which is possible at higher transverse momenta,
due to a more complicated production mechanism. This state is in principle particularly interesting as it is connected to the massive
character of the gauge boson \cite{Bern:2011ie}. The measurements \cite{Chatrchyan:2011ig,ATLAS:2012au} by ATLAS and CMS experiments 
established that $W$ bosons produced in $pp$ collisions with large transverse momenta are predominantly left-handed,
 as expected in the Standard Model.   

In the standard notation of the helicity fractions, the following relations with $A_i$'s of  Eq.~(\ref{Eq:master2}) are valid
\begin{equation}
 \label{Eq:helicity1}
f_L = \frac{1}{4} (2 - A_0 - A_4); \ \ \ f_R = \frac{1}{4} (2 - A_0 +  A_4); \ \ \ f_0 = \frac{1}{2} A_0 \ \ .
\end{equation}
The difference between left- and right-handed fraction is proportional to $A_4$  
\begin{equation}
 \label{Eq:helicity2}
f_L - f_R = - \frac{A_4}{2} \ .
\end{equation}

Note, that even if  Eq.~(\ref{Eq:master2}) is valid for any definition of the W-boson rest frame, 
the $A_i(p_T, Y)$ are frame dependent. The relations  Eq.~(\ref{Eq:helicity1}) and ~(\ref{Eq:helicity2}),
hold  in the helicity frame.

Very similar arguments can be made also for the case of $Z$ production. However, the different characteristic of couplings have 
to be considered: the coupling of Z-boson to quarks does not involve the chirality projector $\frac{1}{2} (1 -\gamma^5)$, 
but asymmetric between left and right handedness. Contrary,
the analysing power of $Z$ leptonic decays is severely
affected by the coupling to right-handed leptons, being similar to the coupling to left handed leptons. 
As a consequence the angular coefficients $f_L, f_R, f_0$ can no longer be interpreted directly as polarisation fractions 
of the Z boson. The respective matrix transformation, involving left- and right couplings of $Z$ boson 
to fermions, relates them to the  $Z$-boson polarisation fractions ~\cite{Peruzzi:2011mqa}.  

For the case of $Z \to \ell \ell$ channel the measurement of the complete set of $A_i$'s coefficients in the 
Collins-Soper frame was recently performed  at 8 TeV pp collisions by the CMS Collaboration~\cite{Khachatryan:2015paa} 
and the ATLAS Collaboration~\cite{Aad:2016izn}. 
The precision of the measurement by the ATLAS Collaboration allowed to clearly show that the violation of the Lam-Tung sum 
rule \cite{Lam:1978pu}  i.e. $A_0 = A_2$, is much stronger than predicted by NLO calculations. 
It has shown also an evidence of $A_5, A_6, A_7$ being not equal to zero. 

As of today, the situation with the measurement of $A_i$ coefficients for $W \to \ell \nu$ production in hadronic collisions is  
far from satisfactory. 
Measuring only some coefficients like $\alpha_2$ in the Collins-Soper frame or $f_L, f_R, f_0$ in the helicity-frame 
as function of W-boson transverse momenta is not giving a complete picture on the QCD dynamics of the production process. 
Already in the first papers \cite{Mirkes:1992hu, Mirkes:1994dp} the point was made,
that measurement  of the complete set of coefficients is not possible, due to limitations related to the reconstruction
of lepton neutrino momenta, leading to a two-fold ambiguity in the determination 
of the sign of $\cos \theta$.

In the present paper we  argue, that following the strategy outlined in~\cite{Aad:2016izn}, 
one can design a measurement which allows to measure the complete set of coefficients also in the case of $W \to \ell \nu$ 
in $pp$ collision. Then, we move to the discussion of the reference frames used for $W \to \ell \nu$ decay
and demonstrate that the {\tt Mustraal} \cite{Berends:1983mi} frame introduced and detailed for LHC 
in \cite{Richter-Was:2016mal} will be interesting in the case of $W \to \ell \nu$ production as well. 

Our paper is organized as follows. Section~\ref{sec:Wenu} is devoted to the presentation of the strategy
which allows to measure complete set of the $A_i$'s coefficients in case of $W \to \ell \nu$ process.
We follow this strategy and show a {\it proof of concept} for such measurement. In Section~\ref{sec:frames},
we discuss variants for the frames of the  $\theta, \phi$ angles definition. 
In Section~\ref{sec:Numerical} we collect numerical results for the  $A_i$'s coefficients in 
the case   of $pp  \to \ell \nu + 1j$ generated with QCD LO {\tt MadGraph5\_aMC@NLO} Monte Carlo generator~\cite{Alwall:2014hca}
and QCD NLO  {\tt Powheg+MiNLO} Monte Carlo generator~\cite{Nason:2004rx,Alioli:2010xd}.
We elaborate on possible choices of the coordinate frame orientation. We recall
arguments for introducing the {\tt Mustraal} frame~\cite{Richter-Was:2016mal}, (where the orientation of axes
is optimized thanks to  matrix element 
and next-to-leading logarithm calculations) and compare the Collins-Soper and {\tt Mustraal} frames. 
We demonstrate that, similarly to the $Z \to \ell \ell$ case discussed in~\cite{Richter-Was:2016mal},
with the help of probabilistic choice of reference frames for each event, the results of formula (3.4) from \cite{Berends:1983mi}
are reproduced and indeed only one non-zero coefficient in the decomposition 
of the angular distribution is needed. Finally, in Section~\ref{sec:summary} we conclude the paper.

To avoid proliferation of the  figures, 
we generally present those  for $W^{-} \to \ell^{-} \nu$ only, while the corresponding ones for  $W^{+} \to \ell^+ \nu$  
are deferred to Appendixes~\ref{App:WpFigs} -~\ref{App:Ais}.

\section{ Angular coefficients in $W \to \ell \nu$ production}
\label{sec:Wenu}

The production of vector bosons at LHC displays new characteristics compatred to the production at Tevatron
due to proton-proton nature of the collision: asymmetries in charge 
and momentum for vector bosons and their decay leptons. Large left-handed polarisation is expected in the transverse plane. 
Contrary to the case of $p \bar p$ collisions, the angular coefficients in  $p p$ collisions of the $W^+$ and $W^-$ are not related 
by CP transformation, due to absence of such symmetry in the proton structure functions. Only quarks can be valence, 
while both quarks and anti-quarks may be non-valence. 

For the numerical results presented in this Section we use a sample of 4M events $p p \to \tau \nu +1j$ generated at QCD LO with
{\tt MadGraph5\_aMC@NLO} Monte Carlo~\cite{Alwall:2014hca}, with minimal cuts on the generation level, i.e. $p_T^j~>~1$~GeV, 
and default initialisation of other parameters\footnote{In principle any other lepton flavour could have been used for presentation 
of numerical results. Our choice to generate $\tau \nu$ final states is motivated by the planned extensions of the work.}. 
The purpose of presented results is not so much to give theoretical predictions
on $A_i$'s but to illustrate the {\it proof of concept} for the proposed measurement strategy. Therefore we will not elaborate therefore 
on the choices of PDF structure functions, QCD factorisation and normalisation scale, or EW scheme used. 
However, numerical results  are sensitive to particular choices.

\subsection{Kinematical selection}

\begin{figure}
  \begin{center}                               
{
   \includegraphics[width=6.5cm,angle=0]{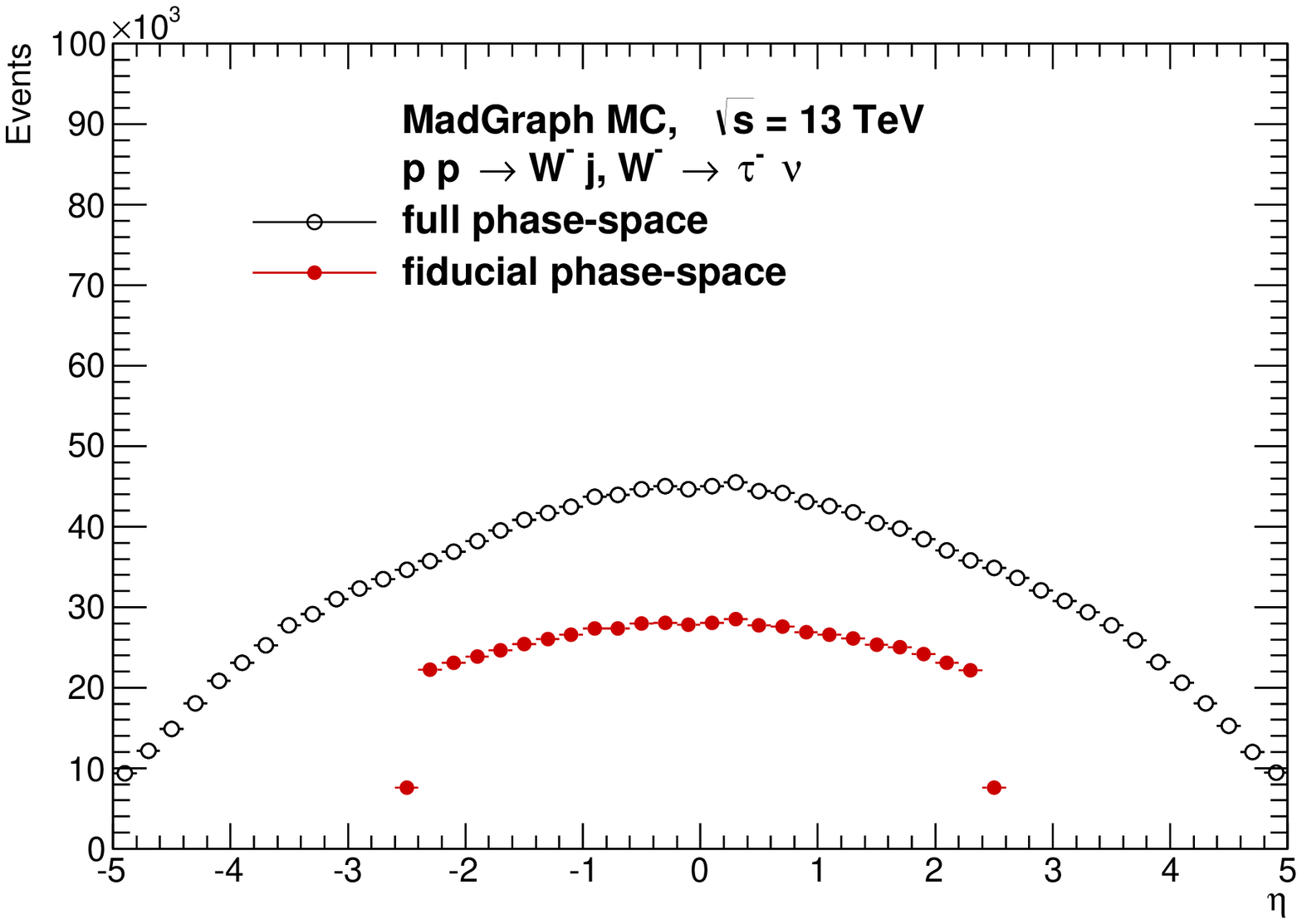}
   \includegraphics[width=6.5cm,angle=0]{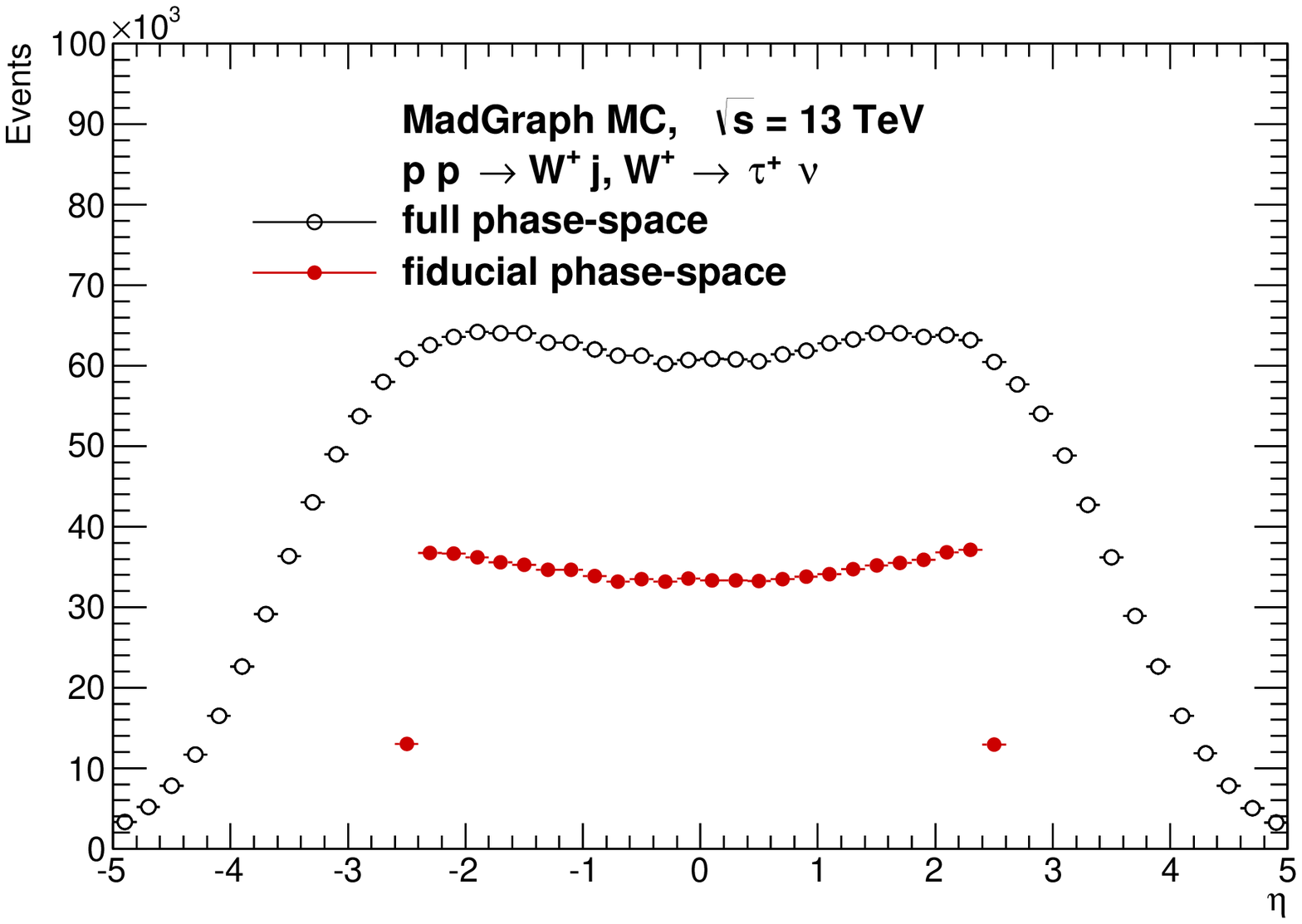}
}
\end{center}
\caption{The pseudorapidity distributions of charged lepton in the laboratory frame, for the full phase-space and fiducial phase-space. 
 Distributions for $W^{-} \to \ell^{-} \nu$ (left) and  $W^{+} \to \ell^{+} \nu$ (right)  are shown. 
\label{Fig:WLab} }
\end{figure}

The kinematical selection needs to be applied in the experimental analysis. The limited coverage in the phase-space is needed
for the efficient triggering, detection and background suppression. It inevitably reshapes angular distributions of the outgoing leptons. 
The minimal set of selection, in the context of LHC experiments is to require that in the laboratory frame, the 
transverse momenta of charged lepton  $p_T^{\ell} > 25$ GeV and pseudorapidity $|\eta^{\ell}| < 2.5$. As the typical selection to  suppress 
background from the multi-jet events, we require  neutrino transverse momenta $p_T^{\nu} > 25$ GeV and the transverse mass 
of the charge-lepton and neutrino system $m_T~>~40$~GeV. This set of selection will define {\it fiducial} phase-space of the measurement.
Similar selection was used e.g. in measurement ~\cite{Aad:2016naf}.
In Fig.~\ref{Fig:WLab} we show as an example the pseudorapidity distribution of the charged lepton from $W^{\pm} \to \ell^{\pm} \nu$ decay,
in the {\it full} phase-space and in the {\it fiducial} phase-space as defined above. 
Clearly, the distributions are different between  $W^+ \to \ell^+ \nu$ and $W^- \to \ell^- \nu $ processes.

\subsection{Solving equation for neutrino momenta } \label{sec:neutrino}
 
For the leptonic decay mode, $W$-bosons have the disadvantage with respect to the $Z$-bosons because the decay kinematics cannot be 
completely reconstructed due to the unobservability of the outgoing neutrino. On the other hand, we can profit from a
simplification: the electroweak interaction does not depend on the virtuality of the intermediate state.
The transverse components 
of the neutrino's momentum $p_x^{\nu}, p_y^{\nu}$ can be approximated from missing transverse momentum balancing the event. 
The longitudinal component $p_z^{\nu}$ can only be calculated up to a twofold ambiguity when solving the quadratic 
equation on the invariant mass of the lepton-neutrino system $m_W$, assuming its value is known. 

Let us recall the corresponding simple formulas: 

\begin{equation}
p_z^{\nu}  = \frac{-b \pm \sqrt{b^2 - 4 a \cdot c}}{2a},
\label{Eq:pznu}
\end{equation}
where
\begin{eqnarray}
a         &=&  4 \cdot p_z^{\ell} -  4 \cdot E^{\ell}, \nonumber \\
b         &=&  4 \cdot (m_W^2 + (p_x^{\ell} + p_x^{\nu})^2 +  (p_y^{\ell} + p_y^{\nu})^2 - (E^{\ell})^2 + (p_z^{\ell})^2  - (p_T^{\ell})^2) \cdot p_z^{\ell},\nonumber \\ 
c         &=&  (m_W^2 + (p_x^{\ell} + p_x^{\nu})^2 +  (p_y^{\ell} + p_y^{\nu})^2 - (E^{\ell})^2 + (p_z^{\ell})^2  - (p_T^{\ell})^2  )^2 
               - 4  \cdot (E^{\ell})^2 \cdot (p_T^{\ell})^2, \nonumber\\
p_T^{\ell}  &=& \sqrt{(p_x^{\ell})^2 + (p_y^{\ell})^2}. \nonumber 
\end{eqnarray}

Eq.~(\ref{Eq:pznu}) has two solutions. Moreover solutions exist only if $\Delta= (b^2 - 4a \cdot c)$ is positive. It requires also,
that the mass of the $W$ boson, $m_W$, is fixed, usually $m_W^{PDG}$ is 
taken (no smearing due to its width).
The solution for the neutrino momentum allows to calculate its energy, completing the kinematics of massless neutrino 
\begin{equation}
E^{\nu}  = \sqrt{ (p_x^{\nu})^2 + (p_y^{\nu})^2 + (p_z^{\nu})^2}.
\label{Eq:Enu}
\end{equation}

Some studies of the past~\cite{StrologasPhD}, investigated if a better option can be designed than taking one of the 
two $p_z^{\nu}$ solutions randomly, with equal probabilities. In particular 
in case that solutions do not exist, if replacing the $m_W^{PDG}$ by e.g. 
transverse mass $m_T^{W}$ can be beneficial. 
No convincing alternative was found. Replacing  $m_W^{PDG}$ with the transverse mass was creating  spikes in 
shapes of angular distributions that are difficult to control. Similar effect, i.e. spiky distortions of the angular 
distributions was caused by favoring some solutions of the neutrino momenta e.g. by selecting the one in the most populated 
regions of the multi-dimensional  phase space, or taking the bigger of the two, etc. 

In the analysis which will be outlined below, we propose to:
\begin{itemize}
\item
Use nominal PDG value for $m_W$ to solve the equation for the neutrino momenta $p_z^{\nu}$.
\item
Drop the event if  $\Delta = (b^2 - 4a \cdot c)$ is negative.
\item
Choose randomly, with equal probabilities, one of the two solutions for the neutrino momenta $p_z^{\nu}$. This solution will be 
called a {\it random} solution.
\end{itemize}

We estimated, that the loss of events due to $\Delta < 0$ is on the level of 10\% and in the experimental analysis can be considered as a part 
of other events losses due to kinematical selection cuts, like thresholds on the lepton transverse momenta or pseudorapidity bounds due to 
limited detector acceptances. 

\subsection{\bf Collins-Soper rest-frame}

For the Drell-Yan productions of the lepton-pair in hadronic collisions, the well known and broadly used 
Collins-Soper reference frame \cite{Collins:1977iv} is defined as a rest-frame of the lepton-pair, with the polar and azimuthal 
angles constructed using proton directions 
in that frame. Since the intermediate resonance, the $W$- or $Z$- boson are produced with non-zero transverse momentum, 
the directions of initial 
protons are not collinear in the lepton-pair rest frame.  The polar axis ($z$-axis) is defined in the lepton-pair rest-frame such that it is 
bisecting the angle between the momentum of one of the proton and inverse of the momentum of the second one. 
The sign of the $z$-axis is defined by the sign of the lepton-pair momentum with respect to $z$-axis 
of the laboratory frame. To complete the coordinate system the $y$-axis is defined as the normal vector 
to the plane spanned by the two incoming proton momenta in the $W$ rest frame and the $x$-axis is chosen to set a right-handed 
Cartesian coordinate system with the other two axes. 
Polar and azimuthal angles are calculated with respect to the outgoing lepton 
and are labeled  $\theta$ and $\phi$ respectively. In the case of zero transverse momentum of the lepton-pair, 
the direction of the $y$-axis is arbitrary. 
Note, that there is an ambiguity in the definition of the $\phi$
angle in the Collins-Soper frame.
The orientation of the $x$-axis here follows convention of \cite{Mirkes:1992hu,Karlberg:2014qua,Gavin:2010az}.
  
For the $Z \to \ell^+ \ell^-$ production, the  formula for $\cos\theta$ can be expressed directly in terms of the  momenta of the outgoing 
leptons in the laboratory frame~\cite{CarloniCalame:2007cd}

\begin{equation}
\label{eq:costhetastar}
\cos\theta=  \frac{p_z (\ell^+ \ell^-)}{|p_z (\ell^+ \ell^-)|} \frac{2}{m(\ell^+ \ell^-)\sqrt{m^2(\ell^+ \ell^-)+p_T^2(\ell^+ \ell^-)}}(P_1^+P_2^--P_1^-P_2^+),
\end{equation}

\noindent with  

\begin{equation*}\label{eq:costhetastar:P}
 P_{\textit{i}}^\pm=\frac{1}{\sqrt{2}}(E_{\textit{i}}\pm p_{z,\textit{i}}),
\end{equation*}

\noindent where $E_{\textit{i}}$  and p$_{z,\textit{i}}$ are respectively the energy and longitudinal momentum of
the lepton ($i=1$) and anti-lepton ($i=2$) and $p_z (\ell^+ \ell^-)$ denotes the longitudinal momentum of the lepton system, $ m(\ell^+ \ell^-)$ its 
invariant mass. The $\phi$ angle is calculated as an angle of the lepton in the plane of the $x$ and $y$ axes in the Collins-Soper frame. 
Only the four-momenta of outgoing leptons and incoming proton directions 
are used. That is why the frame is very convenient for experimental purposes.

\begin{figure}
  \begin{center}                               
{
   \includegraphics[width=6.5cm,angle=0]{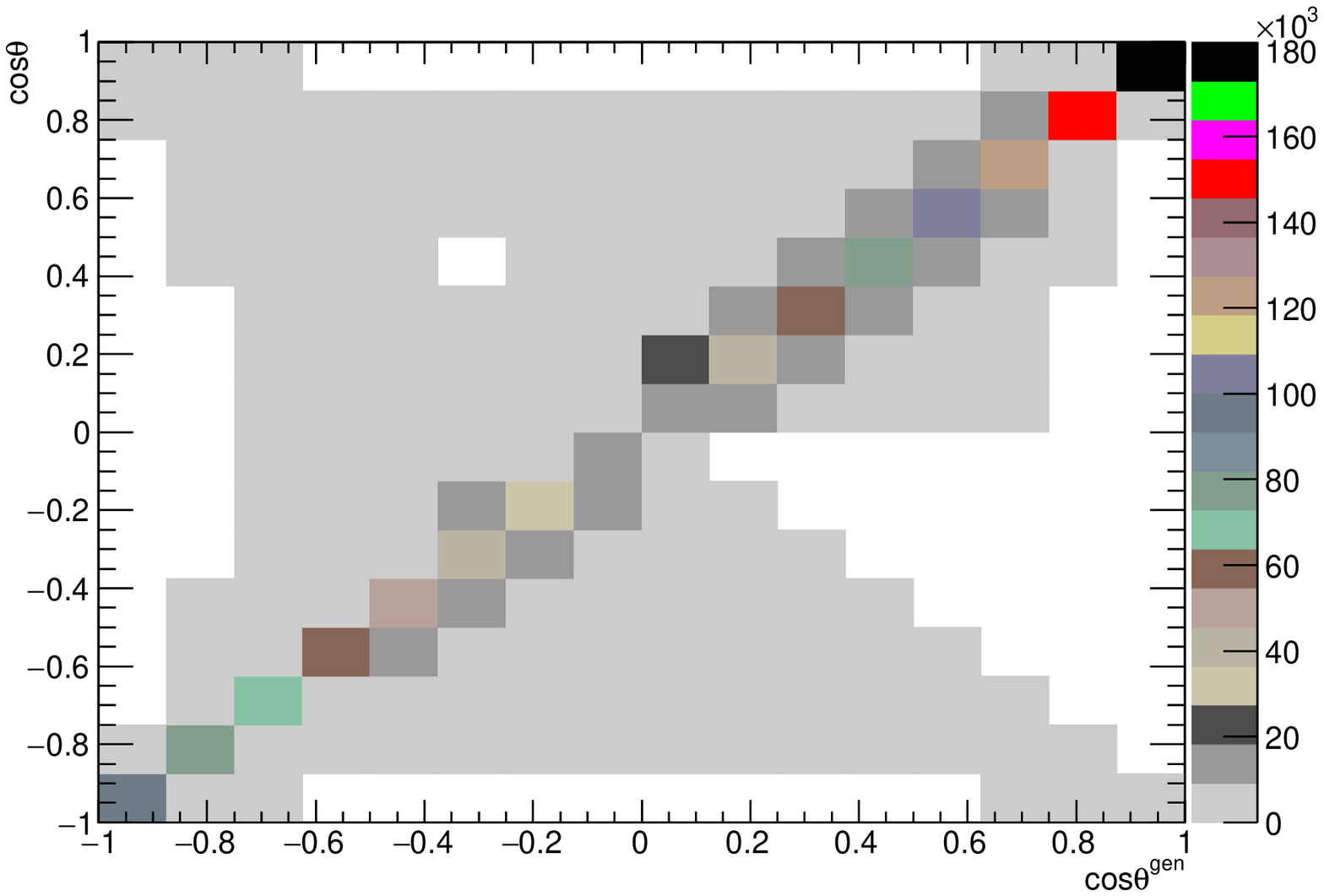}
   \includegraphics[width=6.5cm,angle=0]{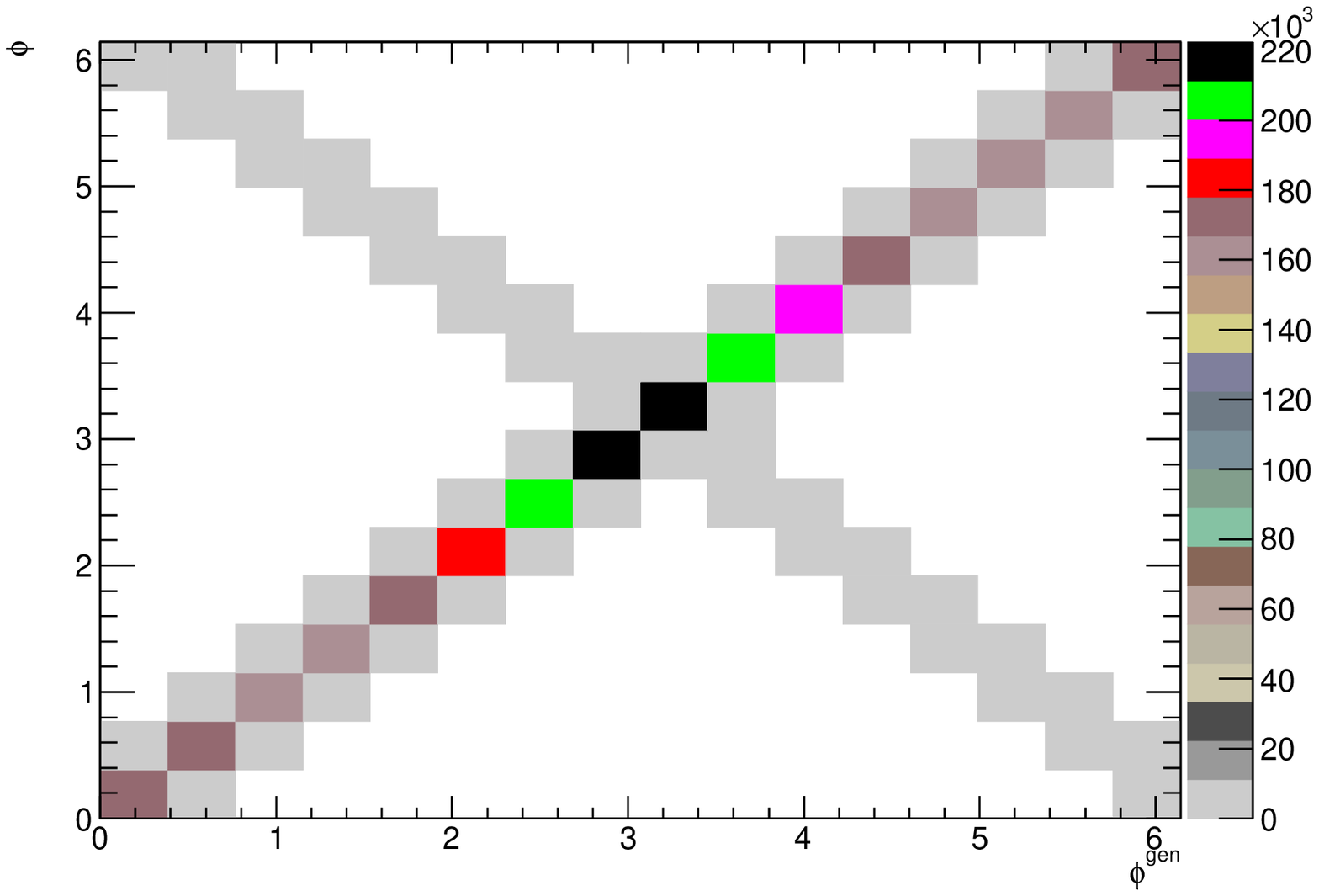}
   \includegraphics[width=6.5cm,angle=0]{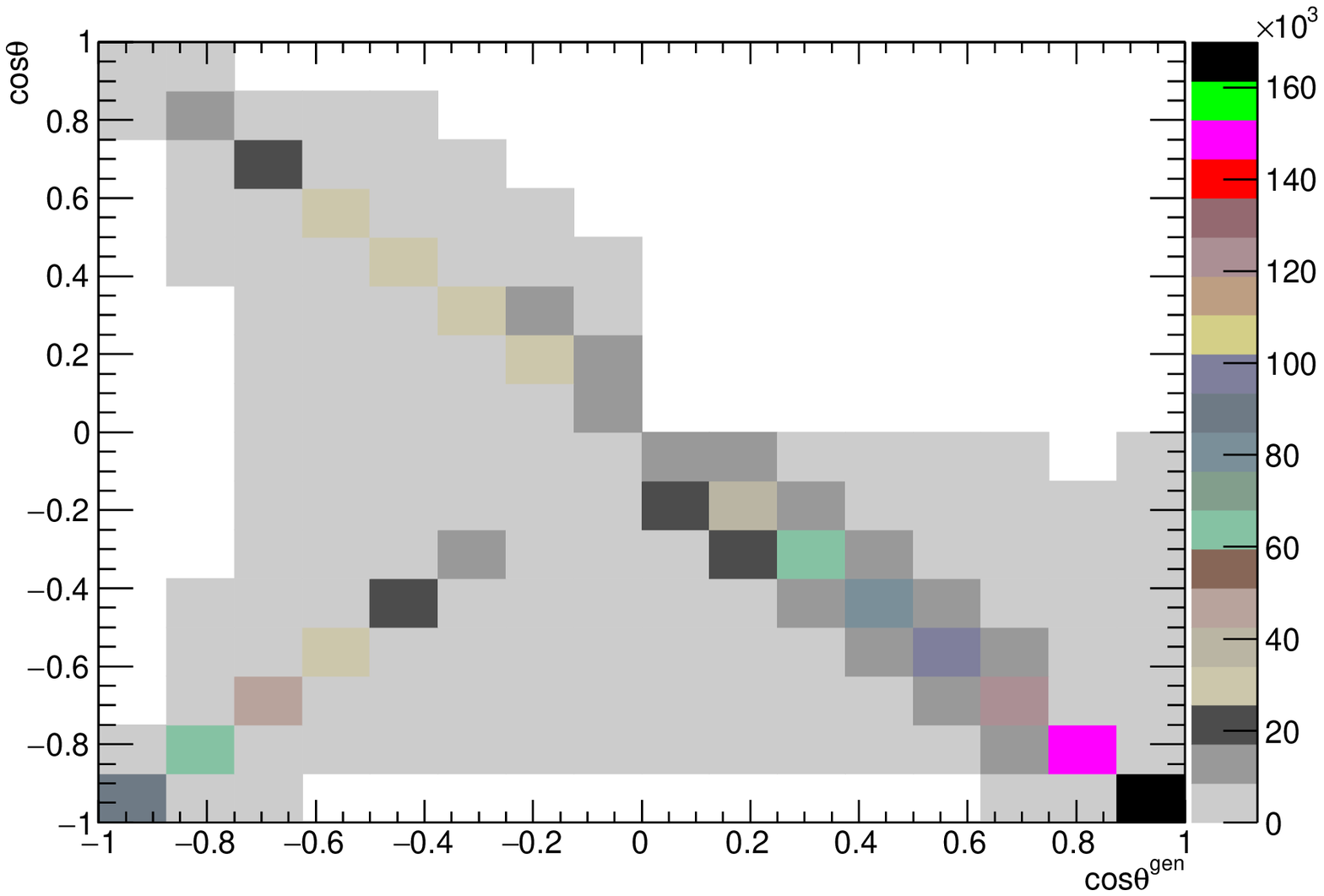}
   \includegraphics[width=6.5cm,angle=0]{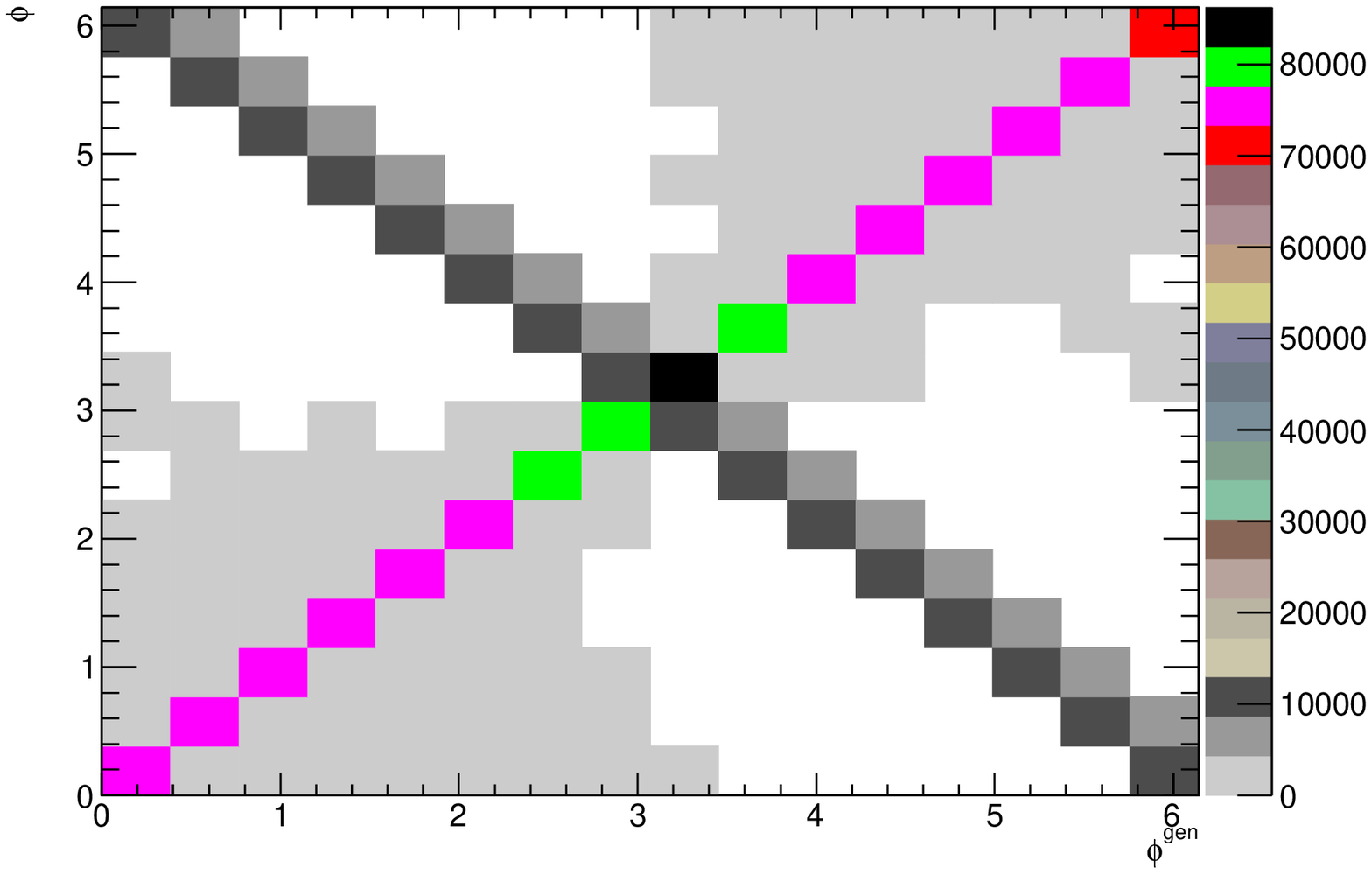}
}
\end{center}
\caption{The correlation plots of $\cos \theta^{gen}$ and $\phi^{gen}$ calculated using generated neutrino momenta and 
$\cos \theta$ and $\phi$ calculated using $m_W = m_W^{PDG}$ for solving Eq.~(\ref{Eq:pznu}). The plots for 
{\it correct} (top) or {\it wrong} (bottom) solution for neutrino momenta
are shown. 
Correlation plots are prepared for the full phase-space of $W^{-} \to \tau^{-} \nu$ process. 
\label{Fig:CScorrel} }
\end{figure}

In case of $W^{\pm} \to \ell^{\pm} \nu$ production we follow the same definition of the frame. 
We use the convention that the $\theta$ and $\phi$ angles define the orientation of the charged lepton, 
i.e. anti-lepton  of $W^+$ production and lepton in case of $W^-$ production. 
We calculate $\cos \theta$ for the chosen solution of neutrino momenta  with formula~(\ref{eq:costhetastar}) and  $\phi$ from the event kinematics as well. 
Figure~\ref{Fig:CScorrel} shows
correlation plots between $\cos \theta^{gen}$, $\phi^{gen}$ calculated using generated neutrino momenta, and 
 $\cos \theta$, $\phi$ calculated using neutrino momenta from formula~(\ref{Eq:pznu}) with $m_W = m_W^{PDG}$. Correlations are shown when 
 {\it correct} or {\it wrong} solution for neutrino momenta\footnote{As {\it correct} we denote solution which is closer to the 
generated  $p_z^{\nu}$ value, as {\it wrong} the other one.} are selected. 
The  $\cos \theta - \cos \theta^{gen}$ and $\phi - \phi^{gen}$, can be anti-correlated in case of {\it correct} and 
{\it wrong} solutions, the effect is much stronger for  {\it wrong} solutions and the $\cos \theta$ variable. We observe also inevitable 
migrations between bins due to the  approximation $m_W = m_W^{PDG}$ used for solving Eq.~(\ref{Eq:pznu}).

\begin{figure}
  \begin{center}                               
{
   \includegraphics[width=6.5cm,angle=0]{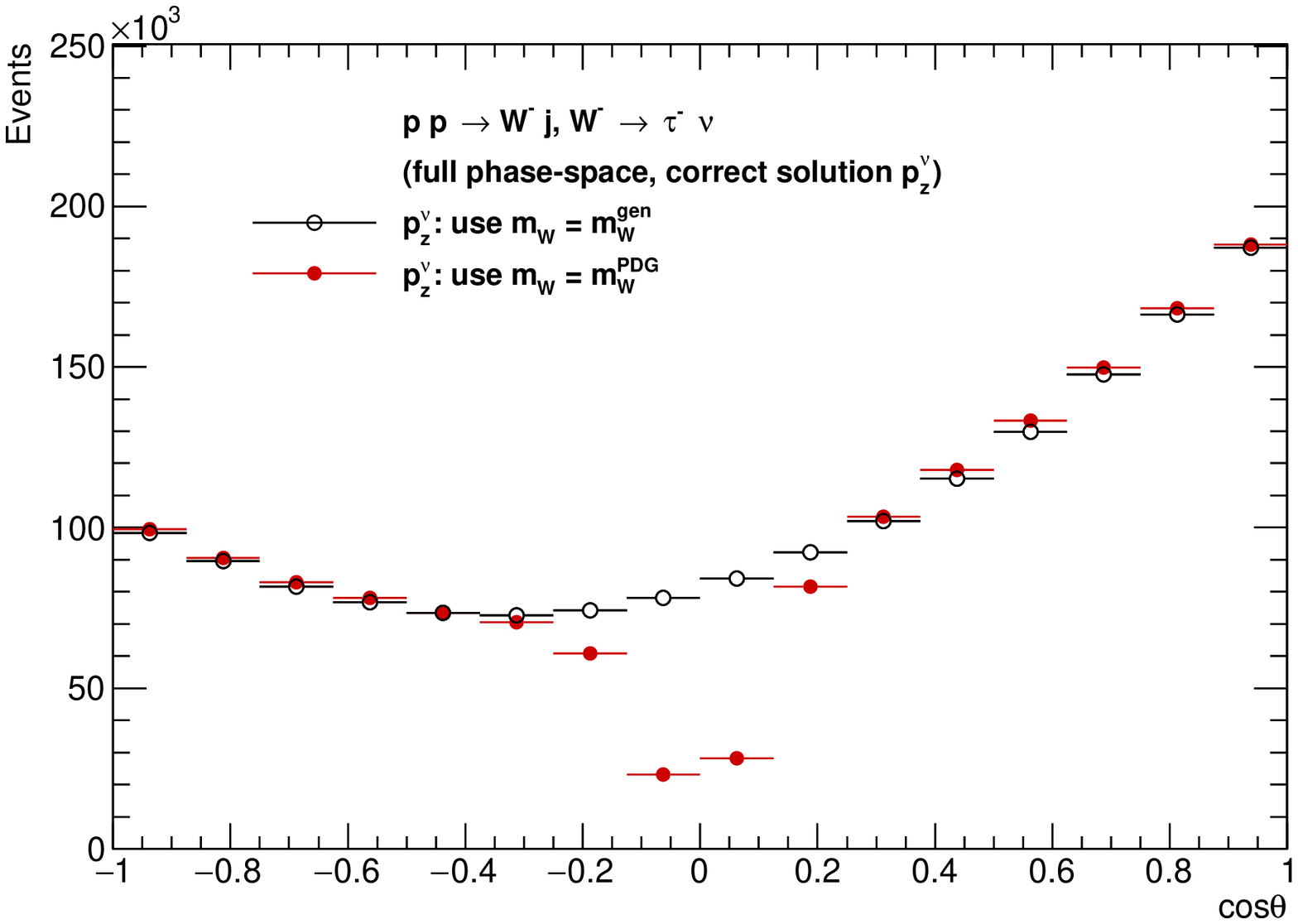}
   \includegraphics[width=6.5cm,angle=0]{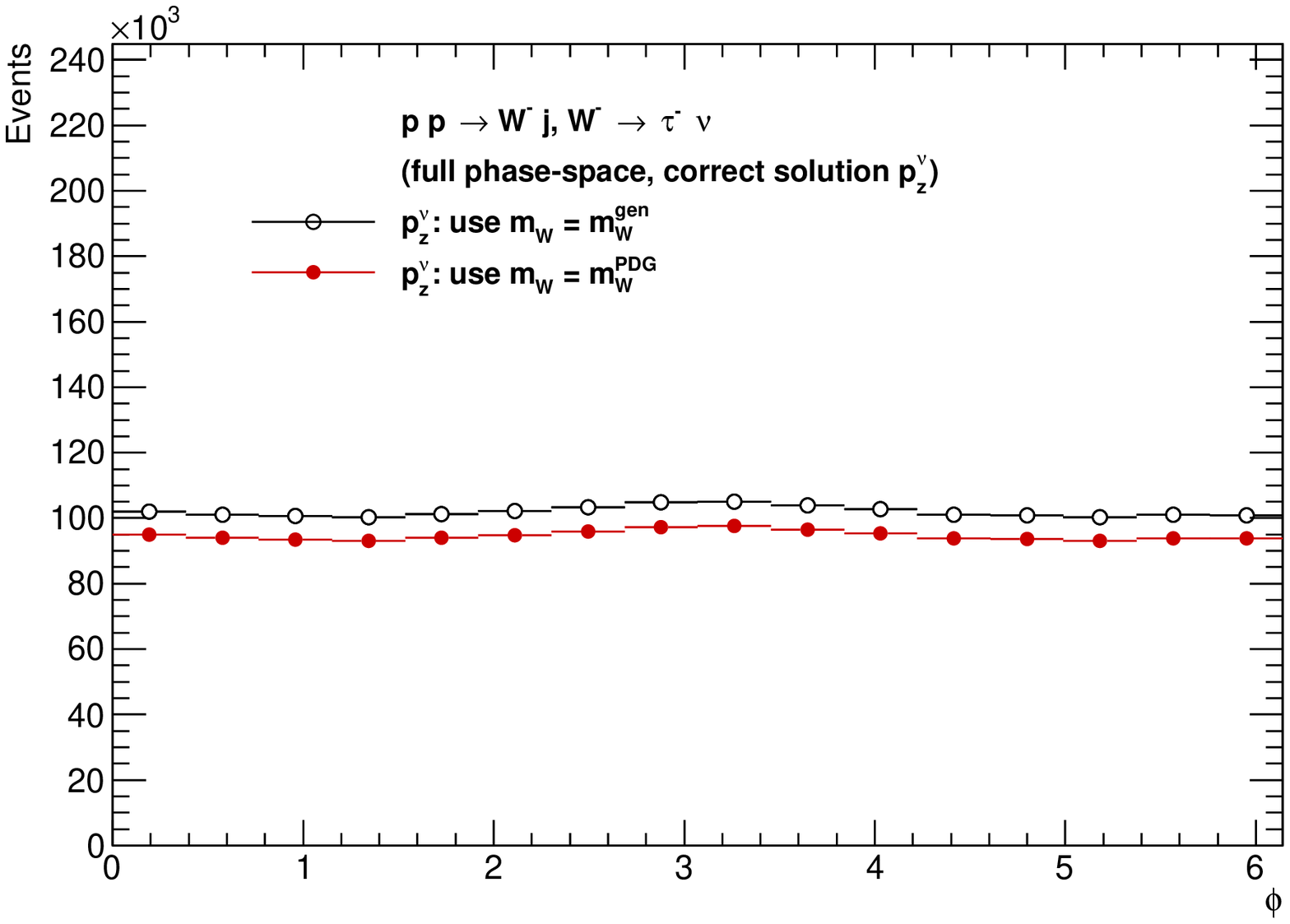}
}
\end{center}
\caption{The $\cos \theta$ and $\phi$ distributions of charged lepton from $W^{-} \to \tau^{-} \nu$, 
in the Collins-Soper rest frame. 
Effect from events loss due to non-existing solution for the neutrino momenta, 
when $m_W = m_W^{PDG}$ is used for  Eq.~(\ref{Eq:pznu}) is concentrated in the 
central bins of the left plot.
\label{Fig:Wmevtloss} }
\end{figure}

Figure~\ref{Fig:Wmevtloss} shows  $\cos \theta$ and $\phi$ distributions of the charged lepton from $W \to \ell \nu$ decays in the Collins-Soper 
rest frame. We use the generated $W$ boson mass $m_W^{gen}$ of a given event  or the fixed PDG value $m_W^{PDG}$ for calculating neutrino
momenta $p_z^{\nu}$, taking the {\it correct} solution for $p_z^{\nu}$. We compare the two results.
The losses due to the non-existence of a  solution of Eq.~(\ref{Eq:pznu}) are concentrated around $\cos \theta = 0$ but are uniformly 
distributed over the full $\phi$ range. 

\begin{figure}
  \begin{center}                               
{
   \includegraphics[width=7.0cm,angle=0]{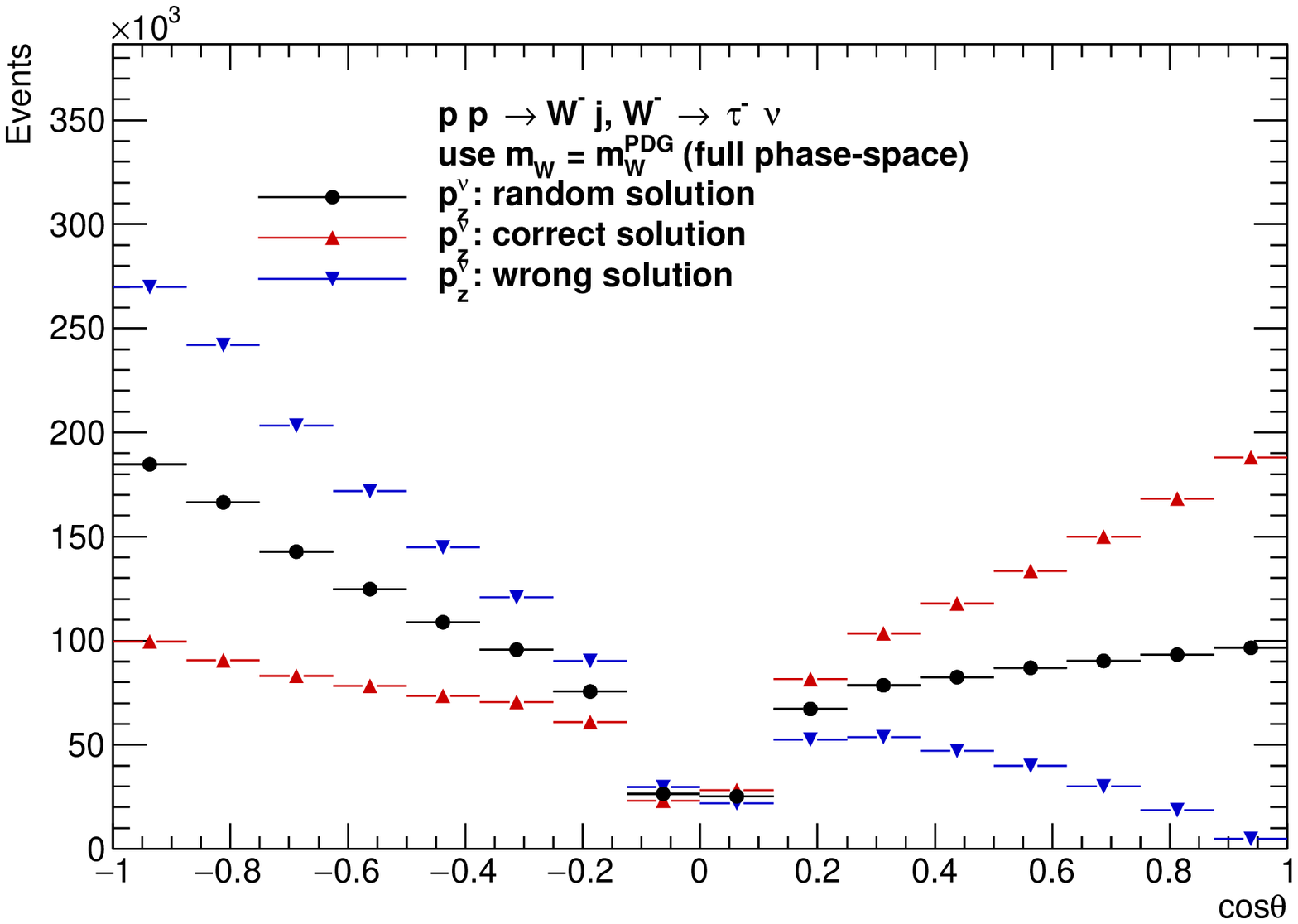}
   \includegraphics[width=7.0cm,angle=0]{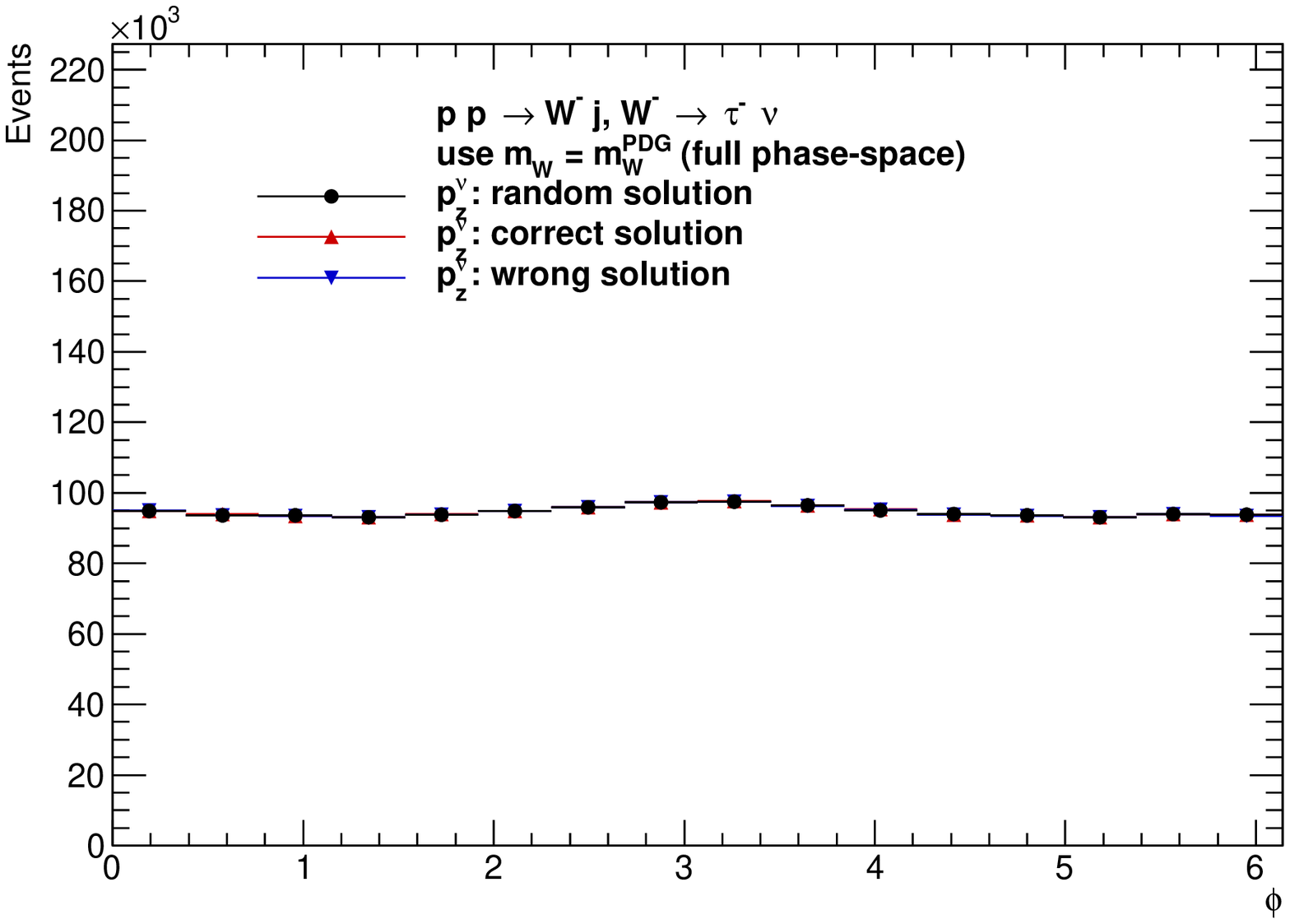}
   \includegraphics[width=7.0cm,angle=0]{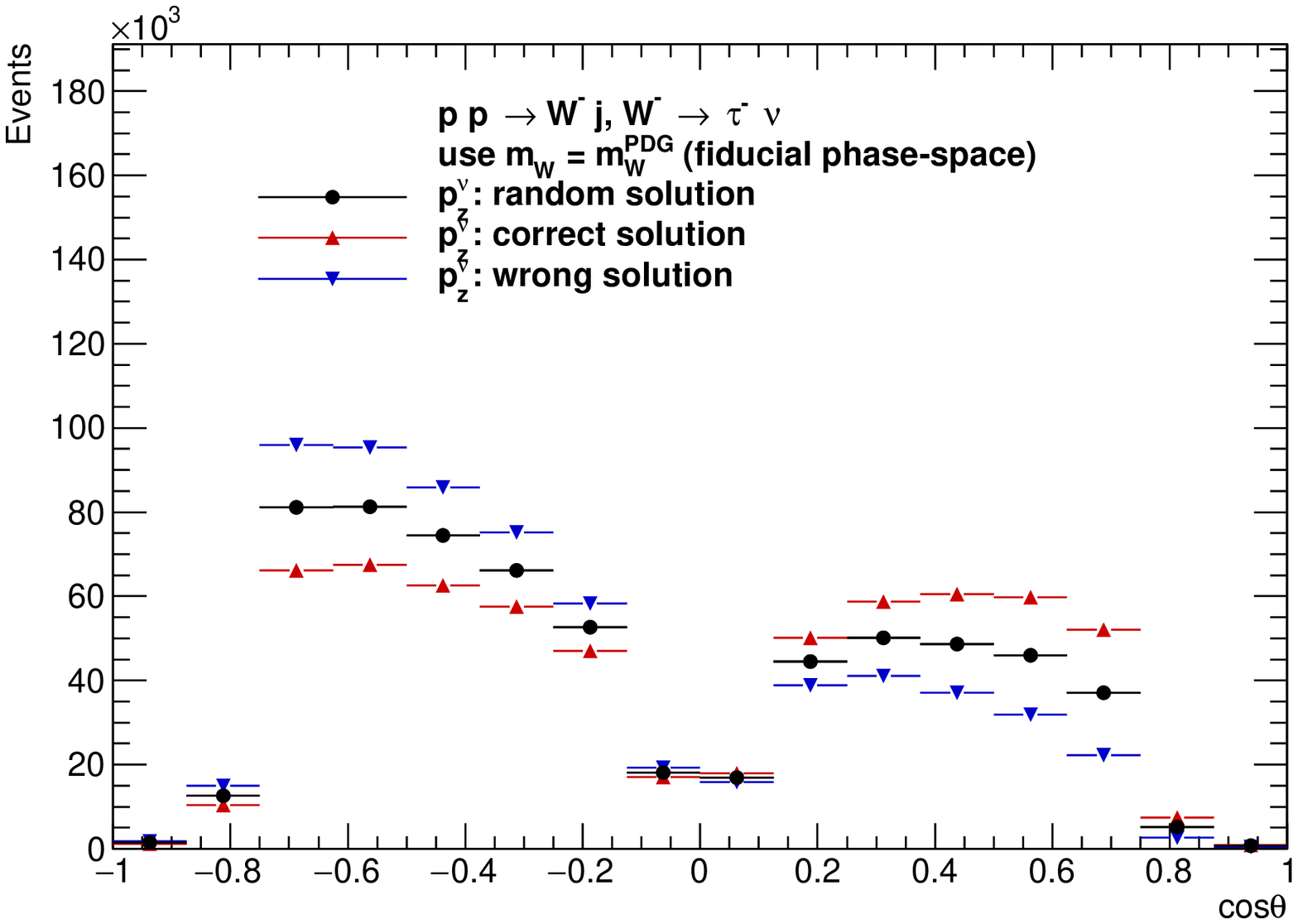}
   \includegraphics[width=7.0cm,angle=0]{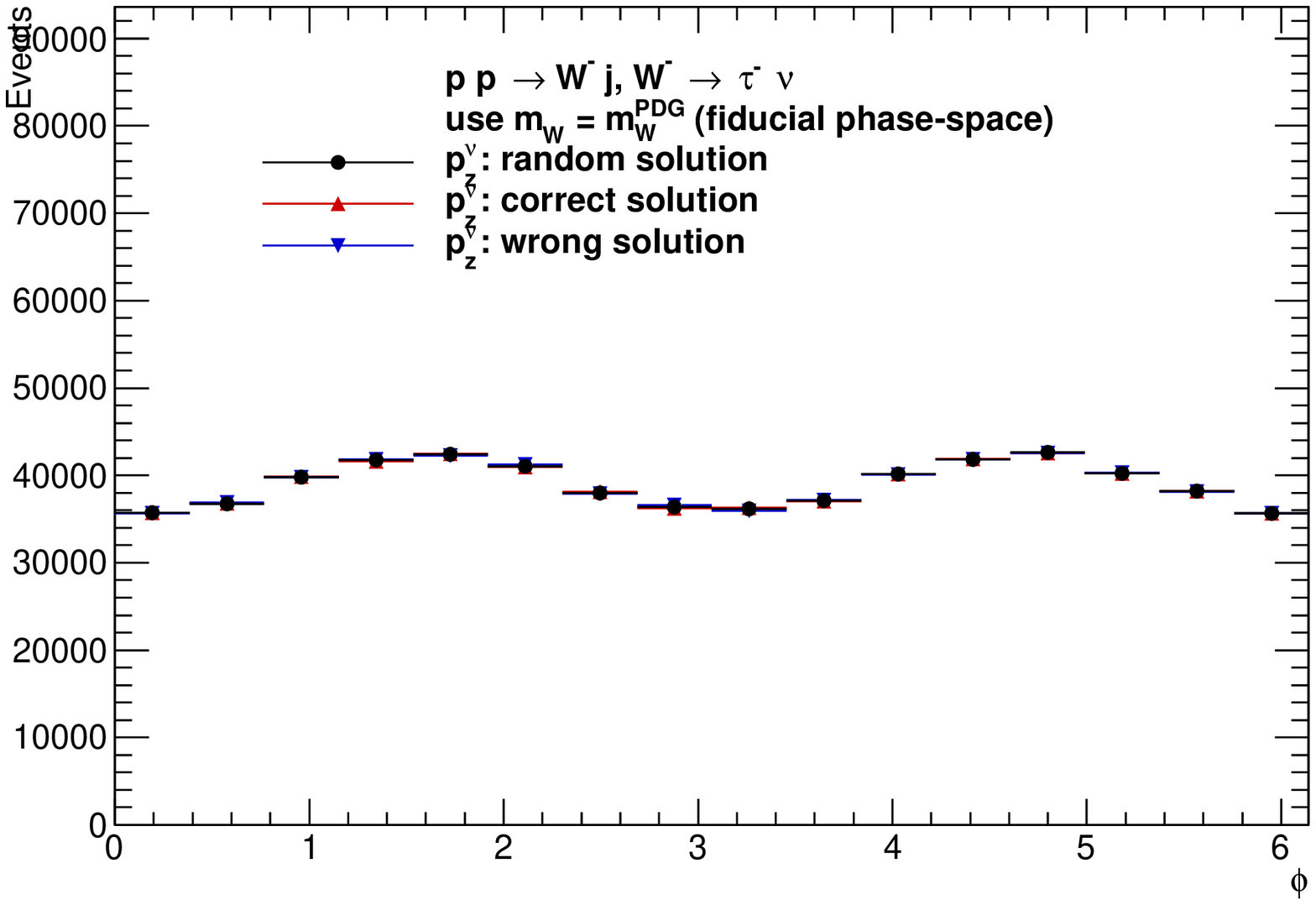}
}
\end{center}
\caption{The $\cos \theta$ and $\phi$ distributions of charged lepton from $W^{-} \to \tau^{-} \nu$ 
in the Collins-Soper rest frame. Cases of $m_W = m_W^{PDG}$ for solving 
Eq.~(\ref{Eq:pznu}) where {\it correct}, {\it wrong} or  {\it random} solution for $p_z^{\nu}$ are taken.
Top plots are for the distributions in the full phase-space, bottom ones for the fiducial phase-space. 
\label{Fig:Wm_cosThetaCSphiCS} }
\end{figure}

Figure~\ref{Fig:Wm_cosThetaCSphiCS} demonstrates the variation of $\cos \theta$ and $\phi$ distributions 
for charged lepton of $W^- \to \ell^- \nu$ decays when $m_W = m_W^{PDG}$ is used for solving  Eq.~(\ref{Eq:pznu}) and the selection 
of the fiducial regions applied. In each case, distributions are shown for {\it correct}, {\it wrong} and  {\it random} solution for $p_z^{\nu}$.
Selection of the fiducial region enhances modulation in the $\phi$ distribution. Corresponding distributions for $W^+ \to \ell^+ \nu$ decay
are shown in Appendix~\ref{App:WpFigs}. 

\begin{figure}
  \begin{center}                               
{
   \includegraphics[width=6.0cm,angle=0]{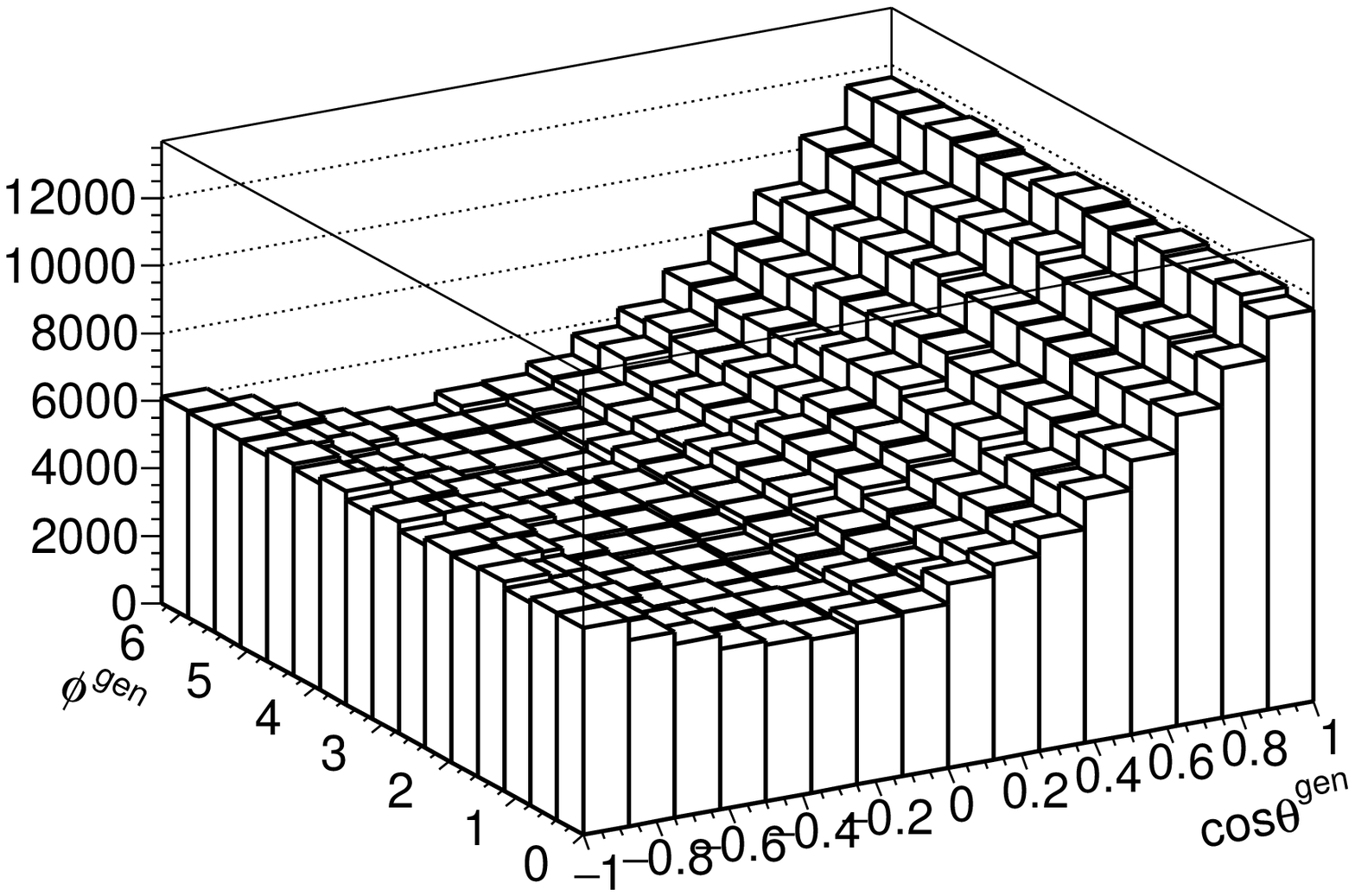}
   \includegraphics[width=6.0cm,angle=0]{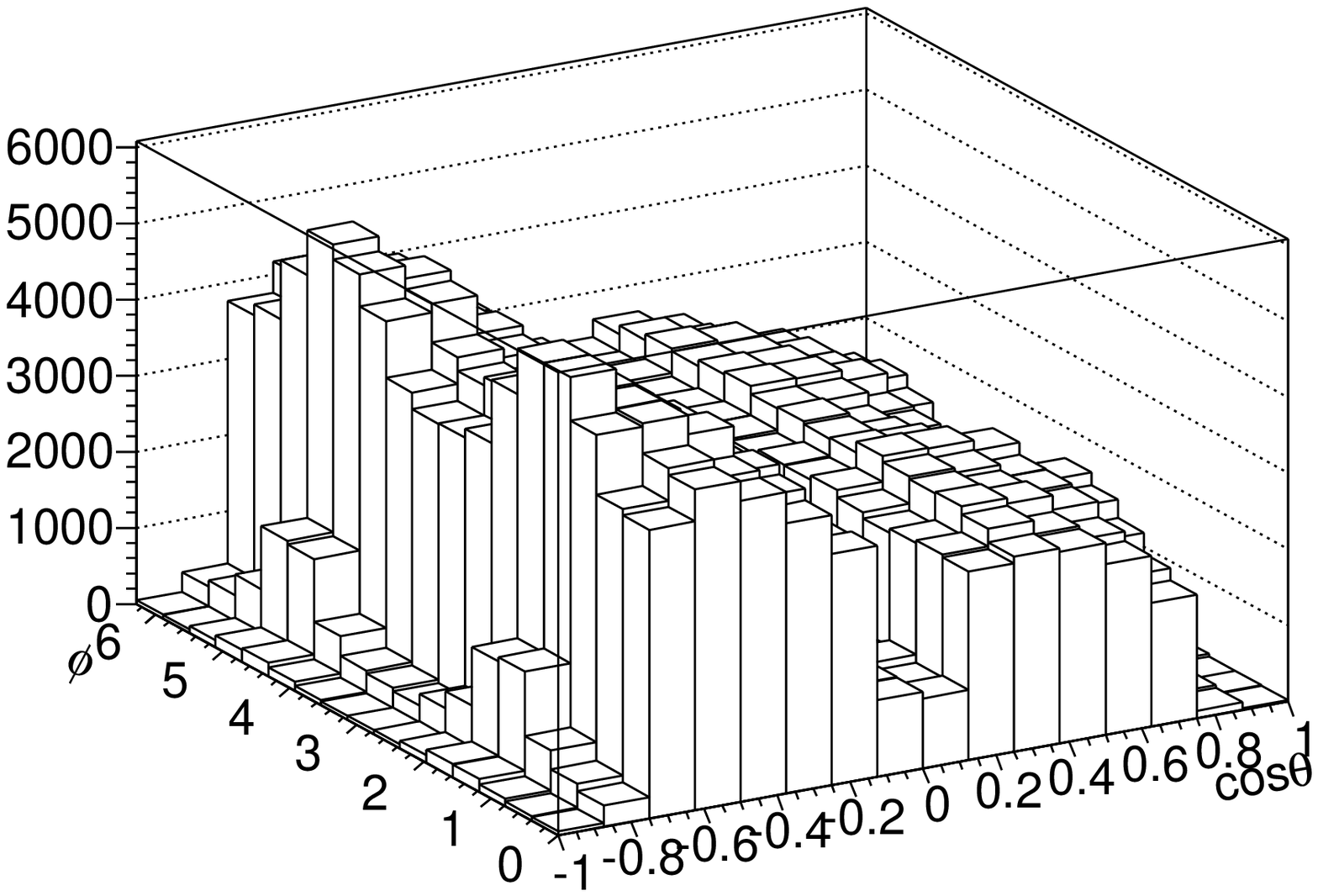}
}
\end{center}
\caption{The 2D distribution of $\cos \theta$, $\phi$ of charged lepton from $W^{-} \to \tau^{-} \nu$. 
Case of  the full-phase space with  generated neutrino momentum, left side plot.
Case of the fiducial phase-space and  $m_W = m_W^{PDG}$ used in
Eq.~(\ref{Eq:pznu}) and {\it random} solution for  $p_z^{\nu}$, right side plot. 
\label{Fig:2DWm_cosThetaCSphiCS} }
\end{figure}

To illustrate the effect of folding into fiducial phase-space, 
2D distributions of $(\cos \theta, \phi)$ are shown in Figure~\ref{Fig:2DWm_cosThetaCSphiCS}: (i)  for events in the full phase-space,
when  generated neutrino momenta are used, (ii)  in the 
fiducial phase-space when  $m_W = m_W^{PDG}$ is used for Eq.~(\ref{Eq:pznu}) 
and {\it random} solution of neutrino momenta is taken. 
Clearly, original shapes of  distributions  
are significantly distorted, but still, as we will see later, 
basic information on the angular correlation of the outgoing 
charged lepton and the beam direction is preserved. In particular, it is non trivial 
that the information is preserved despite approximate knowledge
of the neutrino momentum. Moreover, the information is carried by both, {\it correct} and {\it wrong}, solutions for neutrino momenta. 
These observations are essential for the analysis presented in our paper.   


\subsection{\bf Templated shapes and extracting $A_i$'s coefficients}
\label{sec:Templ}

The standard experimental technique to extract parameters of complicated shapes is to perform the multi-dimensional fit 
to distributions of experimental data using either analytical functions or {\it templated} shapes. Given what we observed  
in Figure~\ref{Fig:Wm_cosThetaCSphiCS} only the second options seems feasible. 
The technique of {\it templated} shapes constructed from Monte Carlo events,
elaborated in ~\cite{Aad:2016izn} for the $A_i$'s measurement in $Z \to \ell \ell$ case, is followed here and 
shortly described below.

We use the Monte Carlo sample of $W^{\pm} \to \ell^{\pm} \nu$ events and extract angular coefficients of Eq.~(\ref{Eq:master2})
using moments methods~\cite{Mirkes:1994eb}. The first moment of a polynomial $P_i(\cos\theta,\phi)$, integrated over a specific 
range of~$p_T$, $Y$ is defined as follows:

\begin{equation}
\langle P_i(\cos\theta,\phi)\rangle = \frac{\int_{-1}^{1}d \cos\theta \int_{0}^{2 \pi}d\phi \ P_i(\cos\theta,\phi) d\sigma(\cos\theta,\phi)}
{\int_{-1}^{1}d \cos\theta \int_{0}^{2 \pi}d\phi\ d\sigma(\cos\theta,\phi) }.
\label{Eq:moment}
\end{equation}

 Owing to the orthogonality of the spherical polynomials of Eq.~(\ref{Eq:master2}), 
the weighted average of the angular distributions with respect to any specific polynomial, Eq.~(\ref{Eq:moment}), isolates its corresponding coefficient,
averaged over some phase-space region. As a consequence of Eq.~(\ref{Eq:master2}) we obtain:

\begin{equation}
\begin{split}
\langle\frac{1}{2}(1-3 \cos^2\theta)\rangle & = \frac{3}{20} (A_0 - \frac{2}{3} ); \ \ \ 
\langle\sin2\theta \cos\phi\rangle  = \frac{1}{5} A_1; \ \ \
\langle\sin^2\theta \cos2\phi\rangle  = \frac{1}{10} A_2; \\
\langle\sin\theta \cos\phi\rangle & = \frac{1}{4} A_3; \ \ \
\langle\cos\theta\rangle  = \frac{1}{4} A_4; \ \ \
\langle\sin^2\theta \sin 2 \phi\rangle = \frac{1}{5} A_5; \\
\langle\sin 2\theta \sin \phi\rangle & = \frac{1}{5} A_6; \ \ \
\langle\sin \theta \sin \phi\rangle  = \frac{1}{4} A_7 .
\end{split}
\label{Eq:moments}
\end{equation}

\begin{figure}
  \begin{center}                               
{
   \includegraphics[width=6.0cm,angle=0]{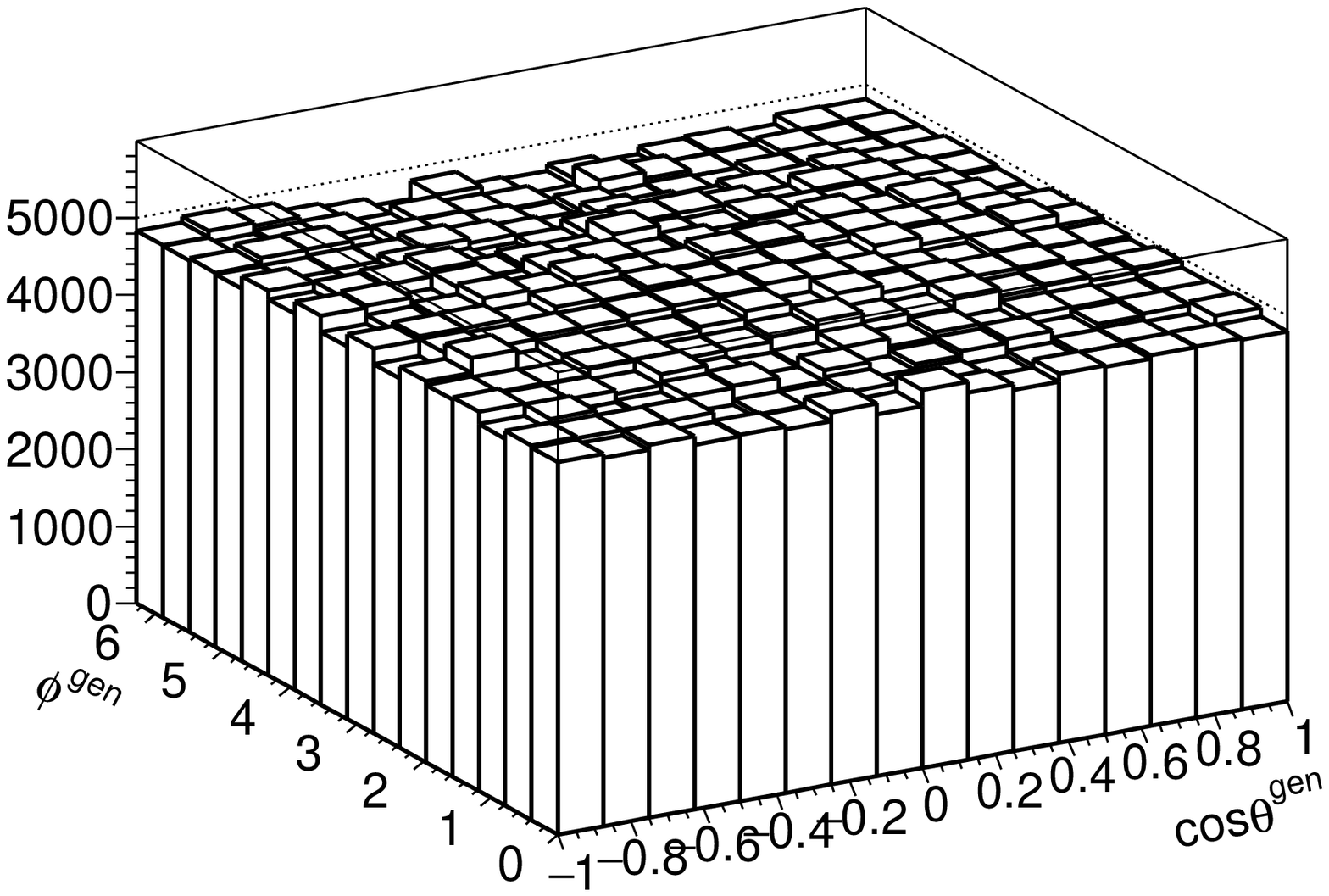}
   \includegraphics[width=6.0cm,angle=0]{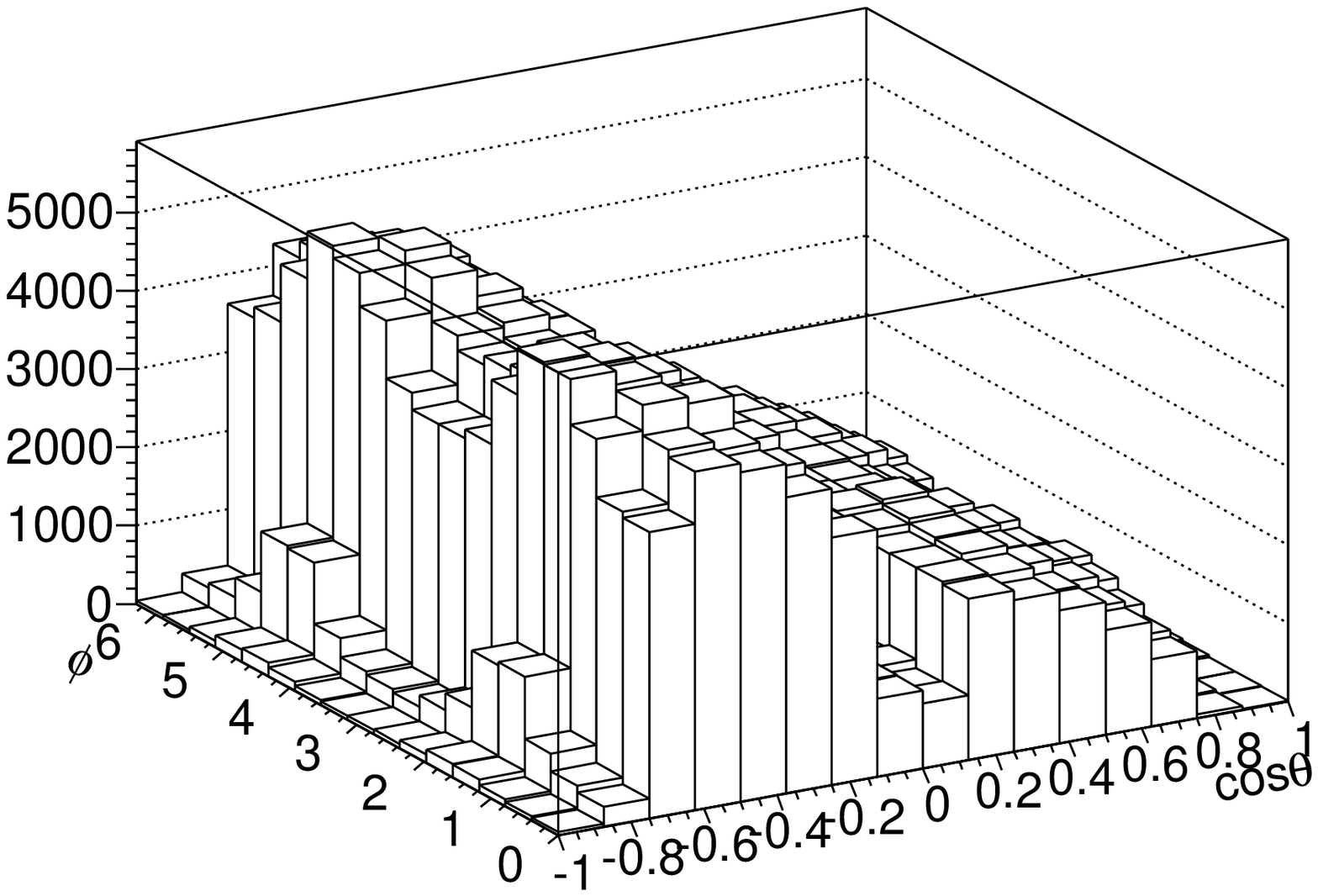}
}
\end{center}
\caption{The 2D distribution of $\cos \theta$ and $\phi$ of charged lepton from $W^{-} \to \tau^{-} \nu$.
On left side distribution of the full phase-space, with generated neutrino momentum used, and  events  weighted $wt_{\Sigma AiPi}$. 
On right, the same distribution is shown, but:  $m_W^{PDG}$ is used for solving Eq.~(\ref{Eq:pznu}), randomly one of the 
solutions for $p_z^{\nu}$ is taken and fiducial selection is applied. The  weight $wt_{\Sigma AiPi}$ is 
calculated with generated neutrino momenta, as it should be.
\label{Fig:Wm_2Dclo} }
  \begin{center}                               
{
   \includegraphics[width=5.0cm,angle=0]{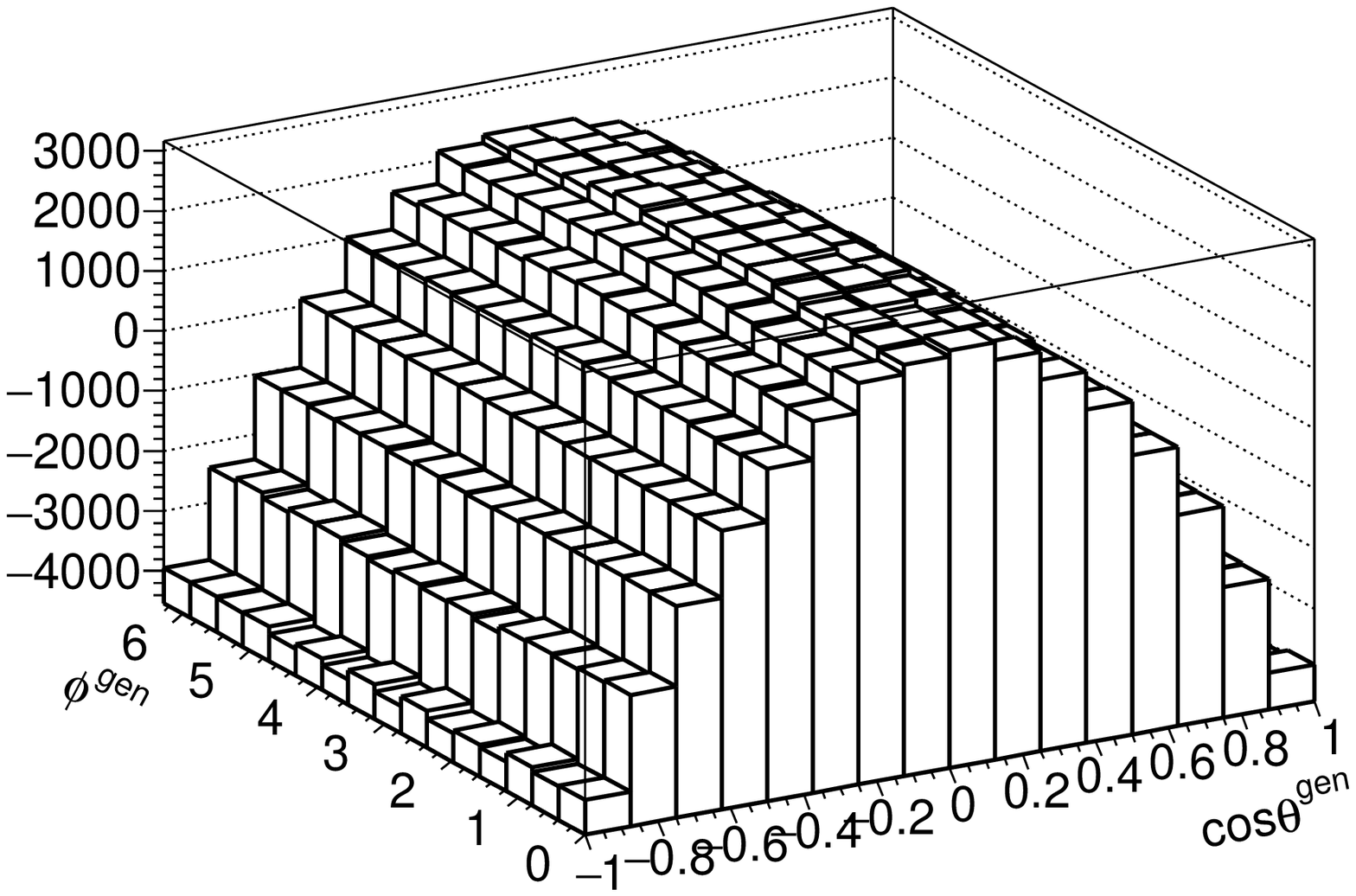}
   \includegraphics[width=5.0cm,angle=0]{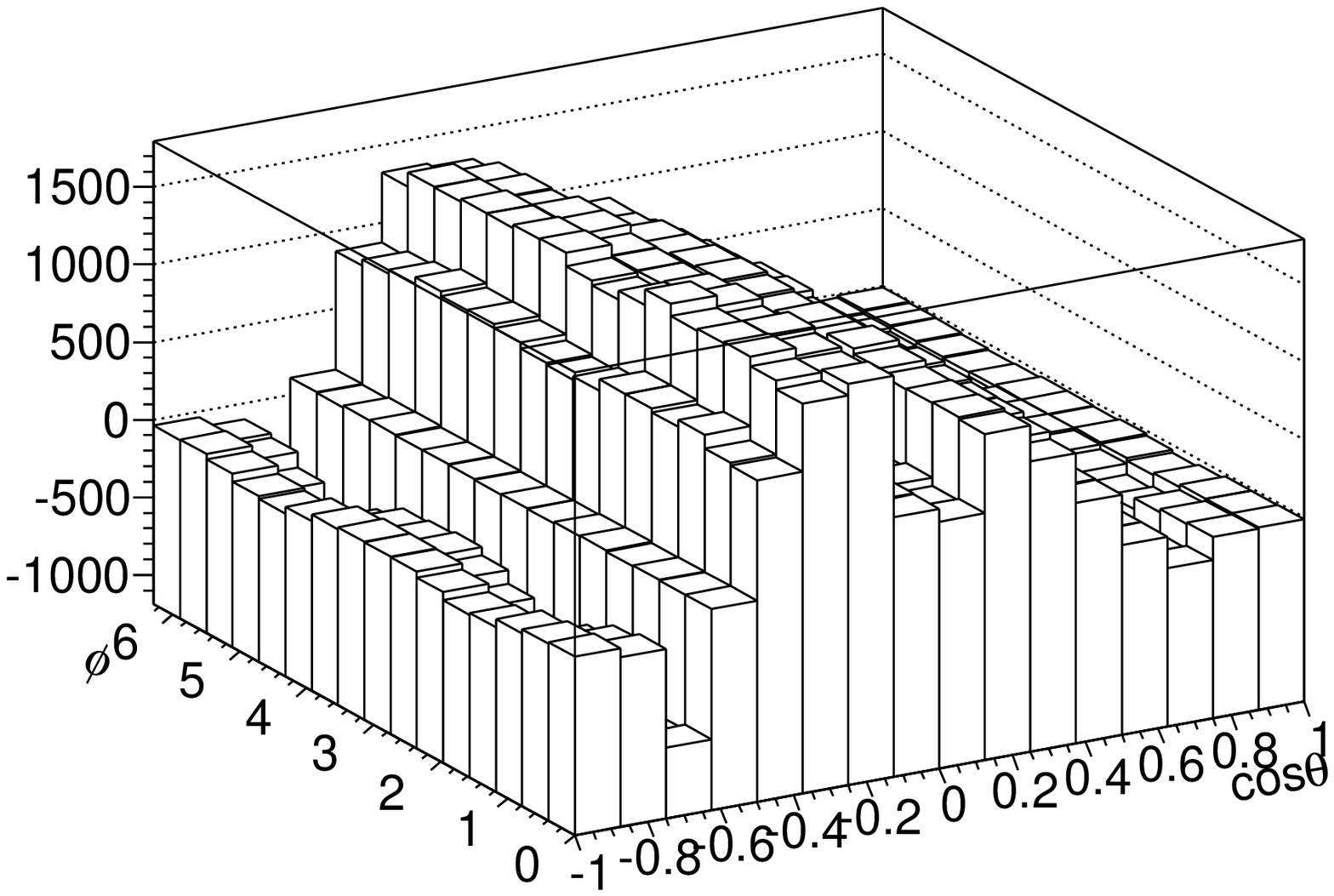}
   \includegraphics[width=5.0cm,angle=0]{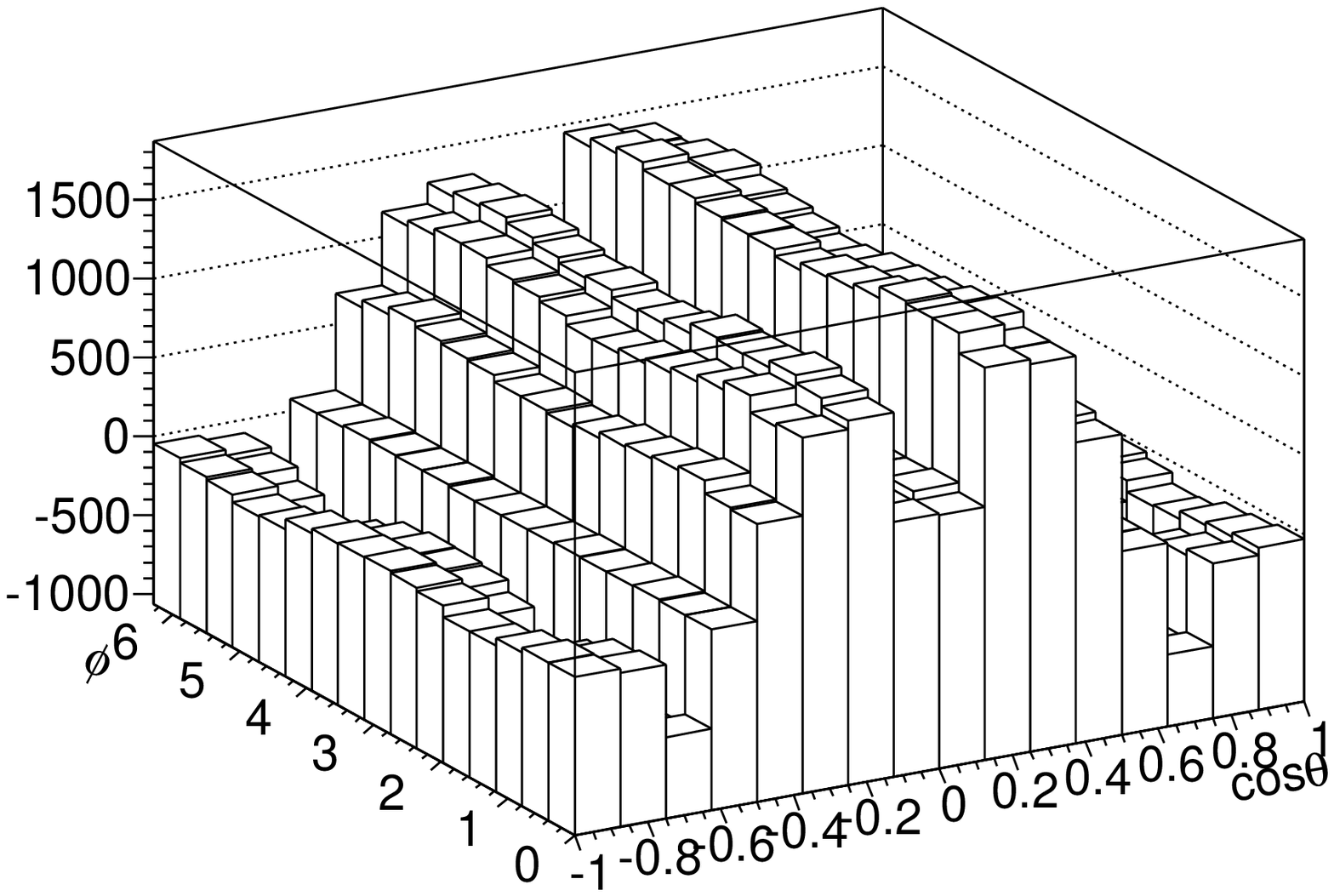}
   \includegraphics[width=5.0cm,angle=0]{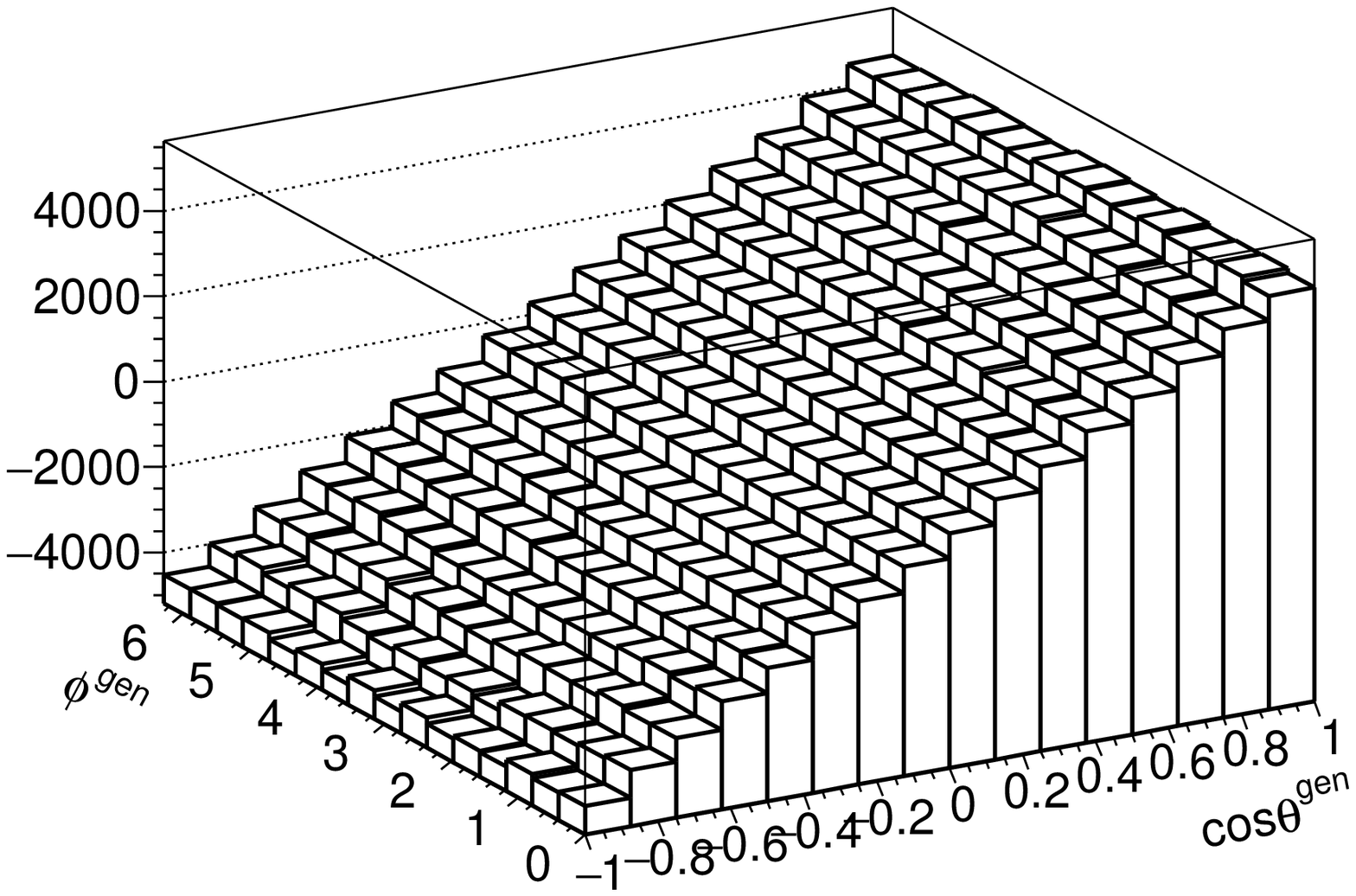}
   \includegraphics[width=5.0cm,angle=0]{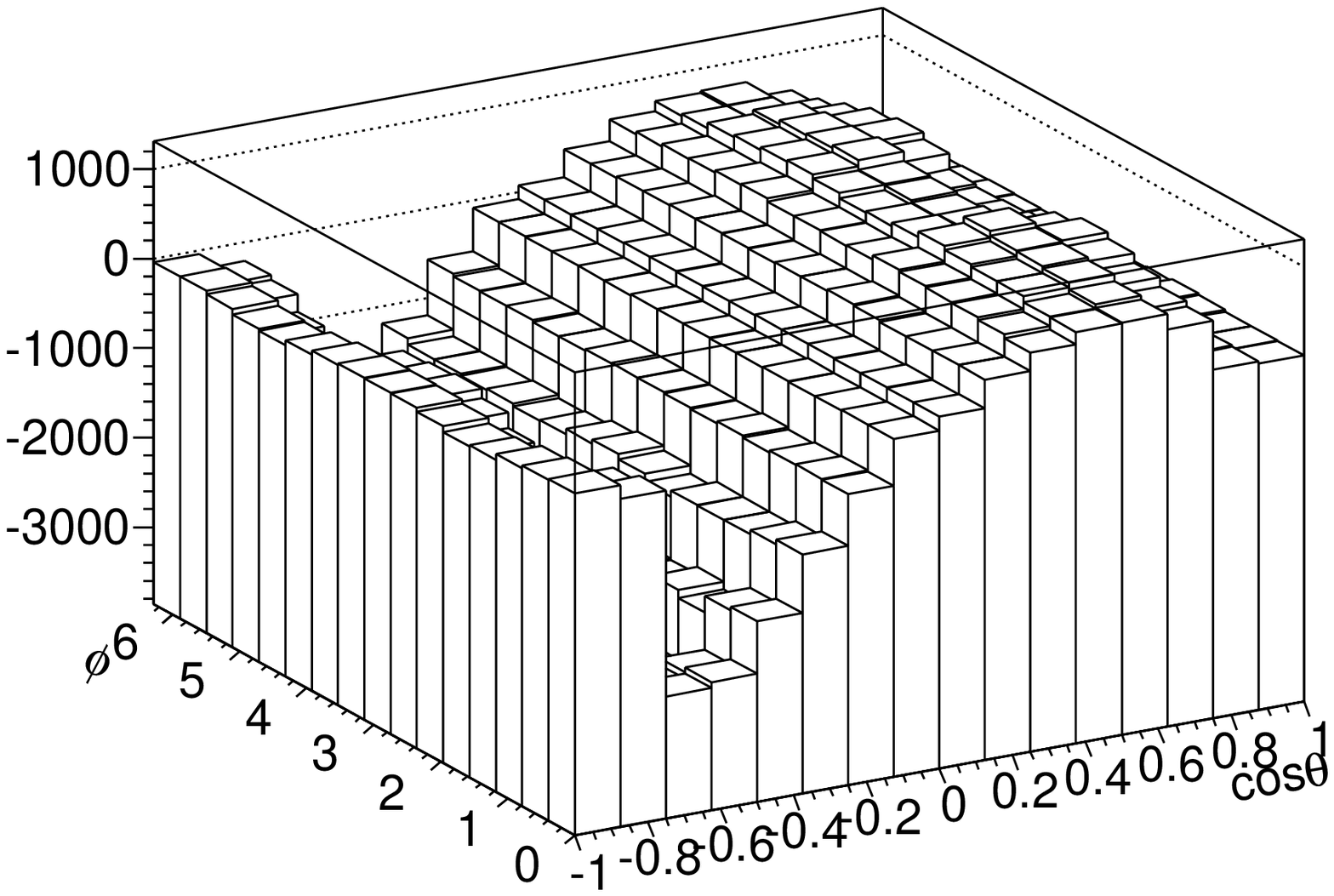}
   \includegraphics[width=5.0cm,angle=0]{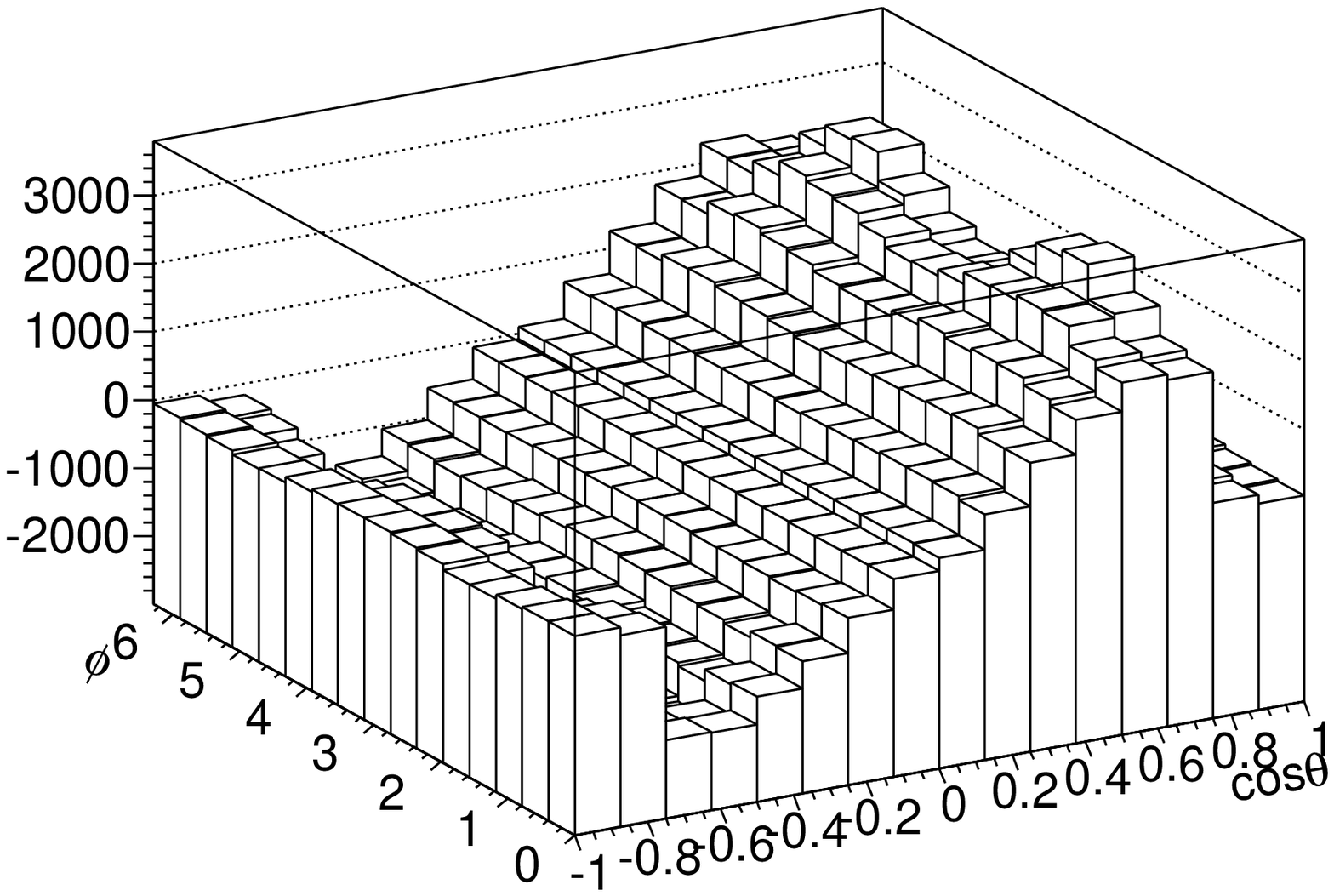}
}
\end{center}
\caption{Analytical shape of the polynomial $P_{0}$ (top) and  $P_{4}$ (bottom) in the full phase-space (left) and templates for polynomials
after reconstructing $p_Z^{\nu}$ and  fiducial selection for: $W^{-}$ (middle) and  $W^{+}$ (right). 
\label{Fig:Wmp_2Dtempl} }
\end{figure}

We extract coefficients $A_i$ using generated neutrino momenta to calculate $\cos \theta$ and $ \phi$. 
As a technical test, we histogram 2D distribution in $(\cos \theta, \phi)$ using
our events weighted with 
\begin{equation}
wt_{\Sigma AiPi} = \frac{1}{\Sigma_{i=0}^{i=8} A_i P_i (\cos\theta,\phi)}
\end{equation}
where $A_8 = 1.0$ and $P_8 = 1 + \cos^2 \theta$. 
We obtain a completely flat distribution in $(\cos \theta, \phi)$, where $\theta, \phi $ 
are calculated using the generated neutrino momentum, see left plot of Figure~\ref{Fig:Wm_2Dclo}. 
This completes our technical test and we can continue the construction of templates. 

We fold now
events weighted with $wt_{\Sigma AiPi}$ into fiducial phase-space of the measurement:  
for the neutrino momentum reconstruction we use $m_W = m_W^{PDG}$ and take 
one of the solutions at random, then we recalculate  $\theta, \phi$ angles 
and  finally we apply the kinematical selection of the fiducial phase-space. 
Right plot of Figure~\ref{Fig:Wm_2Dclo} shows how the initially flat 
distribution is distorted by this folding procedure.

We can now model any desired analytical polynomial shape of the generated full phase-space 
folded into fiducial phase-space of experimental measurement.
It is  enough to apply  $wt_i = P_i \cdot wt_{\Sigma AiPi}$ 
to our events, to model the shape of the  $P_i(\cos\theta,\phi)$ polynomial in the measurement 
fiducial phase-space. In Figure~\ref{Fig:Wmp_2Dtempl}, we show 2D distributions 
modeling polynomials $P_0(\cos\theta,\phi)$ and  $P_4(\cos\theta,\phi)$ in the full and fiducial phase-space as an example. 
Distributions for the remaining ones are shown in the Appendix~\ref{App:Ais}.

\begin{figure}
  \begin{center}                               
{
   \includegraphics[width=6.0cm,angle=0]{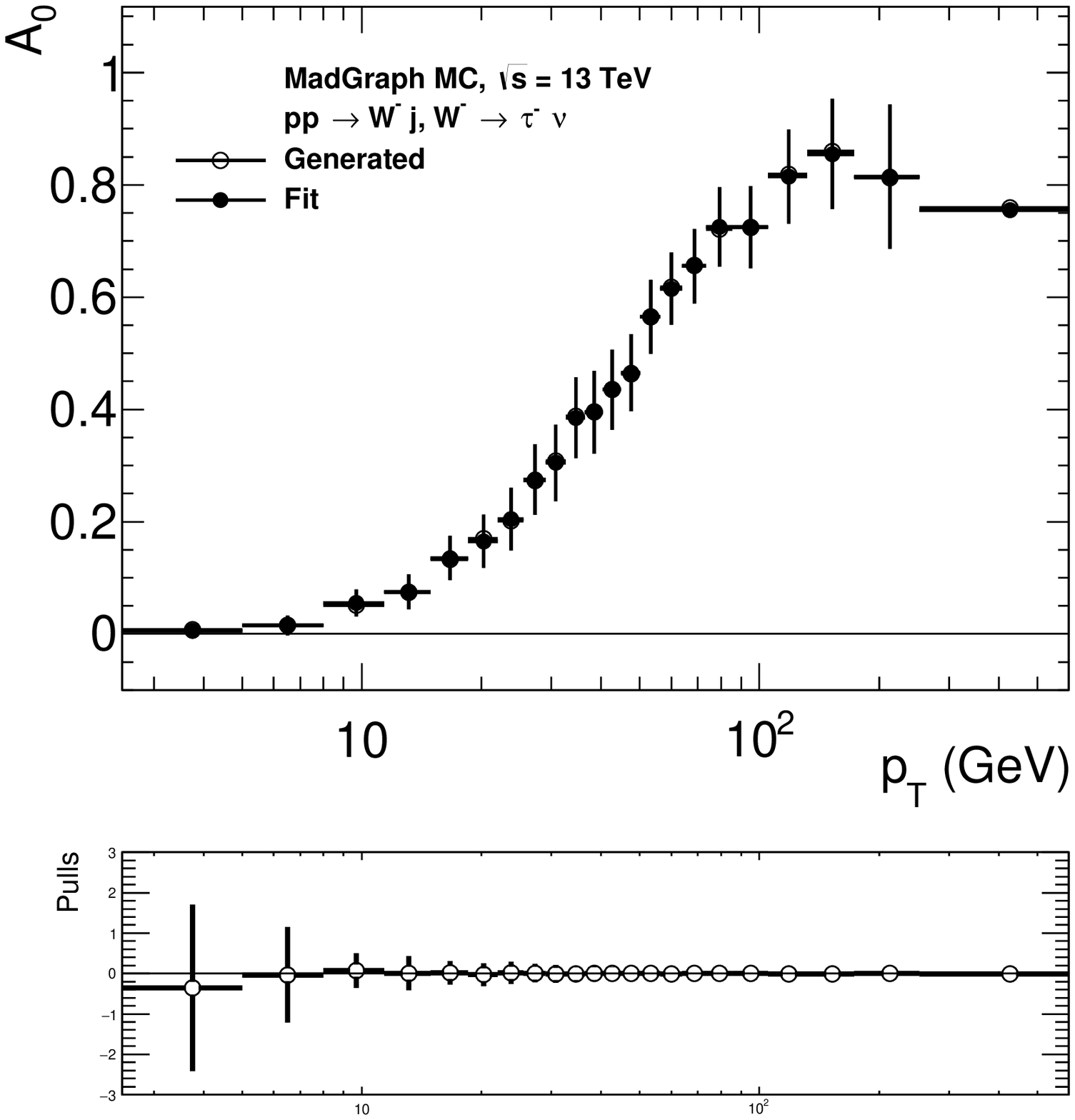}
   \includegraphics[width=6.0cm,angle=0]{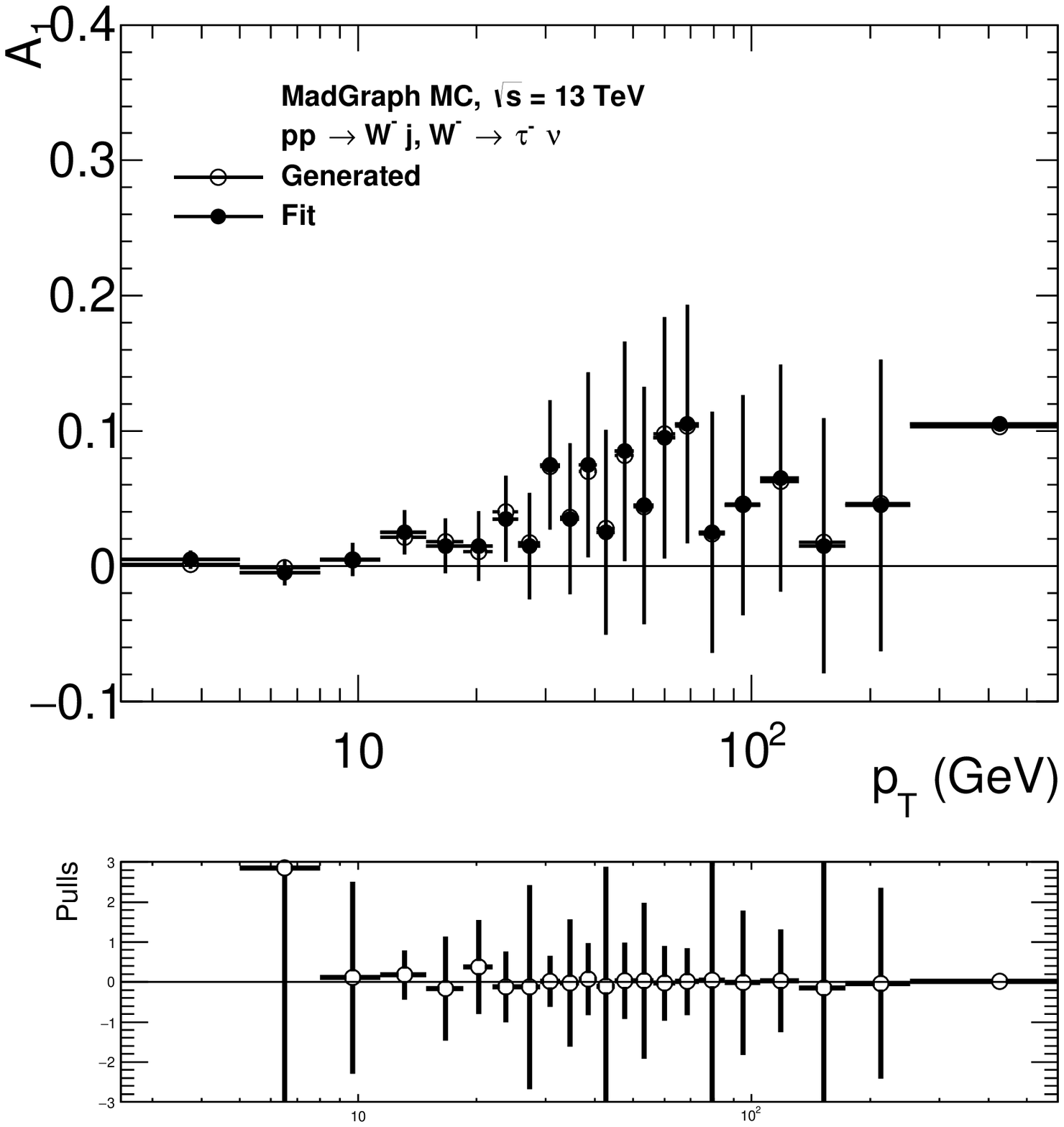}
   \includegraphics[width=6.0cm,angle=0]{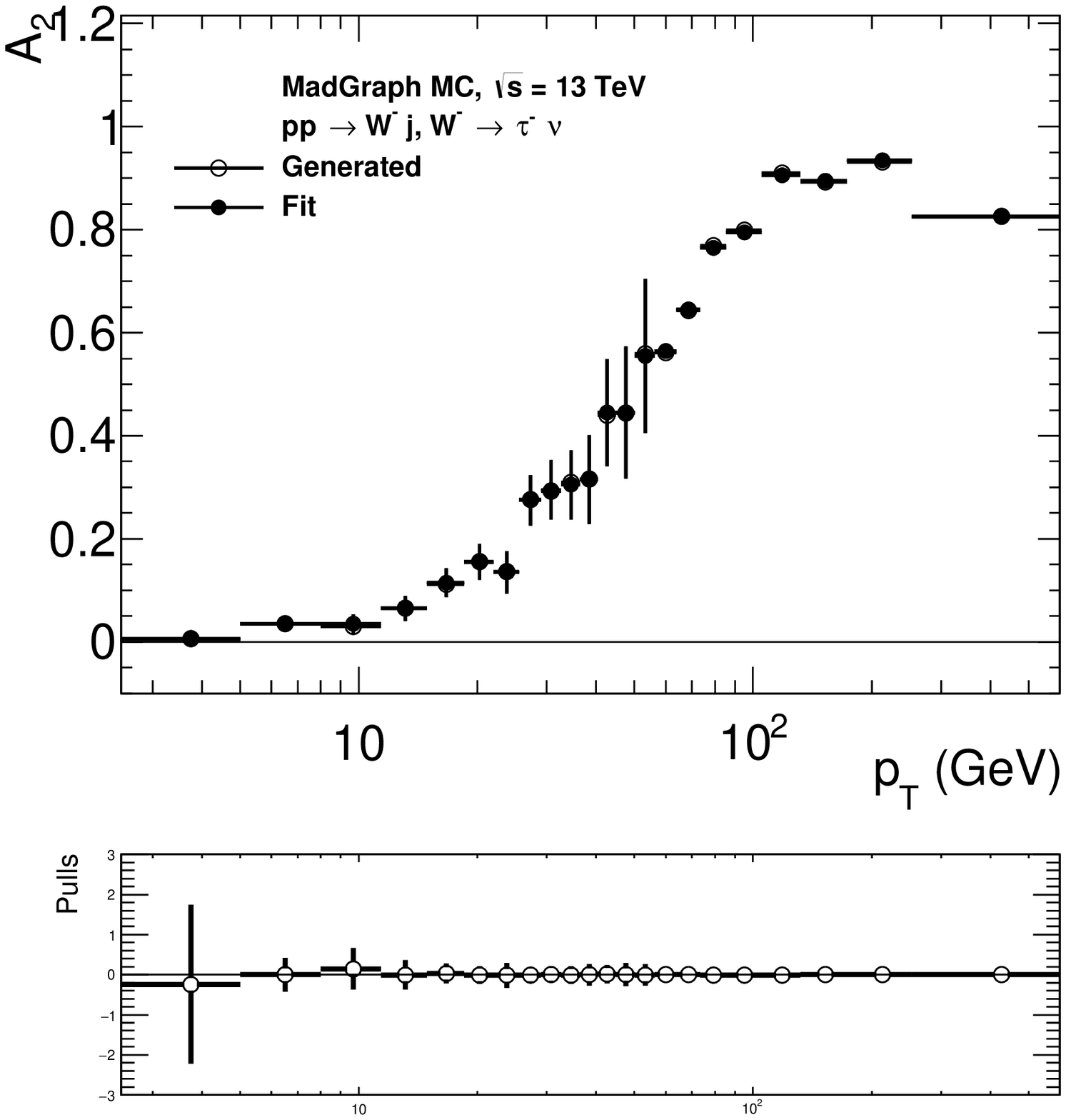}
   \includegraphics[width=6.0cm,angle=0]{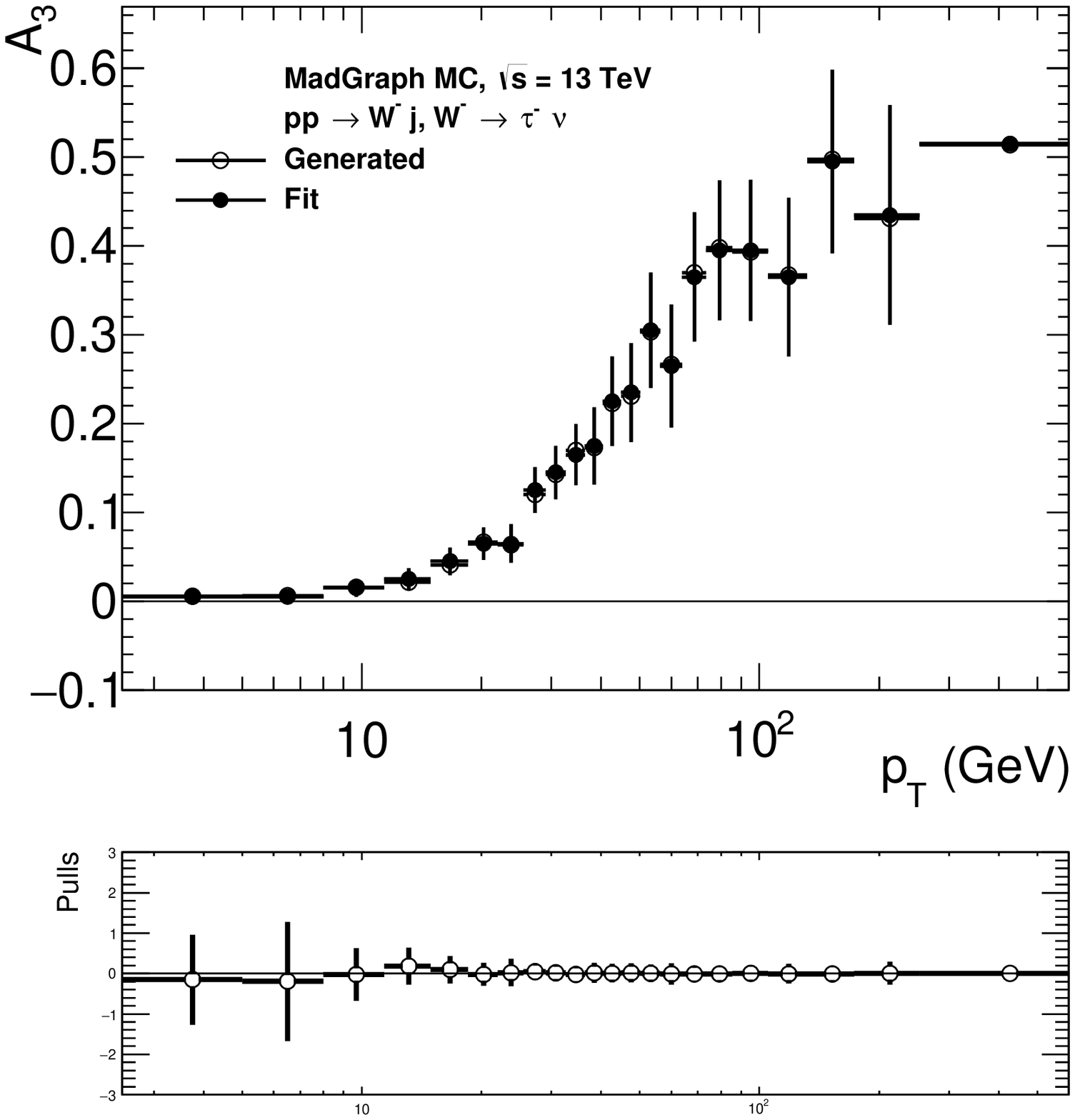}
   \includegraphics[width=6.0cm,angle=0]{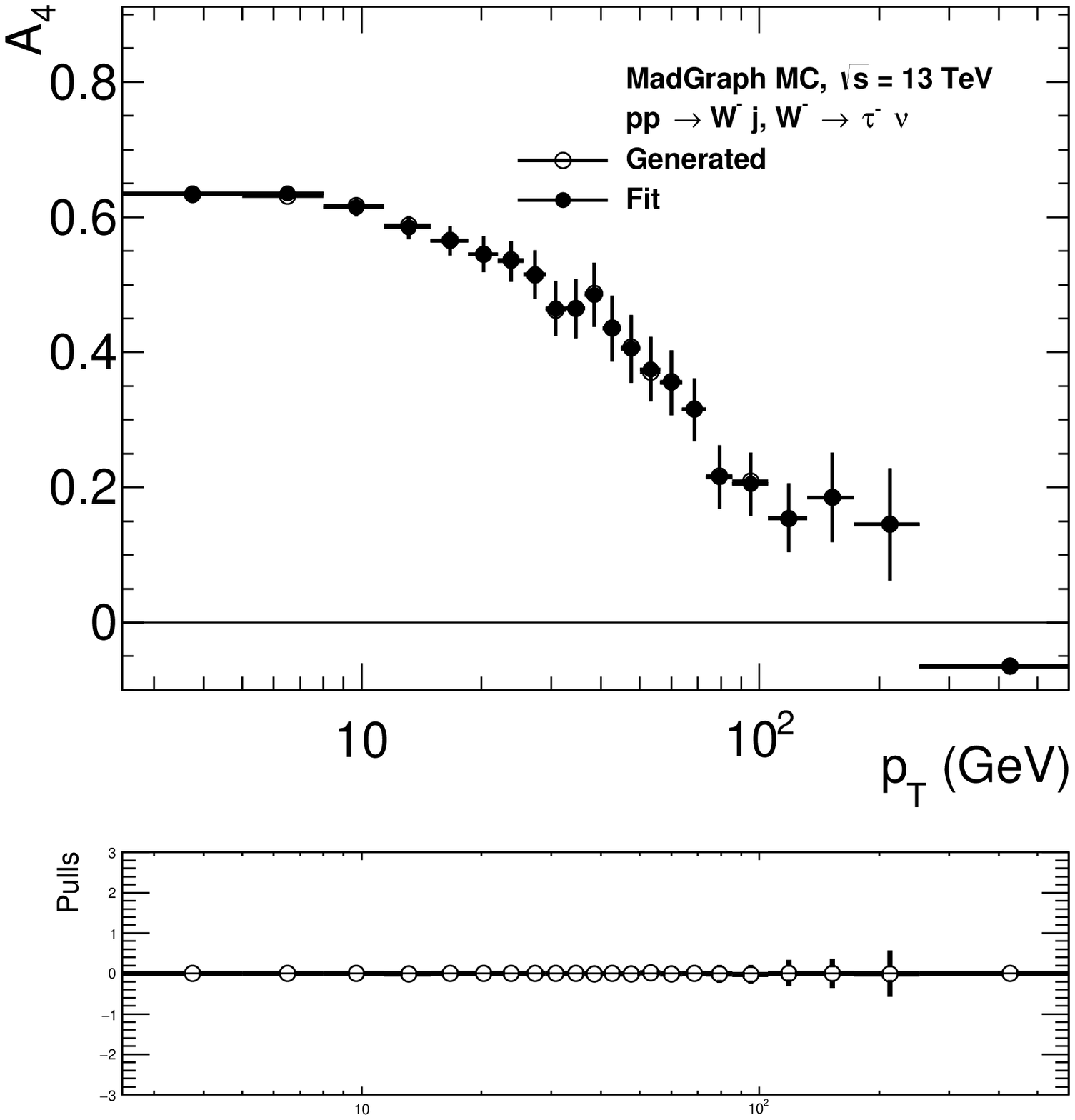}
}
\end{center}
\caption{Closure test on the fitting of angular coefficients $A_i$'s for  $W^- \to \ell^- \nu$. Fit is performed in the fiducial phase-space. 
Shown are generated $A_i$'s coefficients (open circles) and their fitted values (black points). In the bottom panels shown are pulls 
(difference between generated and fit value, divided by the statistical error of the fit). 
Pulls are smaller than one could expect. This is because events of pseudodata and templates are statistically correlated.
\label{Fig:WmLLfits} }
\end{figure}

We can now proceed with the fit of a {\it linear} combination of templates to distributions 
of the fiducial phase-space pseudo-data. We bin in $p_T^{W}$ both  templates shown in Fig.~\ref{Fig:Wmp_2Dtempl}, ~\ref{FigApp:Wmp_2Dtempl}
and pseudo-data distributions shown in Fig.~\ref{Fig:2DWm_cosThetaCSphiCS}.  
We perform a multi-parameter log-likelihood fit in each $p_T^W$ bin; parameters of the fit are the angular coefficients $A_i(p_T^W)$.
Results of the fitting procedure are shown in Figure~\ref{Fig:WmLLfits}. The black points represent fitted 
values of $A_i$'s with their fit error, black open circles are the generated values of the $A_i$'s (which we extracted with moments method 
described above). Bottom panels show difference between fitted and true values divided by their errors (so called pulls distributions). 
Pulls are small because of the samples correlations. We confirm {\it closure} of 
the method, i.e. extracted coefficients are equal to their nominal value for analysed events sample. 
The same procedure has been repeated for $W^+\to \ell^+ \nu$ and results are shown in Appendix~\ref{App:Ais}.

We have also performed the fit using templates and pseudo-data distributions prepared with only {\it correct} or only 
{\it wrong} solutions for the neutrino momenta\footnote{With experimental data one can use {\it random} solution for $p_z^{\nu}$ only,
{\it correct} or {\it wrong}  option is  for technical tests.}. In both cases the fit returned nominal values of the $A_i$ coefficients,
just confirming that both solutions of neutrino momenta carry the same information on the angular correlations.  

In this {\it proof of concept} for the proposed measurement strategy we have not discussed possible experimental effects like resolution of the 
missing transverse energy which will be used to reconstruct neutrino transverse momenta or e.g. background subtraction which can
limit precision of the  measurement. 
\begin{itemize}
\item
We have shown, that despite the neutrino escaping detection, one can define under some assumptions 
the equivalent to the rest frame of lepton-neutrino system and preserve sensitivity for the
complete set of angular coefficients of the decomposition. 
\item
Then, we have shown that simplified version of the method used  in~\cite{Aad:2016izn} to measure 
$A_i$ coefficients in case 
of $Z \to \ell \ell$ decays, can  be applied in case of $W \to \ell \nu$ decays 
and allows  for the measurement of 
a complete set of angular coefficients in this case as well. 
\end{itemize}

\section{\bf Angular coefficients and reference frames}
\label{sec:frames}

In this paper, we  concentrate on the numerical analysis of  tree level  parton-parton collisions 
into a lepton pair and accompanying jets, convoluted with parton distributions, but without parton showers.
Even though such  approach is limited, it provides input for general discussions.
Such configurations constitute parts of the higher order corrections, or can be seen as the lowest order terms
but for observables of tagged high $p_T$ jets. 

For the choices of the reference frames to be discussed here, let us point out that 
in the limit of zero transverse momenta, all coefficients
except  $A_4$ vanish. 
The $(p_T, Y)$ dependence of the $A_i$ coefficients differs with the choice of the reference frame. 

So far, we have introduced and discussed angular coefficients in the Collins-Soper frame only. 
Let us present now the variant of the reference frame definition we are also going to use, i.e. the {\tt Mustraal} frame.

\subsection{The {\tt Mustraal} reference frame}

The {\tt Mustraal} reference frame is also defined as a rest frame of the lepton pair. 
It has been proposed and used for the first time in the {\tt Mustraal} Monte Carlo program~\cite{Berends:1983mi} 
for the parametrization of the phase space for muon pair production at LEP. 
The resulting  optimal frame was minimising higher order corrections from initial state radiation to the 
$e^+ e^- \to Z/\gamma^* \to f \bar f$  and was used very successfully for the algorithms implementing genuine
weak effects in the LEP era Monte Carlo program {\tt KORALZ} \cite{koralz4:1994}. A slightly different variant
was successfully used in the {\tt Photos} Monte Carlo program ~\cite{Davidson:2010ew} for simulating QED radiation in decays 
of particles and resonances. The parametrization was useful not only for compact representation of single photon emissions
but for multi-emission configurations as well.

Recently in \cite{Richter-Was:2016mal}, the implementation of the {\tt Mustraal} frame has been extended to the case of $pp$ collisions 
and studied for configurations with one or two partons in the final state accompanying  Drell-Yan production of the lepton pairs. 
The details of the implementation of this phase space parametrization have been discussed in context of $Z \to \ell \ell$ events
and the complete algorithm how to calculate $\cos \theta$ and $\phi$ angles was given.
There is no need to repeat it here. 

Let us point out that unlike the case of the Collins-Soper frame, the {\tt Mustraal} frame requires not only information on
4-momenta of outgoing  leptons but also on outgoing jets (partons). The information on jets (partons), is used to approximate 
the directions and energies of incoming partons for the calculation of weights (probabilities) with which each event contributes to one 
of two possible Born-like configurations. Each configuration requires different $\cos \theta$, $\phi$ definition.
This does not have to be very precise but can introduce  additional experimental systematics, and requires attention.
No dependence on coupling constants or PDF's is introduced in this way.

\subsection{QCD and EW structure of angular correlations} \label{sec:QCD}

The measurement of the angular distribution of leptons from the decay of a gauge boson $V \to \ell \ell$ 
where $V = W, Z$ or $\gamma^*$, produced in  hadronic collisions via a 
Drell-Yan-type process $ h_1 + h_2 \to V + X$ provides a detailed test of the production mechanism, revealing
its QCD and EW structure. 

The predictive power of QCD is based on the factorisation theorem \cite{Collins:1989gx}.
It provides a framework for separating out long-distance effects in 
hadronic collisions. In consequence, it allows for a systematic prescriptions and provides tools to calculate the short-distance
dynamics perturbatively, at the same time allowing for the identification of the leading nonperturbative 
long-distance effects which can be extracted from experimental measurements or from numerical calculations 
of Lattice QCD.   

The question of the input from the Electroweak sector of the Standard Model is important, 
especially for distributions of leptons originating from the intermediate $Z/\gamma^*$ state. We have addressed numerical 
consequences of this point recently in \cite{Kalinowski:2016qcd} in the 
context of $\tau$ lepton polarization in Drell-Yan processes at the LHC.
A wealth of publications was devoted during last years to this issue, see e.g.~\cite{Barze':2013yca,Dittmaier:2014qza,Dittmaier:2015rxo}. 
 We should underline 
limitations of  separating  interactions  into Electroweak and QCD part. Limitations 
are well known,  since more than 15 years now, see e.g.~\cite{Kulesza:1999gm}.

Let us come back now to Eq.~(\ref{Eq:master1}) and (\ref{Eq:master2}) and discuss the structure of cross-section decomposition into 
harmonic polynomials multiplied by angular coefficients. The $A_i$ coefficients represent ratios of the helicity cross-sections and 
following the conventions and notations of \cite{Mirkes:1992hu,Mirkes:1994eb}, the following 
coefficients constructed from couplings, appear 
in $A_i$'s:

\begin{eqnarray}
\sigma ^{U+L }  & \sim & (v_{\ell}^2 + a_{\ell}^2)(v_{\ell}^2 + a_{q}^2), \nonumber \\ 
A_0, A_1, A_2  & \sim & 1, \nonumber \\ 
A_3, A_4       & \sim & \frac{ v_{\ell} a_{\ell} v_q a_q}{(v_{\ell}^2 + a_{\ell}^2)(v_{\ell}^2 + a_{q}^2)},  \label{eq:couplings}\\ 
A_5, A_6       & \sim & \frac{(v_{\ell}^2 + a_{\ell}^2) ( v_q  a_q)}{(v_{\ell}^2 + a_{\ell}^2)(v_{\ell}^2 + a_{q}^2)}, \nonumber \\ 
A_7            & \sim & \frac{ v_{\ell} a_{\ell} ( v_q^2 + a_q^2)}{(v_{\ell}^2 + a_{\ell}^2)(v_{\ell}^2 + a_{q}^2)}. \nonumber
\end{eqnarray}
where $v_i, a_i, i=q,\ell$ denote vector and axial couplings of intermediate boson to quarks and leptons.

In case of $W$ boson the EW sector at leading order is simply a $(V -A)$ coupling only.
At higher order and higher $p_T^W$ the more complicated structure, and of more interesting nature of 
the multi-boson couplings, if such is present, may be revealed.
In case of the $Z/\gamma^*$, the sensitivity to the EW sector is much richer from the physics point of view, in particular 
for $A_3$ and $A_4$ coefficients, and we have discussed it recently in \cite{Richter-Was:2016mal, Kalinowski:2016qcd}.

\section{Numerical results for Collins-Soper and {\it Mustraal} frames} \label{sec:Numerical}

Let us now present numerical results for the angular coefficients $A_i$ and compare predictions in the Collins-Soper 
and {\tt Mustraal} frame for $W^-$ production. Most of results for $W^+$ are delegated to Appendix~\ref{App:Ais}.  

\subsection{Results with LO simulation}
We use samples of events generated with the 
{\tt MadGraph5\_aMC@NLO} Monte Carlo~\cite{Alwall:2014hca} for Drell-Yan production of $W+1j$ with $W \to \tau \nu$ 
and 13 TeV $pp$ collisions. Lowest order spin amplitudes are used in this program for the parton level process. 
To better populate higher $p_T^{W}$ bins we merged
(adjusting properly for relative normalization)
 3 samples, 2M events each, generated with thresholds of 
$p_T^j > 1, 50, 100$ GeV respectively. The incoming partons distributed accordingly to PDFs (using CTEQ6L1 
PDFs \cite{Pumplin:2002vw} linked through LHAPDF v6 interface) remain precisely collinear to the beams. 
At this level, jet (j) denotes outgoing parton of unspecified flavour. 

\begin{figure}
  \begin{center}                               
{
   \includegraphics[width=7.5cm,angle=0]{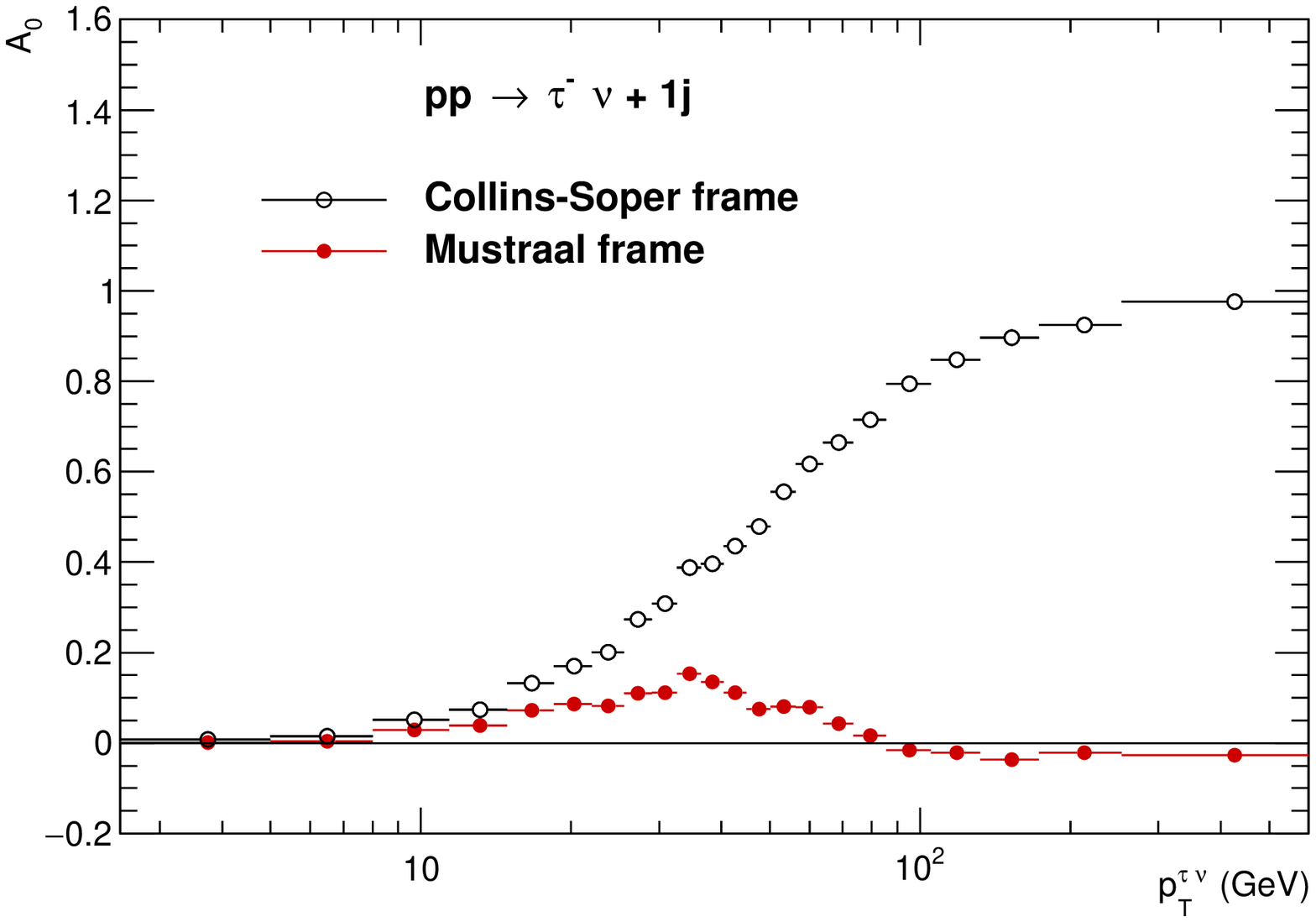}
   \includegraphics[width=7.5cm,angle=0]{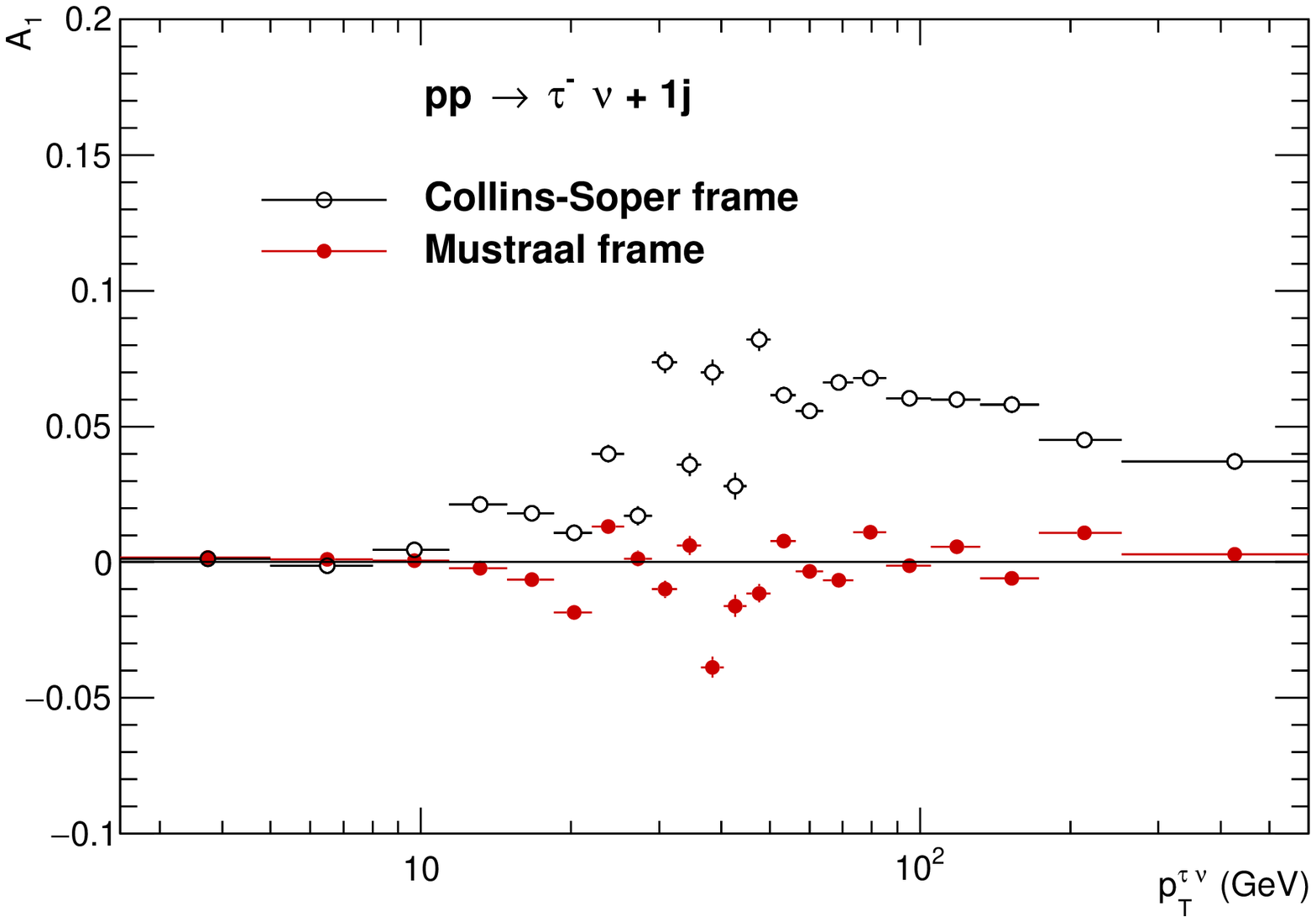}
   \includegraphics[width=7.5cm,angle=0]{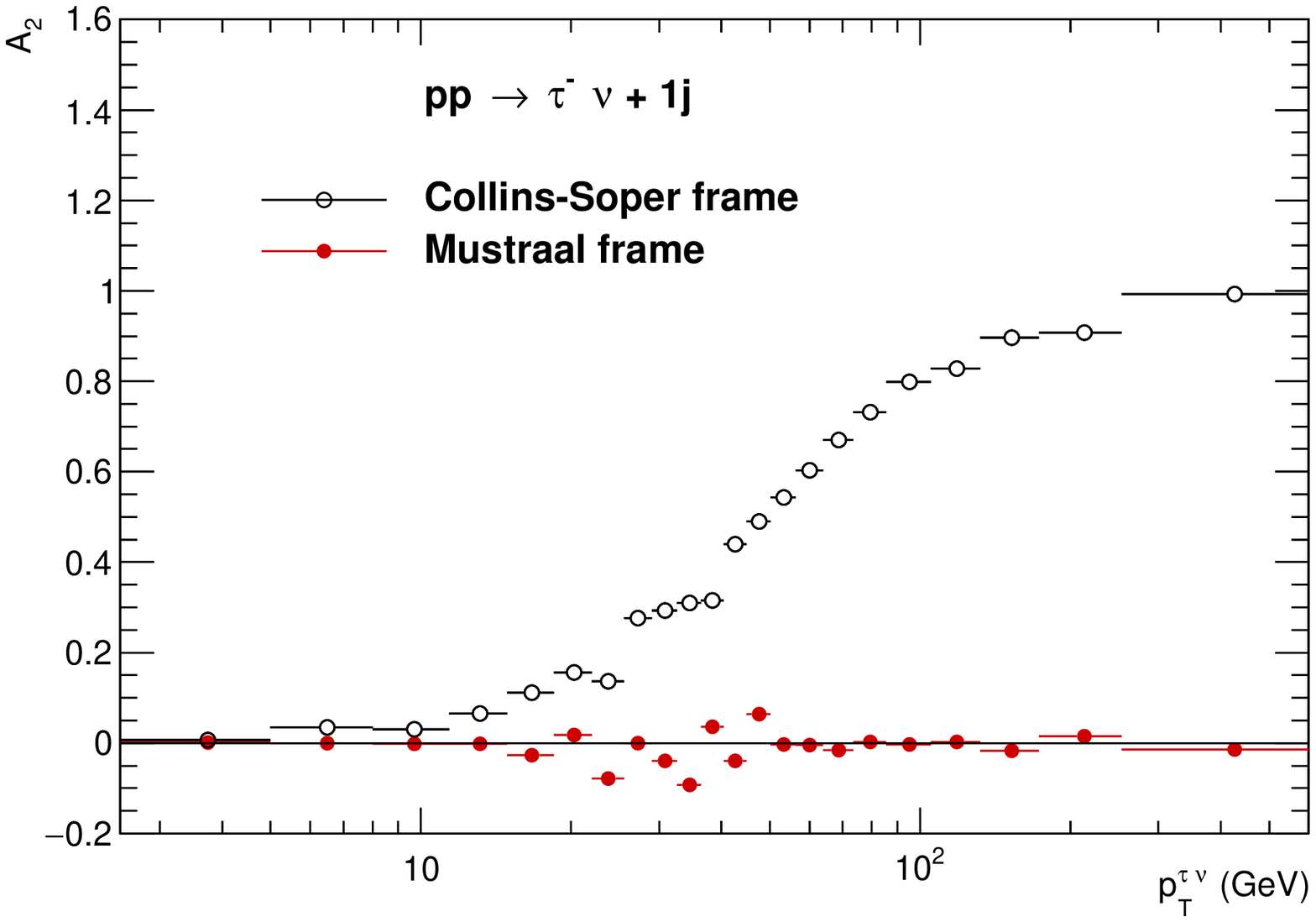}
   \includegraphics[width=7.5cm,angle=0]{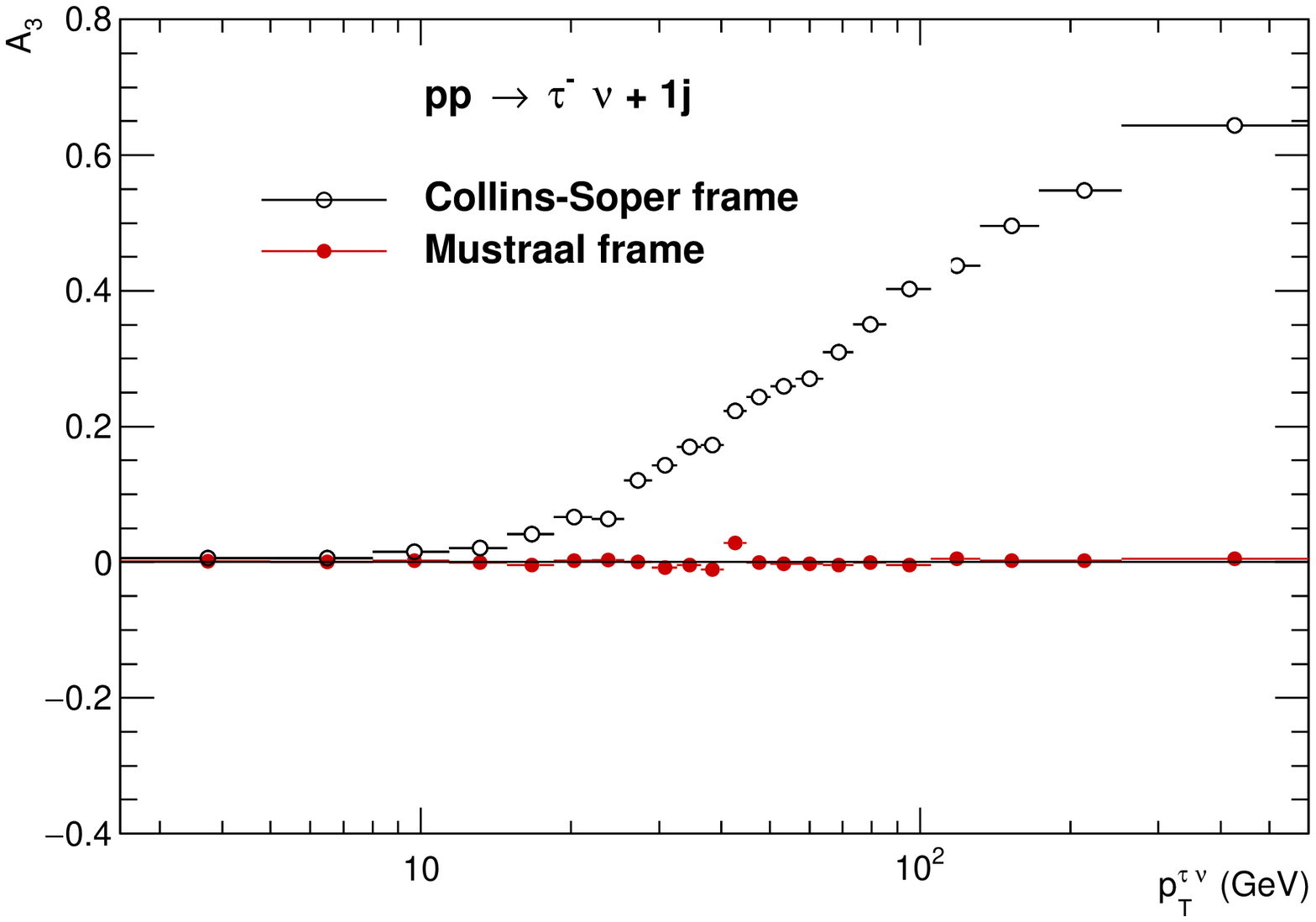}
   \includegraphics[width=7.5cm,angle=0]{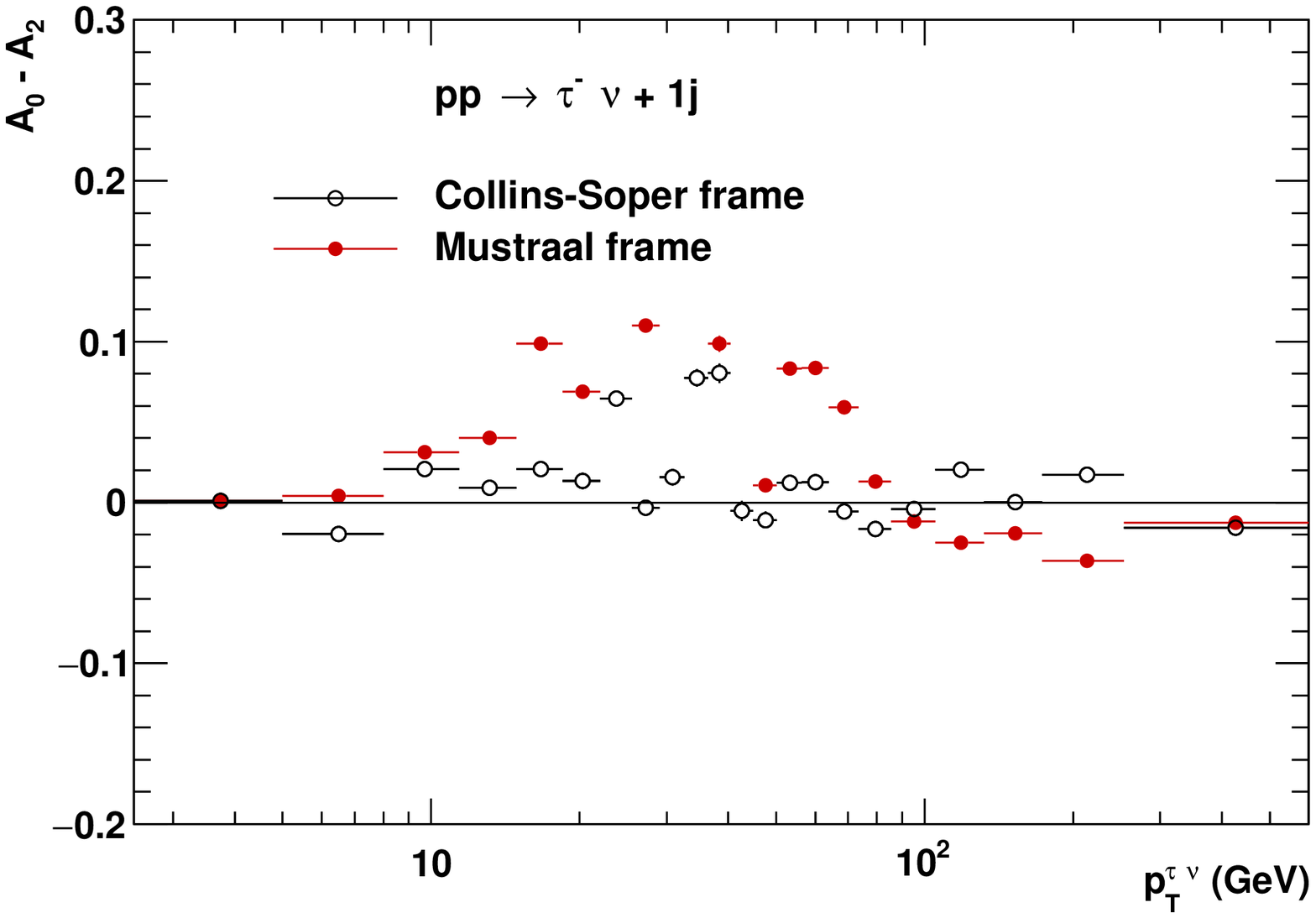}
   \includegraphics[width=7.5cm,angle=0]{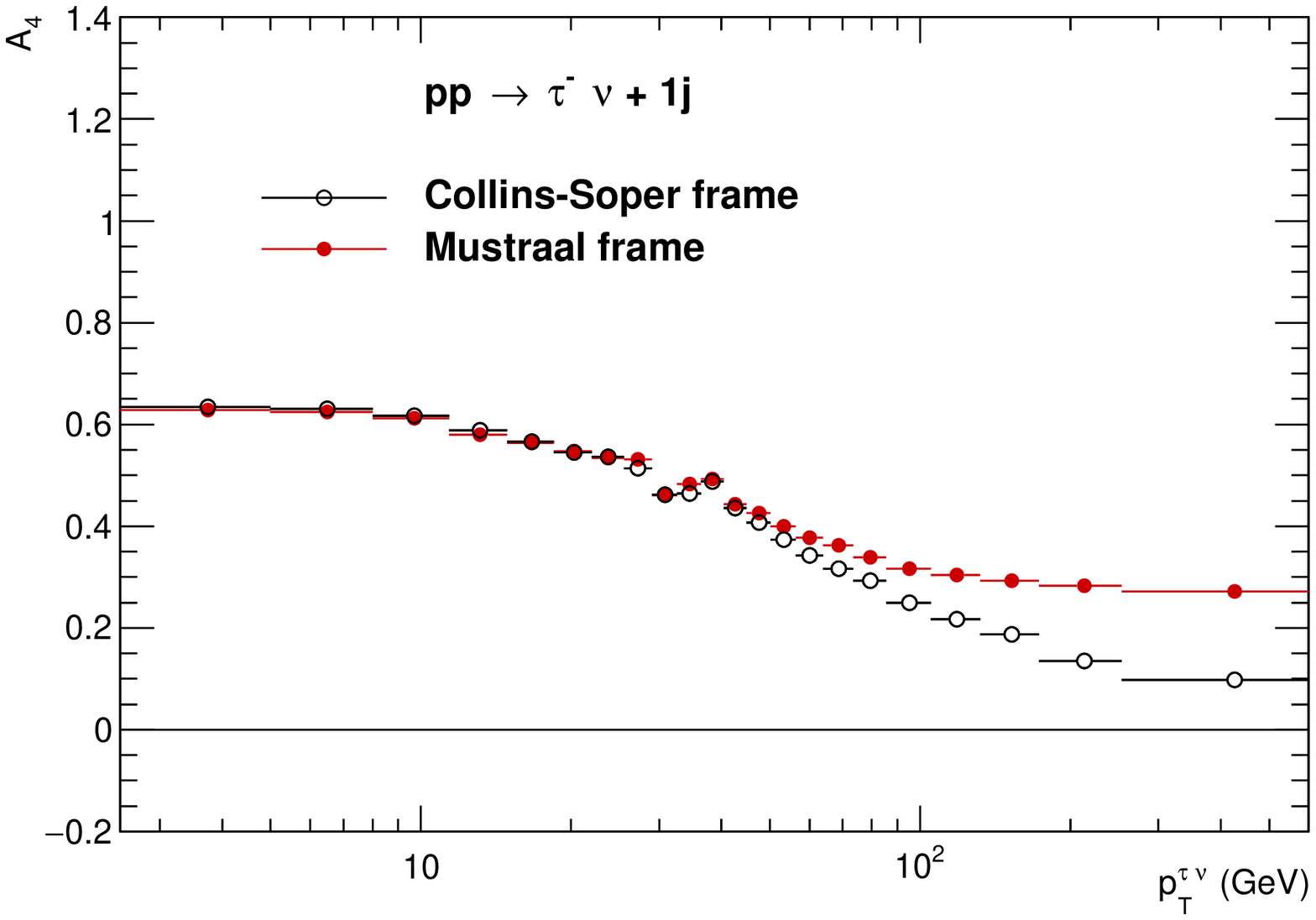}
}
\end{center}
\caption{The $A_i$ coefficients calculated in Collins-Soper (black) and in {\tt Mustraal} (red) frames 
for $p p  \to \tau^{-} \nu\ + 1j$ process generated with {\tt MadGraph}.
\label{Fig:Ai1jet} }
\end{figure}

Figure~\ref{Fig:Ai1jet}, collects results for angular coefficients $A_i$ of the processes with $W^- \to \tau^- \nu$ 
in the final state. We show sets of five angular coefficients $A_0 -A_4$ only; 
the remaining ones $A_5 -A_7$ are close to zero over the full $p_T^{W}$ range  for both definition of frames; 
Collins-Soper and {\tt Mustraal}. For the $A_3$ and $A_4$ coefficients we show their absolute value, as in the convention
we have adopted the signs depend on the sign of the charge of the $W$ boson (so the charge of lepton). 
In case of  $W^+$ production, both $A_3$ and $A_4$ are negative, see Appendix~\ref{App:Ais}.

In both frames, at low $p_T^W$ the only non-zero coefficient is $A_4$, and is of the same value. 
Similarly as we observed in case of Drell-Yan $Z \to \ell \ell$ process \cite{Richter-Was:2016mal}, the only  
significantly non-zero coefficient in the {\tt Mustraal} frame at higher $p_T^W$ remains $A_4$, 
while $A_0$, $A_2$ and $A_3$ are rising steeply for higher $p_T^W$ in the Collins-Soper frame. 

\begin{figure}
  \begin{center}                               
{
   \includegraphics[width=7.5cm,angle=0]{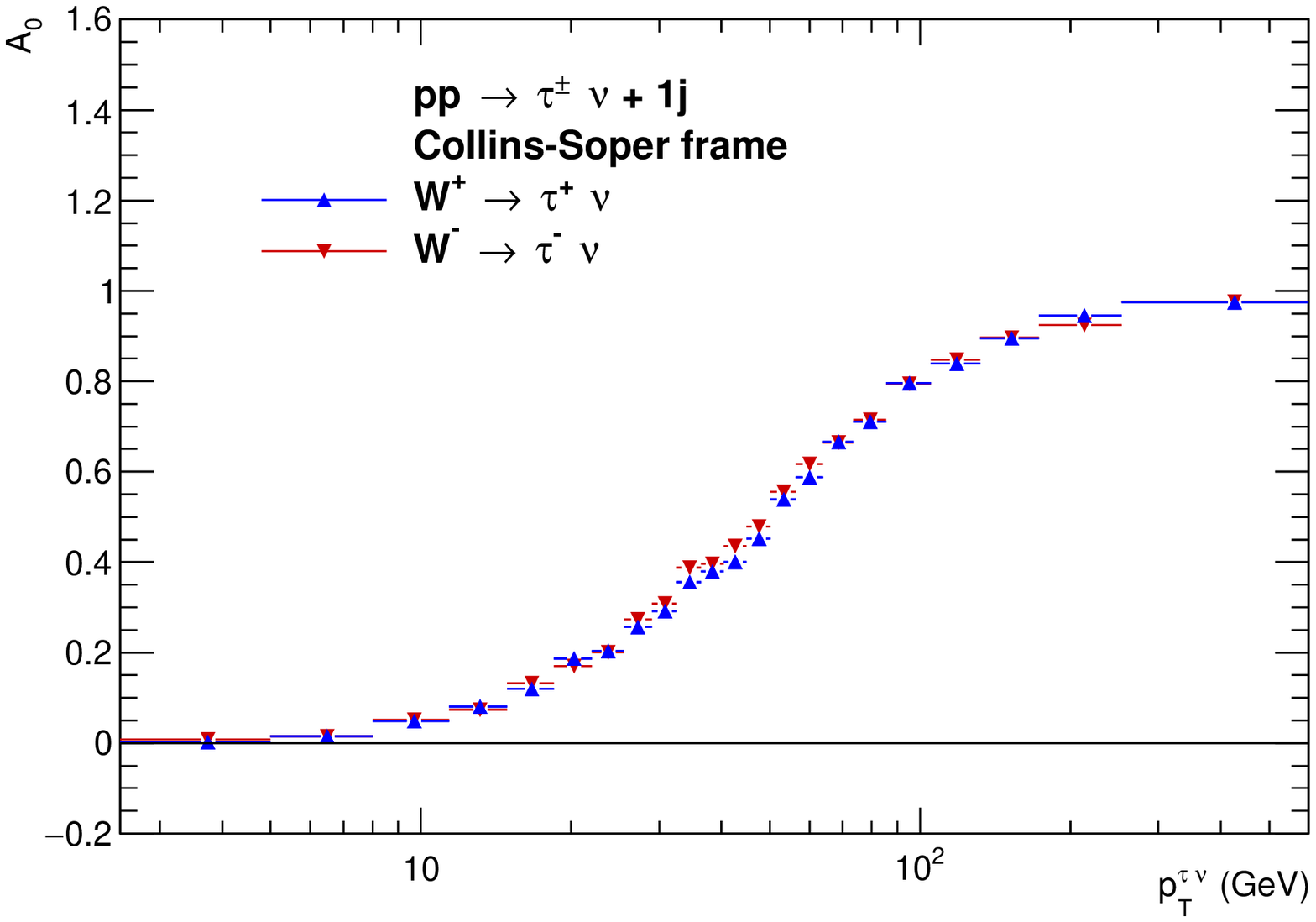}
   \includegraphics[width=7.5cm,angle=0]{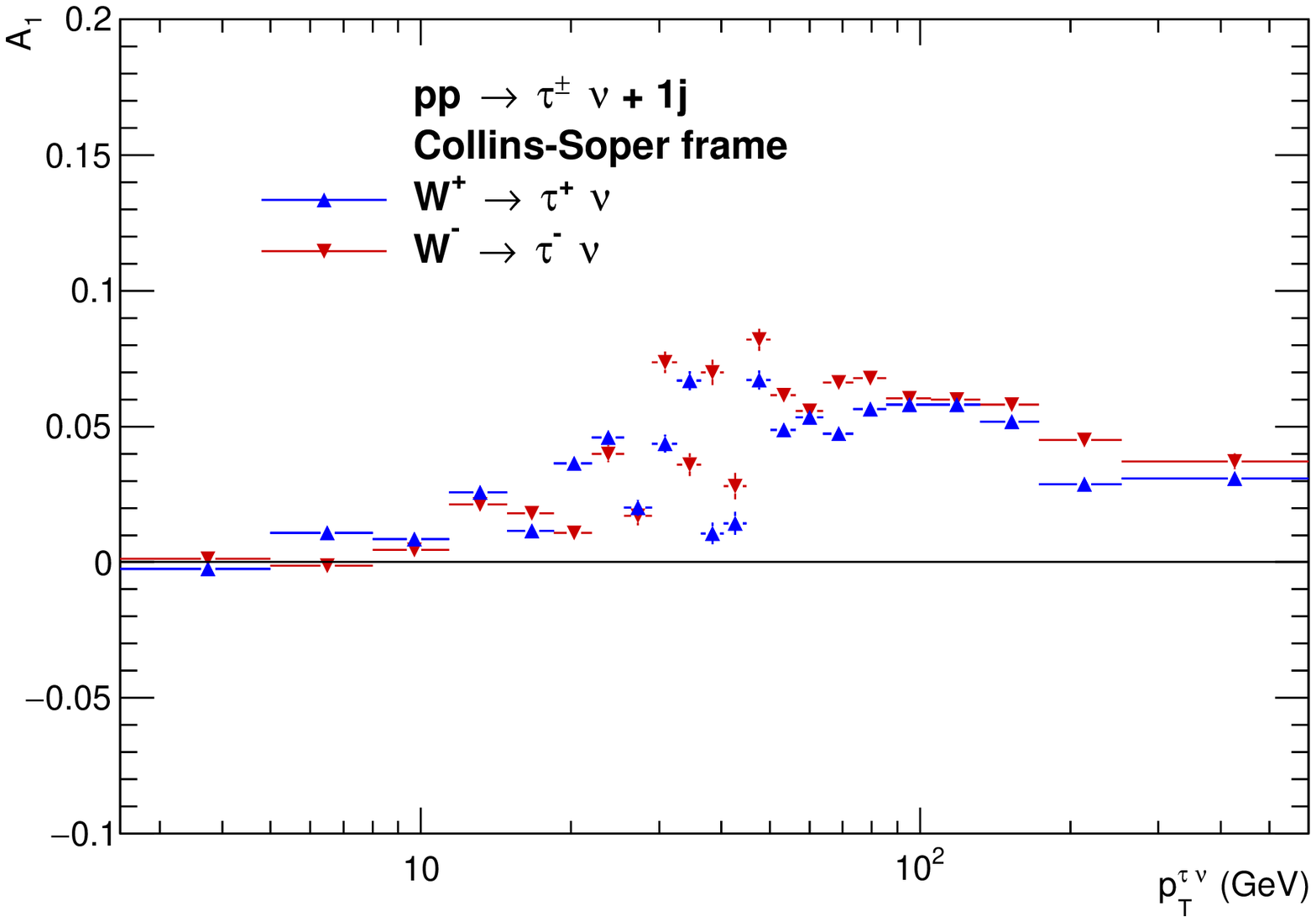}
   \includegraphics[width=7.5cm,angle=0]{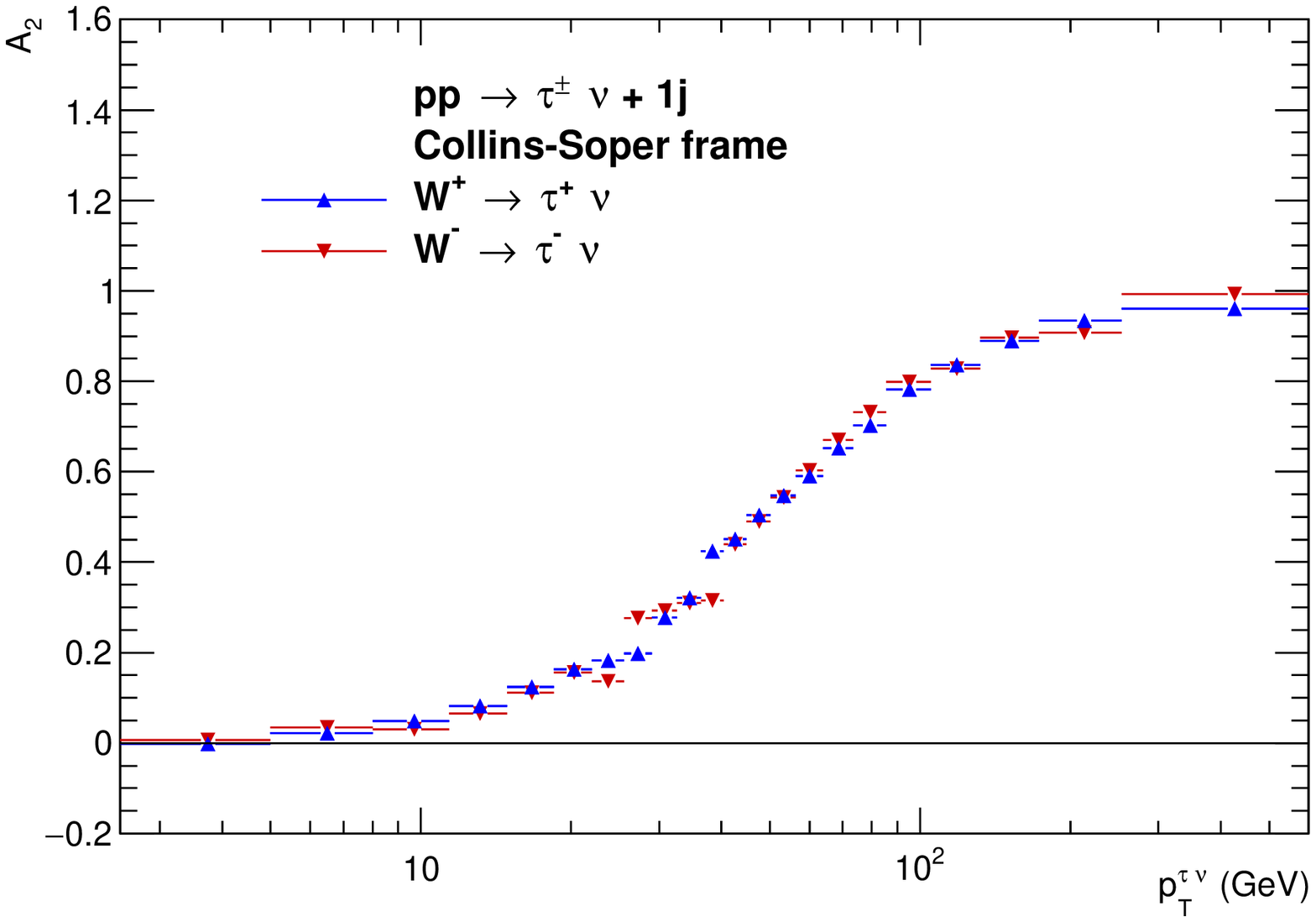}
   \includegraphics[width=7.5cm,angle=0]{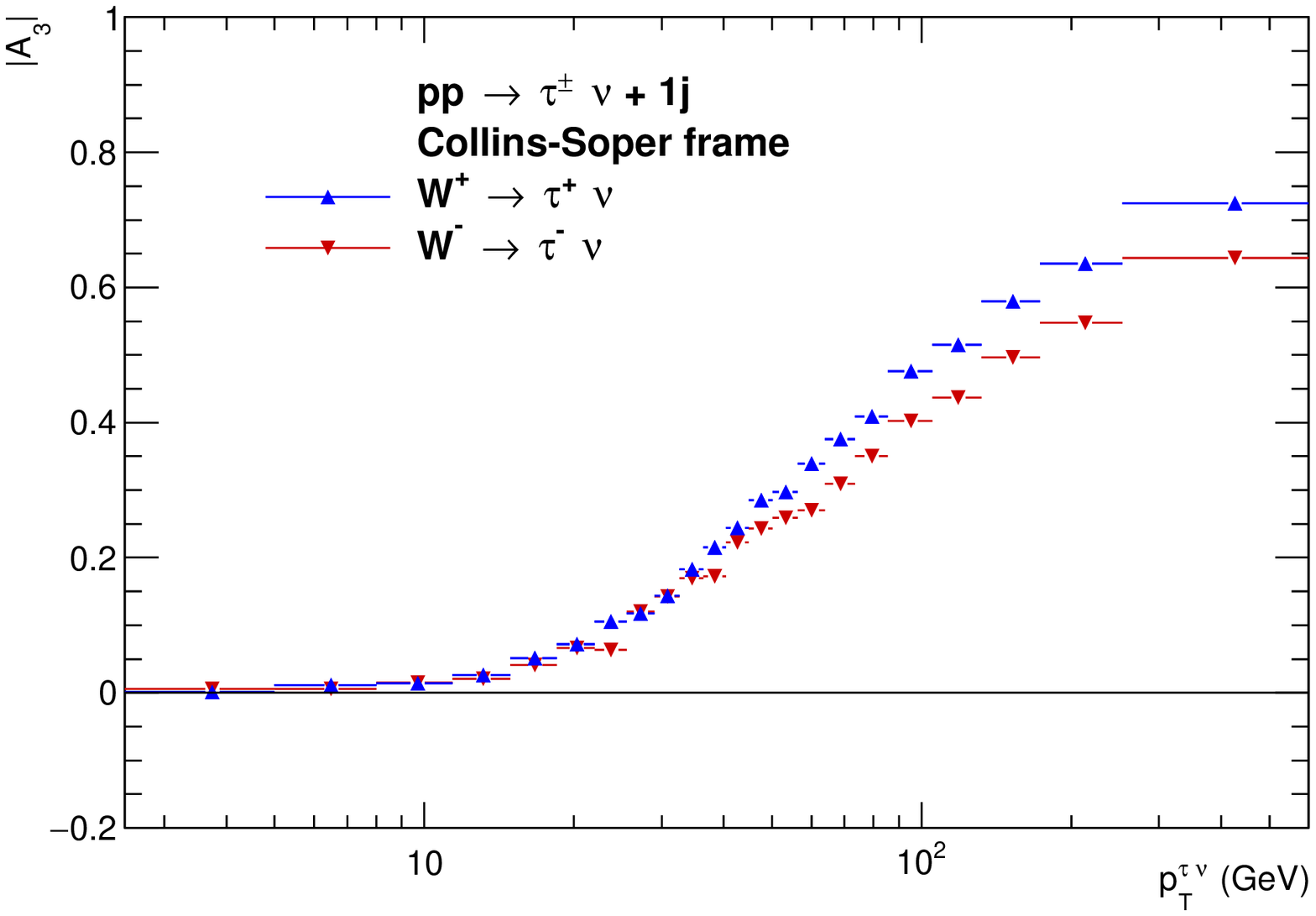}
   \includegraphics[width=7.5cm,angle=0]{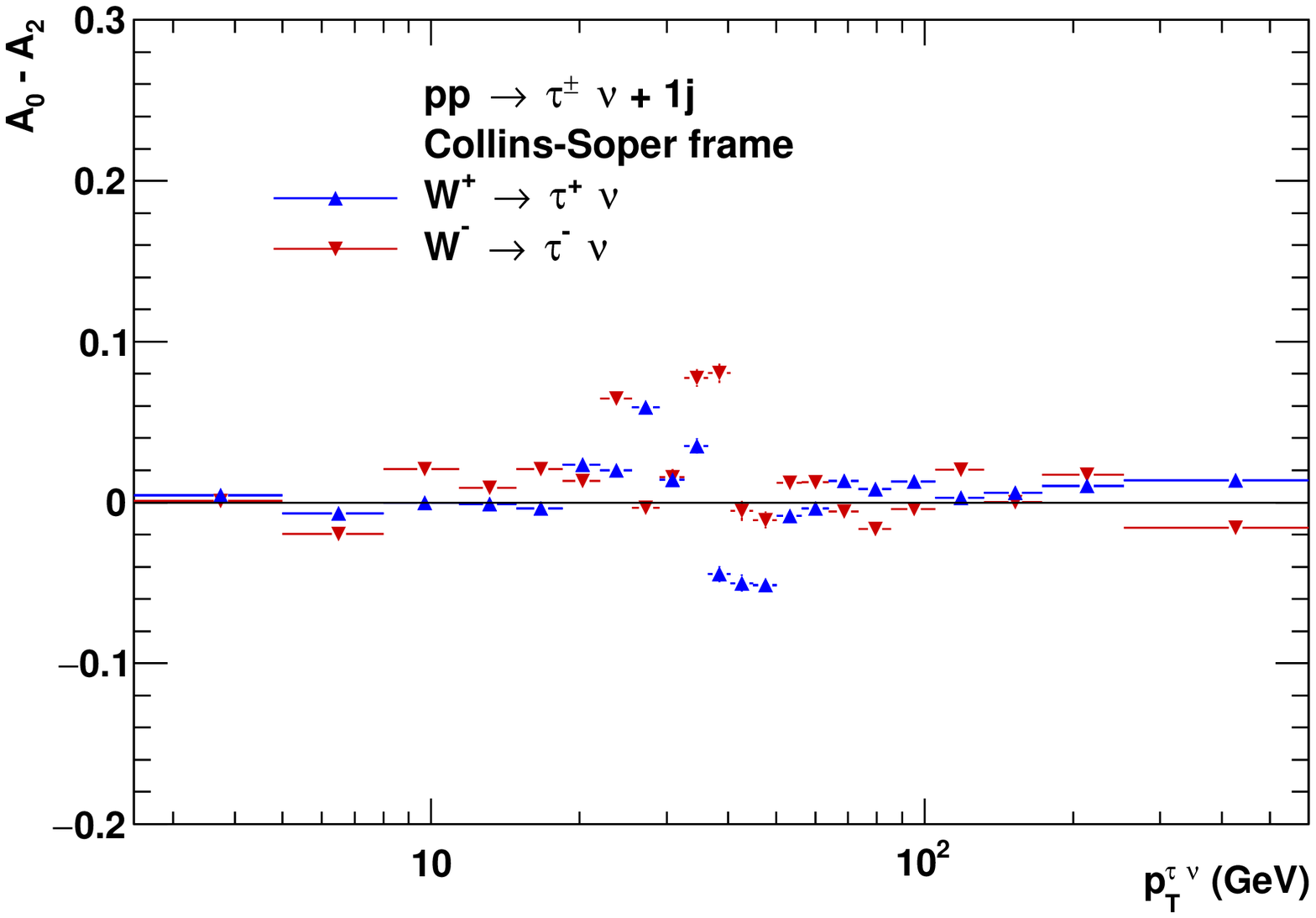}
   \includegraphics[width=7.5cm,angle=0]{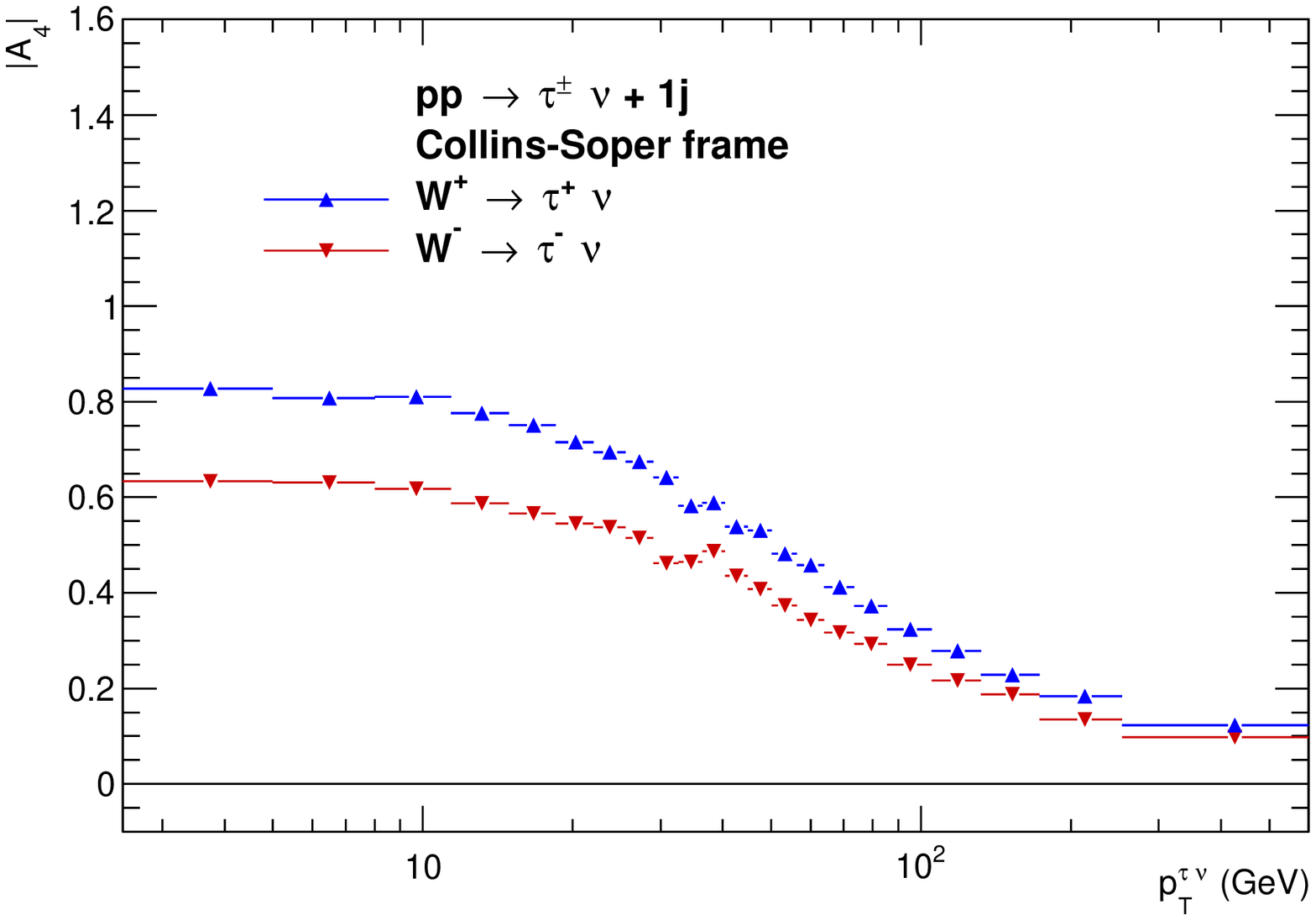}
}
\end{center}
\caption{The $A_i$ coefficients  calculated in Collins-Soper frame for 
 $p p  \to \tau^{\pm} \nu \ + 1j$ processes generated with {\tt MadGraph}.
\label{Fig:Ai1jWpWmCS} }
\end{figure}

\begin{figure}
  \begin{center}                               
{
   \includegraphics[width=7.5cm,angle=0]{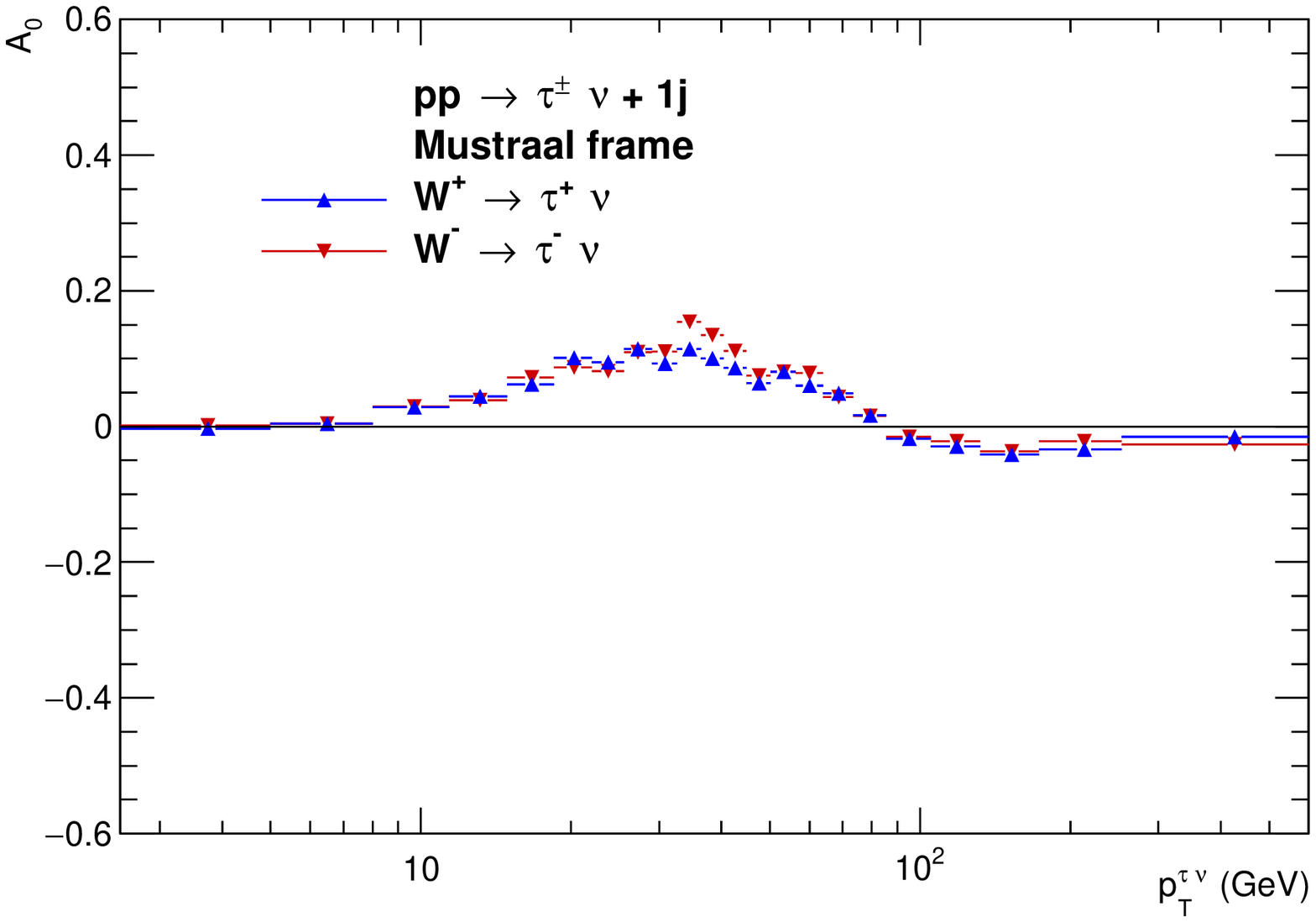}
   \includegraphics[width=7.5cm,angle=0]{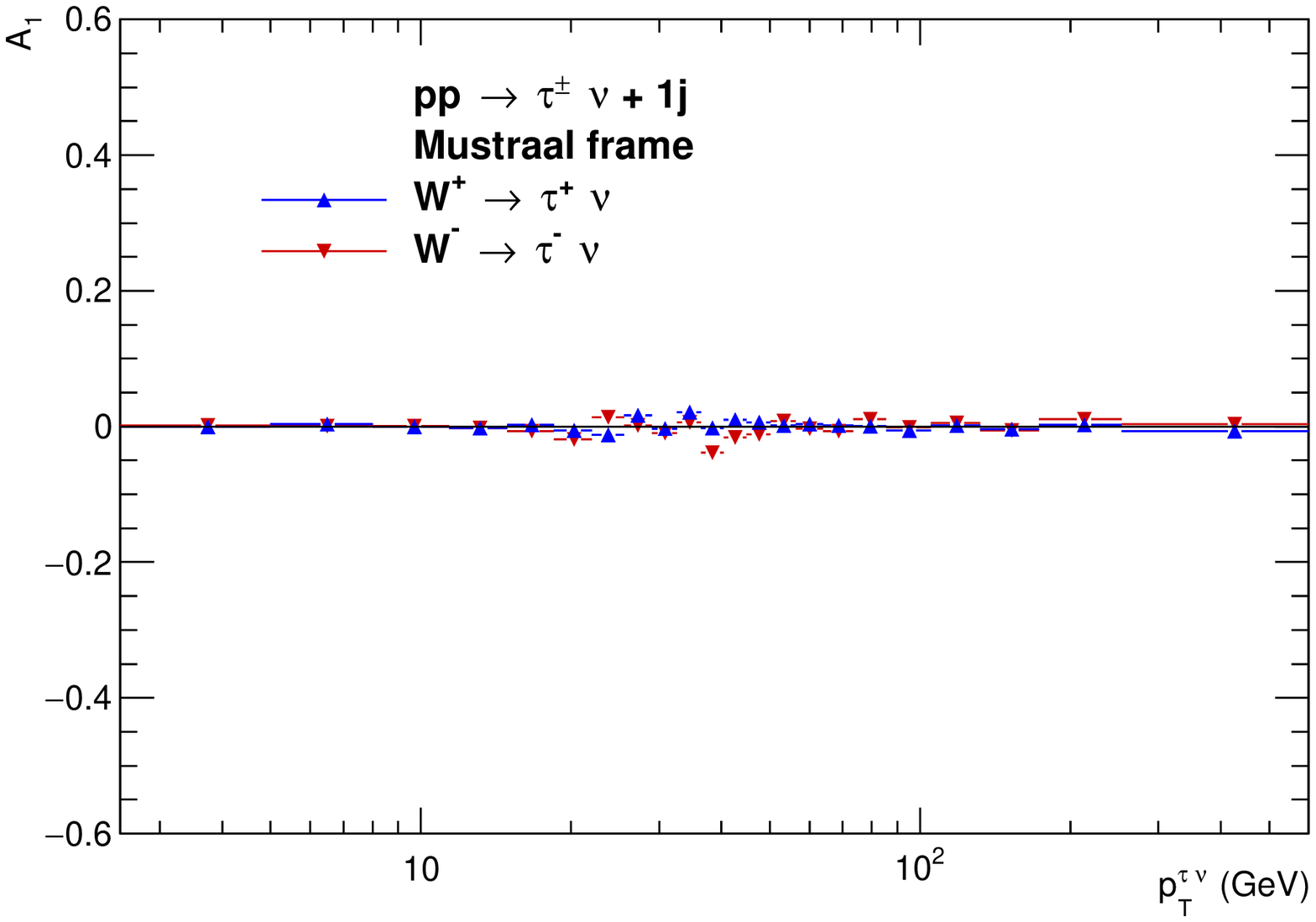}
   \includegraphics[width=7.5cm,angle=0]{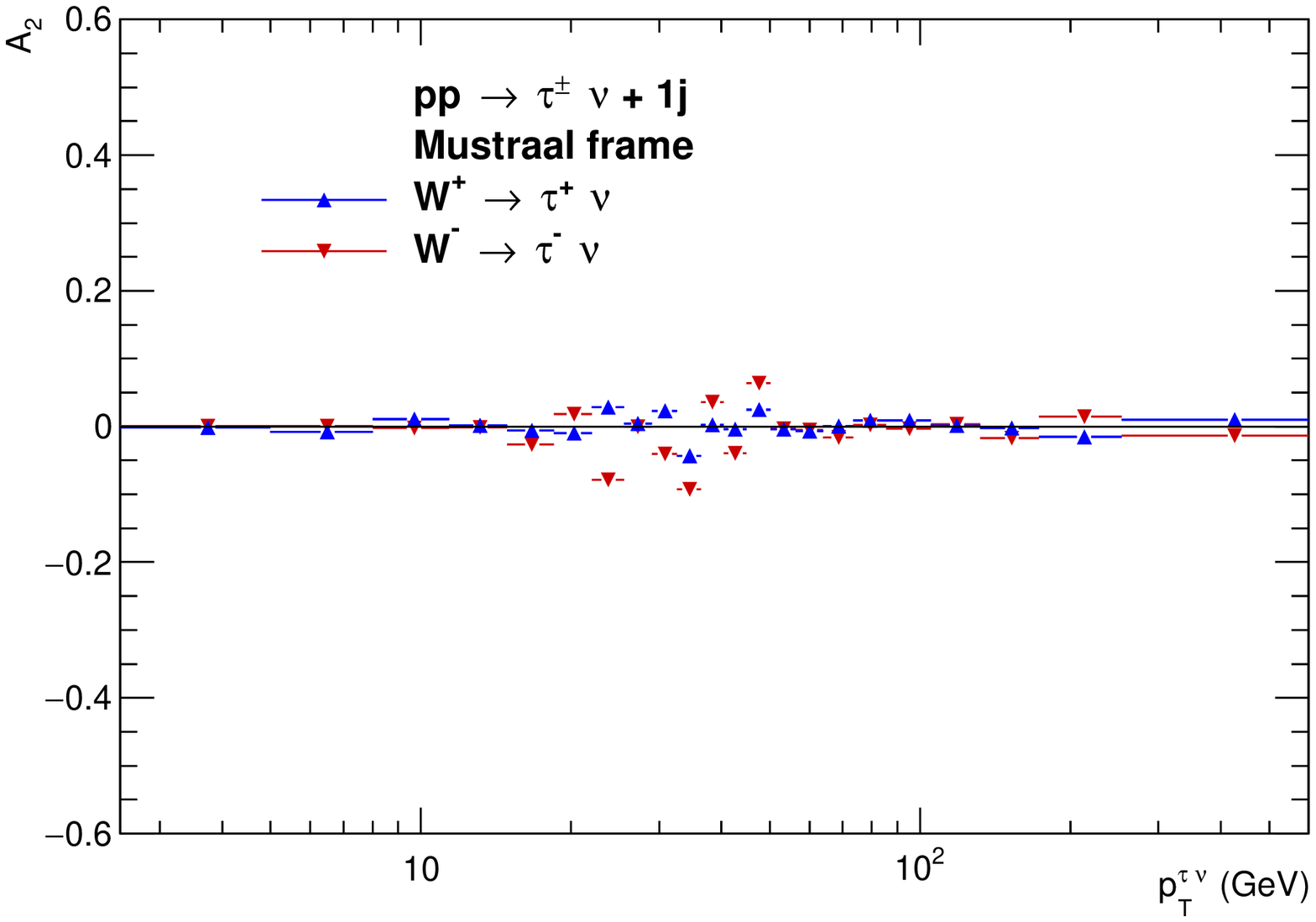}
   \includegraphics[width=7.5cm,angle=0]{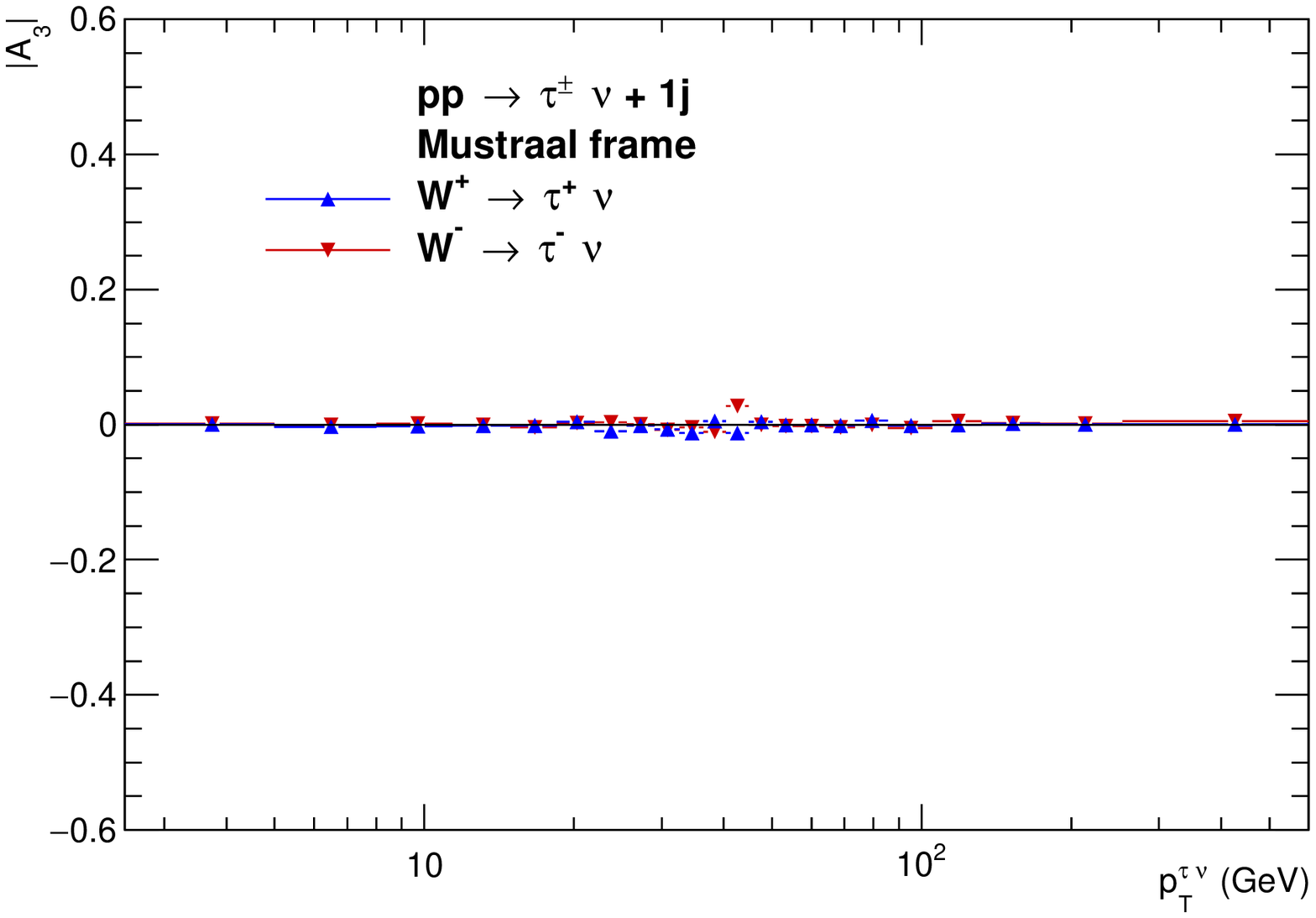}
   \includegraphics[width=7.5cm,angle=0]{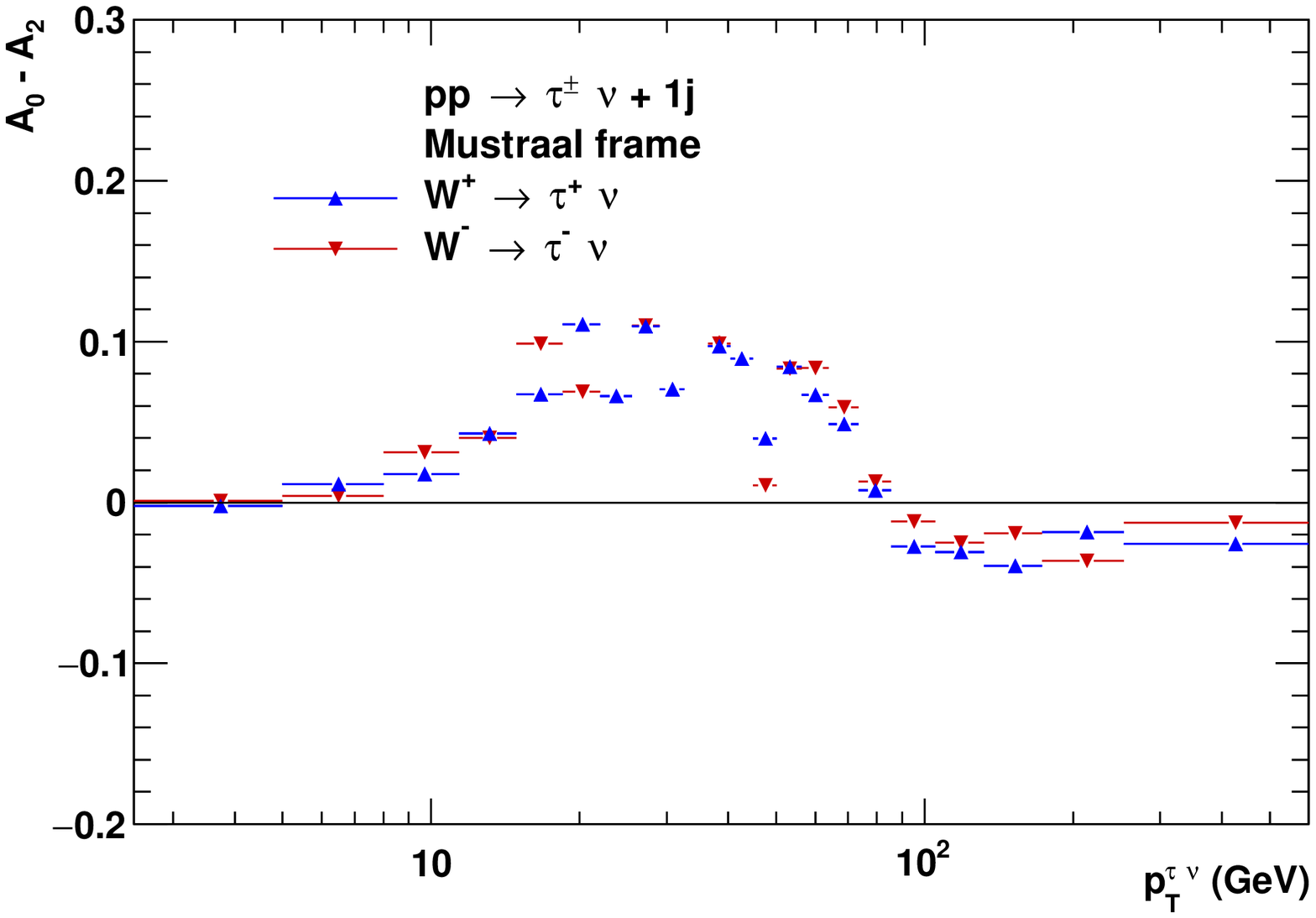}
   \includegraphics[width=7.5cm,angle=0]{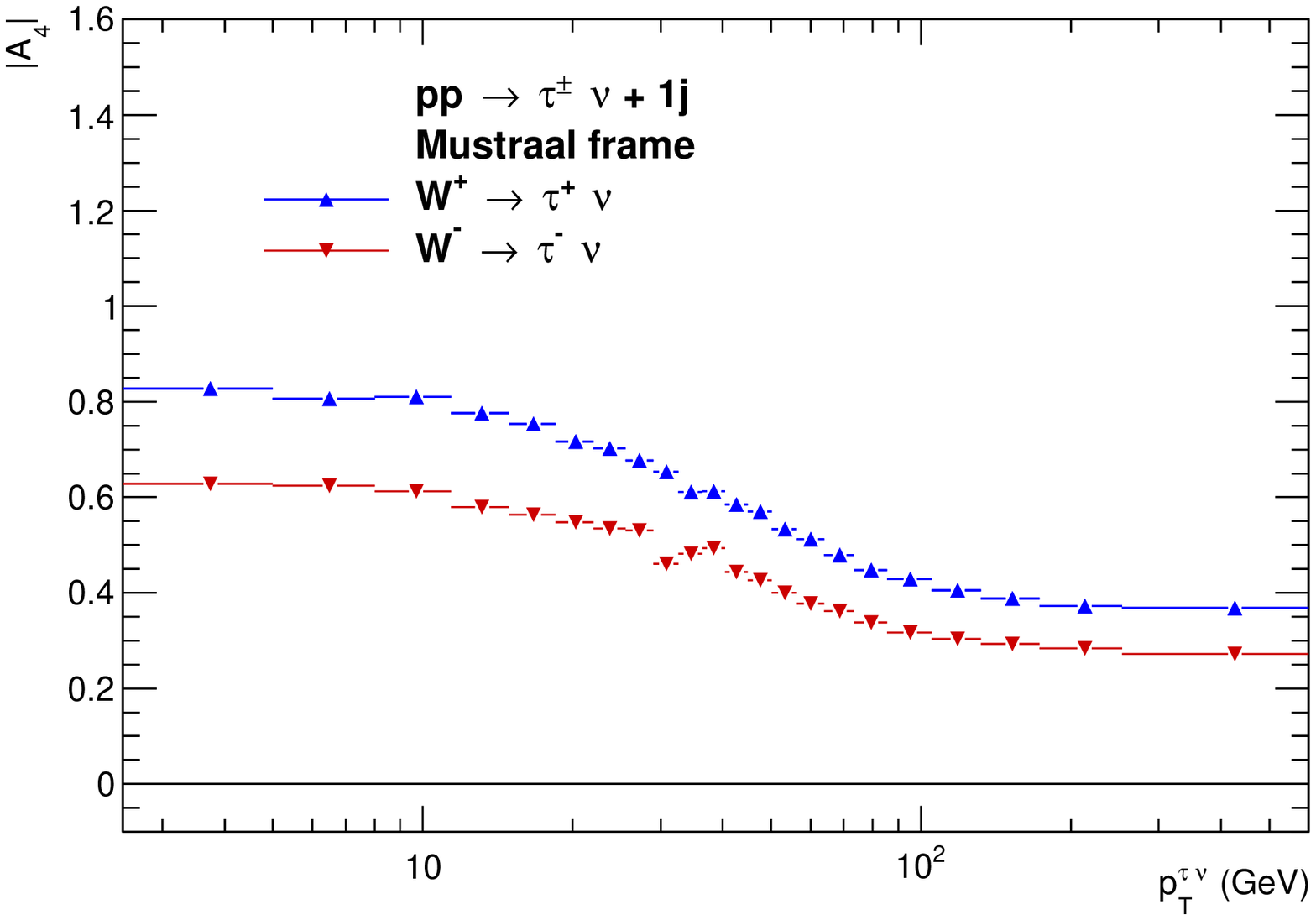}
}
\end{center}
\caption{The $A_i$ coefficients  calculated in {\tt Mustraal} frame 
for $p p  \to \tau^{\pm} \nu \ + 1j$ processes generated with {\tt MadGraph}.
\label{Fig:Ai1jWpWmMustraal}} 
\end{figure}

Figures~\ref{Fig:Ai1jWpWmCS} and~\ref{Fig:Ai1jWpWmMustraal} show the comparison of predicted coefficients
for $W^+$ and $W^-$, respectively in Collins-Soper and {\tt Mustraal} frames. The noticeable difference of the $A_4$
coefficients at low $p_T^W$ directly reflects different compositions of the structure functions in $pp$ collisions 
to produce $W^+$ and $W^-$ which enters the average over couplings shown in Eq.~\ref{eq:couplings}.
This difference is present for both the Collins-Soper and the {\tt Mustraal} frame. For the Collins-Soper
frame we observe also sizable $A_3$ coefficient  above $p_T^W = 100$ GeV.    

As stated already in \cite{Mirkes:1992hu}, the theoretical uncertainties due to the choice of the 
factorisation and renormalization scales are very small for  $A_i$ representing the cross section ratios.
Also most of the uncertainties from the choice of structure functions and factorisation scheme cancel 
in the ratios.

\subsection{Results with NLO simulation}
\label{NNLO}
So far, we have discussed results for samples of fixed order tree level matrix elements and of single parton (jet) emission.
In general, configurations with a variable number of jets and
effects of loop corrections and parton shower of initial state should be used to complete our studies. 
We have performed this task partially only, with the help of 10M weighted  $W^{+}+j$ and  $W^{-}+j$ events, 
with $W^{\pm} \to \tau^{\pm} \nu$  generated with  {\tt Powheg+MiNLO} Monte Carlo,  again for $pp$ collisions at 13 TeV
and the effective EW scheme.
The {\tt PowhegBox v2} generator \cite{Nason:2004rx,Alioli:2010xd}, augmented with {\tt MiNLO} method for choices of 
scales~\cite{Hamilton:2012np} and inclusion of Sudakov form factors~\cite{Hamilton:2012rf}, by construction achieves
NLO accuracy for distributions involving finite non-zero transverse momenta of the lepton system. Two jet configurations
are thus present.

\begin{figure}
  \begin{center}                               
{
   \includegraphics[width=7.5cm,angle=0]{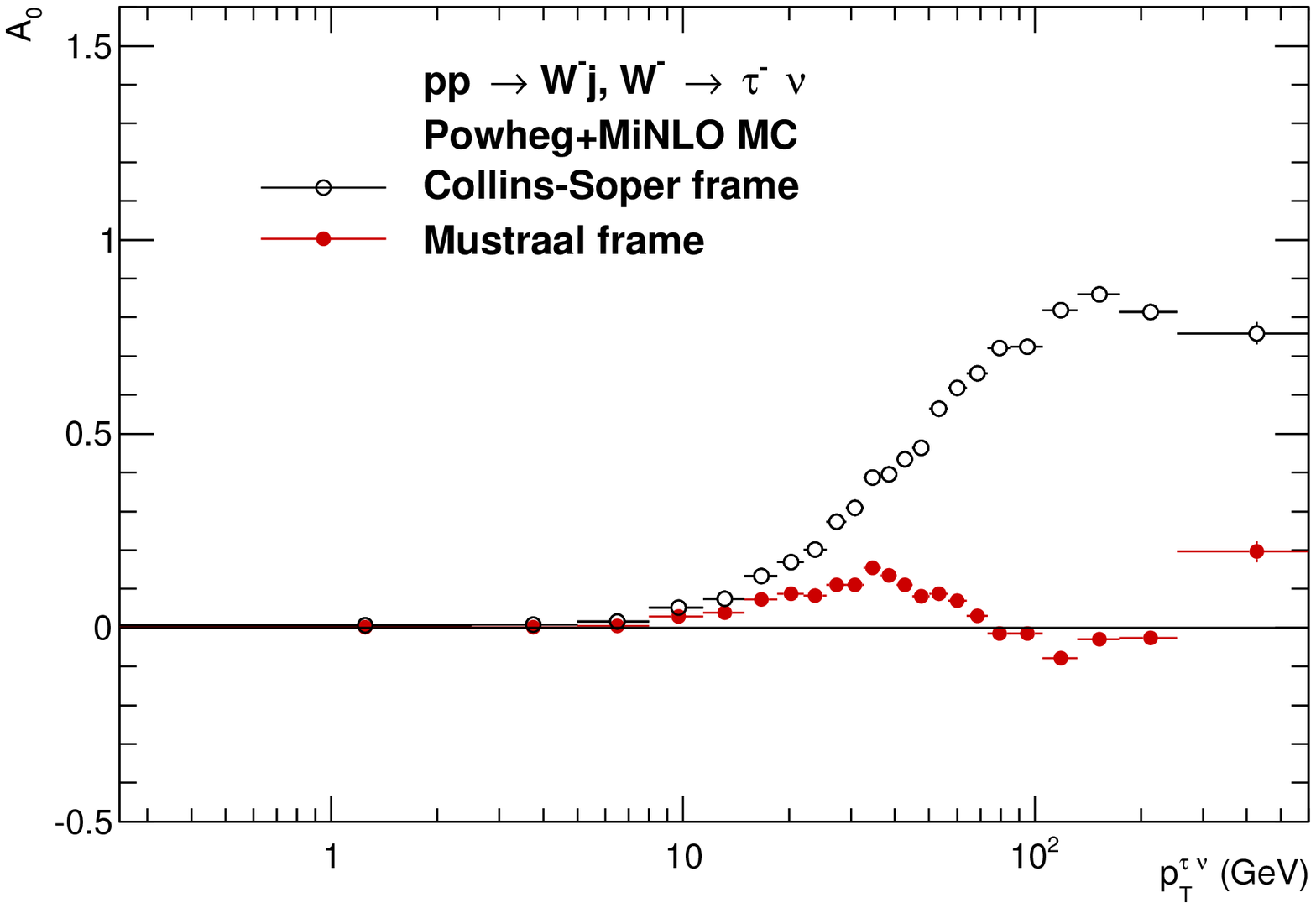}
   \includegraphics[width=7.5cm,angle=0]{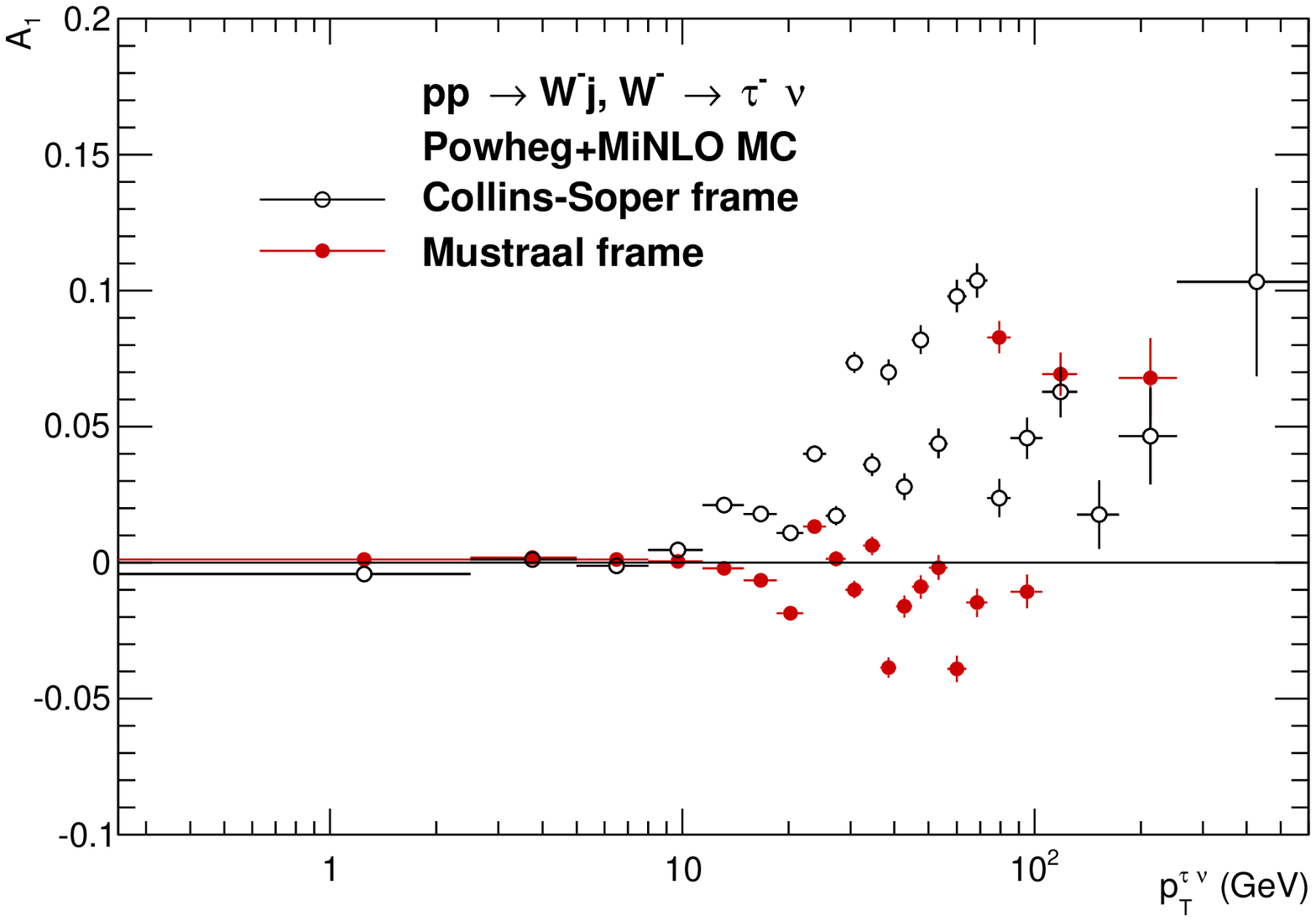}
   \includegraphics[width=7.5cm,angle=0]{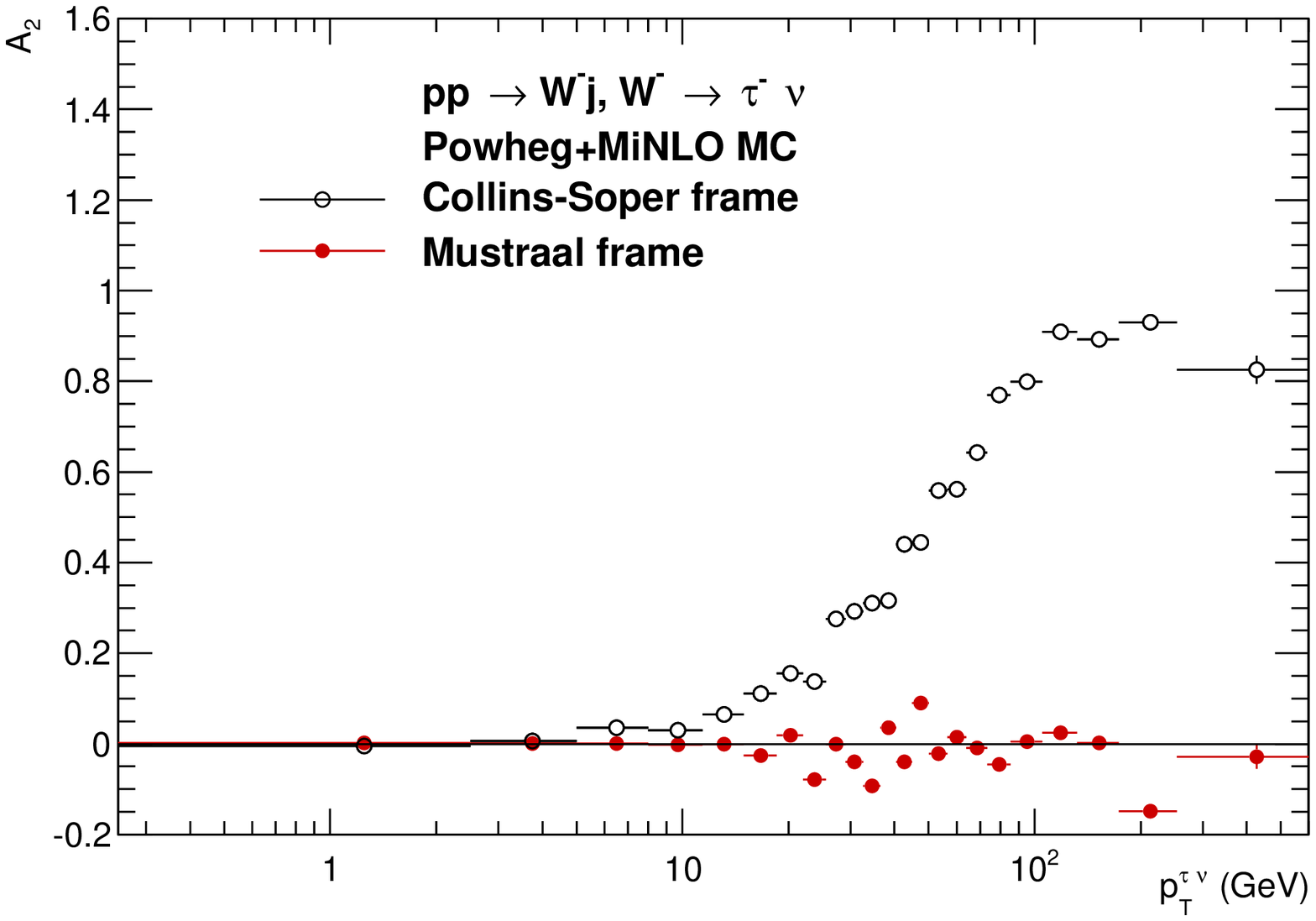}
   \includegraphics[width=7.5cm,angle=0]{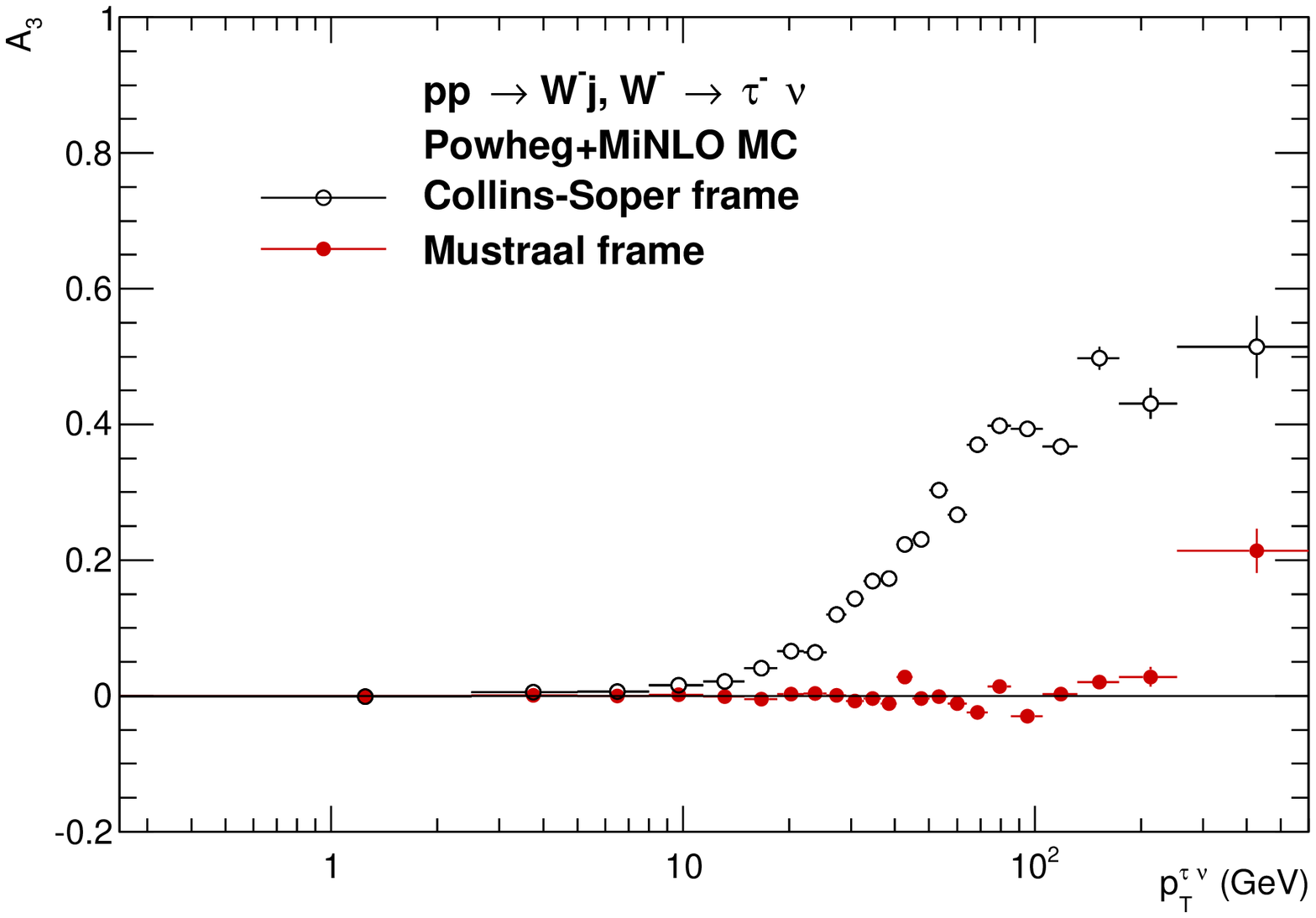}
   \includegraphics[width=7.5cm,angle=0]{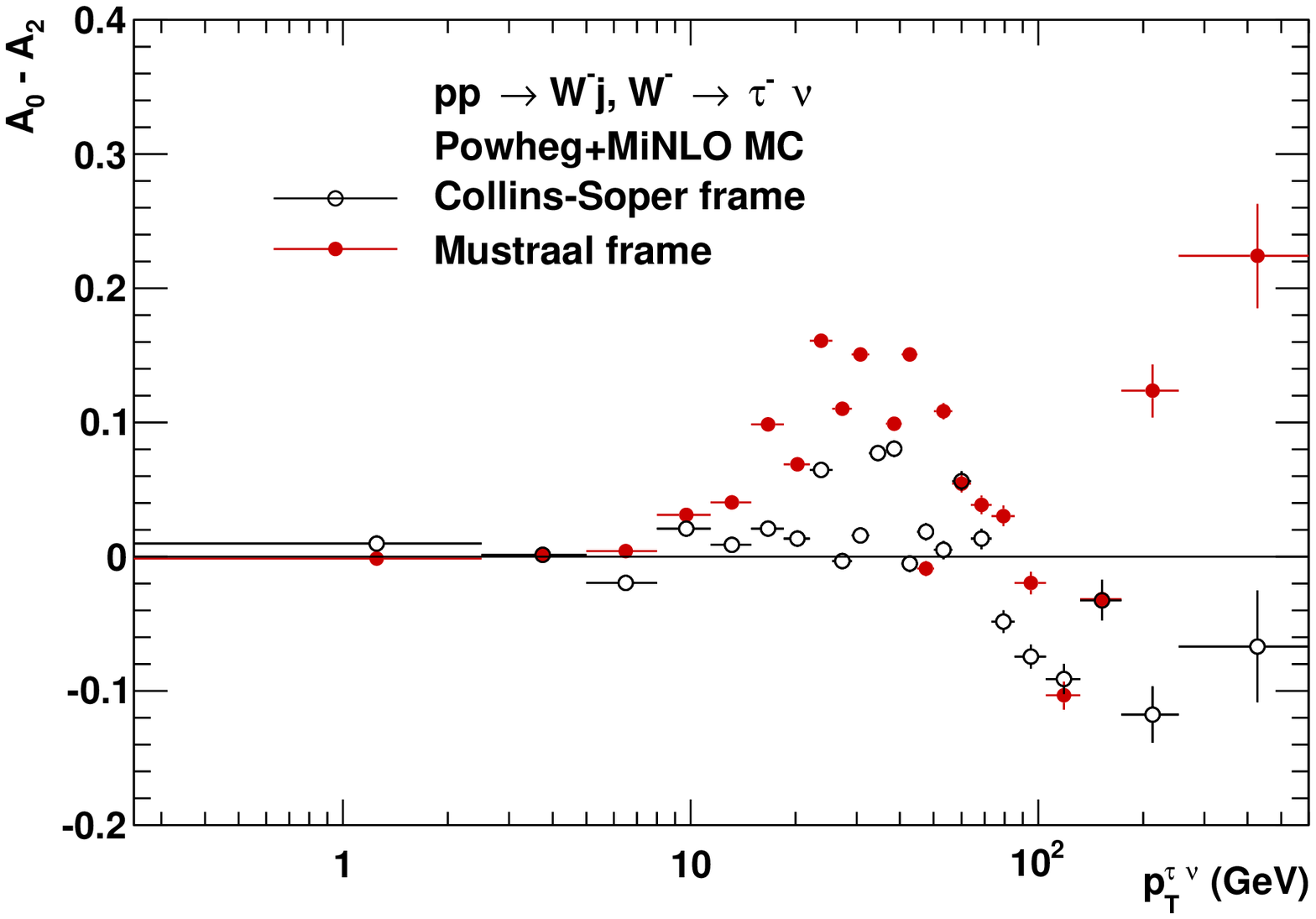}
   \includegraphics[width=7.5cm,angle=0]{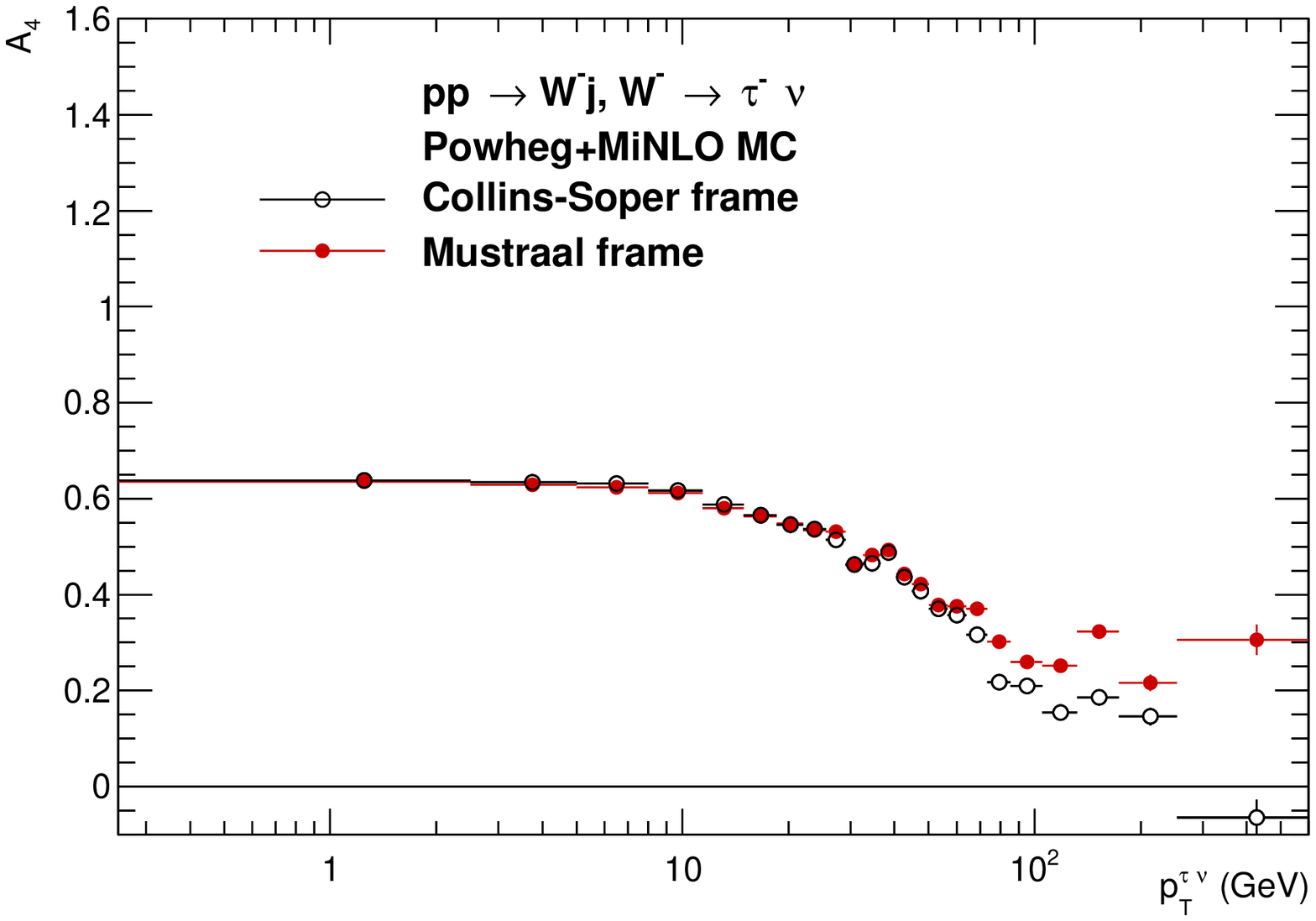}
}
\end{center}
\caption{The $A_i$ coefficients  calculated in Collins-Soper (black) and in {\tt Mustraal} (red) frames 
for $p p  \to \tau^{-} \nu \ + 1j$ process generated with {\tt  Powheg+MiNLO}.
\label{Fig:PoWm} }
\end{figure}

In Fig.~\ref{Fig:PoWm} results for $A_i$'s coefficients for $W^{-} \to \tau^{-} \nu$ are shown, 
extracted using moments method~\cite{Mirkes:1994eb} described in Section~\ref{sec:Templ}. Comparisons of results
using {\tt Mustraal} and the Collins-Soper frames feature again the pattern, observed already for QCD LO $W+1j$ events
generated with {\tt MadGraph}. As predicted for $Z \to \ell \ell$ and higher $p_T^Z$, the Lam-Tung relation 
$A_0 = A_2$ \cite{Lam:1978pu} is again violated at QCD NLO in the Collins-Soper frame. 
This confirms the robustness of our conclusions that only the {\tt Mustraal} frame retains Born like $A_i$ coefficients for high
$p_T^W$ configurations.

\section{Summary}\label{sec:summary} 

The interest in the decomposition of results for measurement of final states in Drell-Yan processes at the LHC
into coefficients of second order spherical polynomials for angular
distributions of leptons in the lepton-pair rest-frame was
recently confirmed by experimental publications for $Z \to \ell \ell$ 
process \cite{Aaltonen:2011nr,Khachatryan:2015paa,Aad:2016izn} and 
$W \to \ell \nu$ processes \cite{Chatrchyan:2011ig,ATLAS:2012au}.

Inspired by those measurements, we have investigated the possibility to apply a strategy similar to \cite{Aad:2016izn}
for $W \to \ell \nu $ production and decay. Thus, to contest the statements which are often  made  in the literature, 
that the neutrino escaping detection makes measurement of the complete set of angular coefficients 
not possible. We have shown a {\it proof of concept} for the proposed strategy, for 
the measurement of the complete set of $A_i$. As an example, Monte Carlo events of simulated 
$W \to \ell \nu$ +$ 1j$ process were analysed and the complete set of $A_i$ coefficients was extracted from a fit to 
pseudo-data distributions in the fiducial phase-space, in agreement with the prediction for this sample calculated
with the moments method. The results were cross-checked to hold when QCD NLO effects were taken into account.

In the second part of this paper, we have discussed the optimal reference frame for such
measurement. Two frames:  Collins-Soper and {\tt Mustraal} \cite{Richter-Was:2016mal} were studied. We have presented predictions 
for the angular coefficients in those frames as function of $W$-boson transverse momenta. 
We have shown that as expected, in case of the {\tt Mustraal} frame,
only one coefficient remains significantly non-zero and constant, almost  up to $p_T^{W}$~=~100~GeV, where it 
starts decreasing. Similarly as we argued in~\cite{Richter-Was:2016mal},  this may help to facilitate the
interpretation of experimental results  into quantities sensitive to strong interaction effects. The longitudinal
$W$ boson polarisation seems to appear predominantly as kinematic consequence of the choice of reference frames even 
in configurations of high $p_T$ jets.

\vskip 1 cm
\centerline{\bf \Large Acknowledgments}
\vskip 0.5 cm
E.R-W. would like to thank Daniel Froidevaux for numerous inspiring discussions on the angular
decomposition and importance of measuring angular coefficients for $W$-boson production at LHC.
We would like to thank W. Kotlarski for providing us with samples generated with {\tt MadGraph} 
which were used for numerical results presented here.

E.R-W. was partially supported by the funds of Polish National Science
Center under decision  UMO-2014/15/B\-/ST2/00049. 
Z.W. was partially supported by the funds of Polish National Science
Center under decision DEC-2012/04/M/ST2/00240.
We  acknowledge  PLGrid Infrastructure of the Academic
Computer Centre CYFRONET AGH in Krakow, Poland, where majority of numerical calculations were performed.

\providecommand{\href}[2]{#2}
\clearpage
\appendix
\section{Distributions for $W^+ \to \ell \nu$}
\label{App:WpFigs}

\begin{figure}
  \begin{center}                               
{
   \includegraphics[width=6.5cm,angle=0]{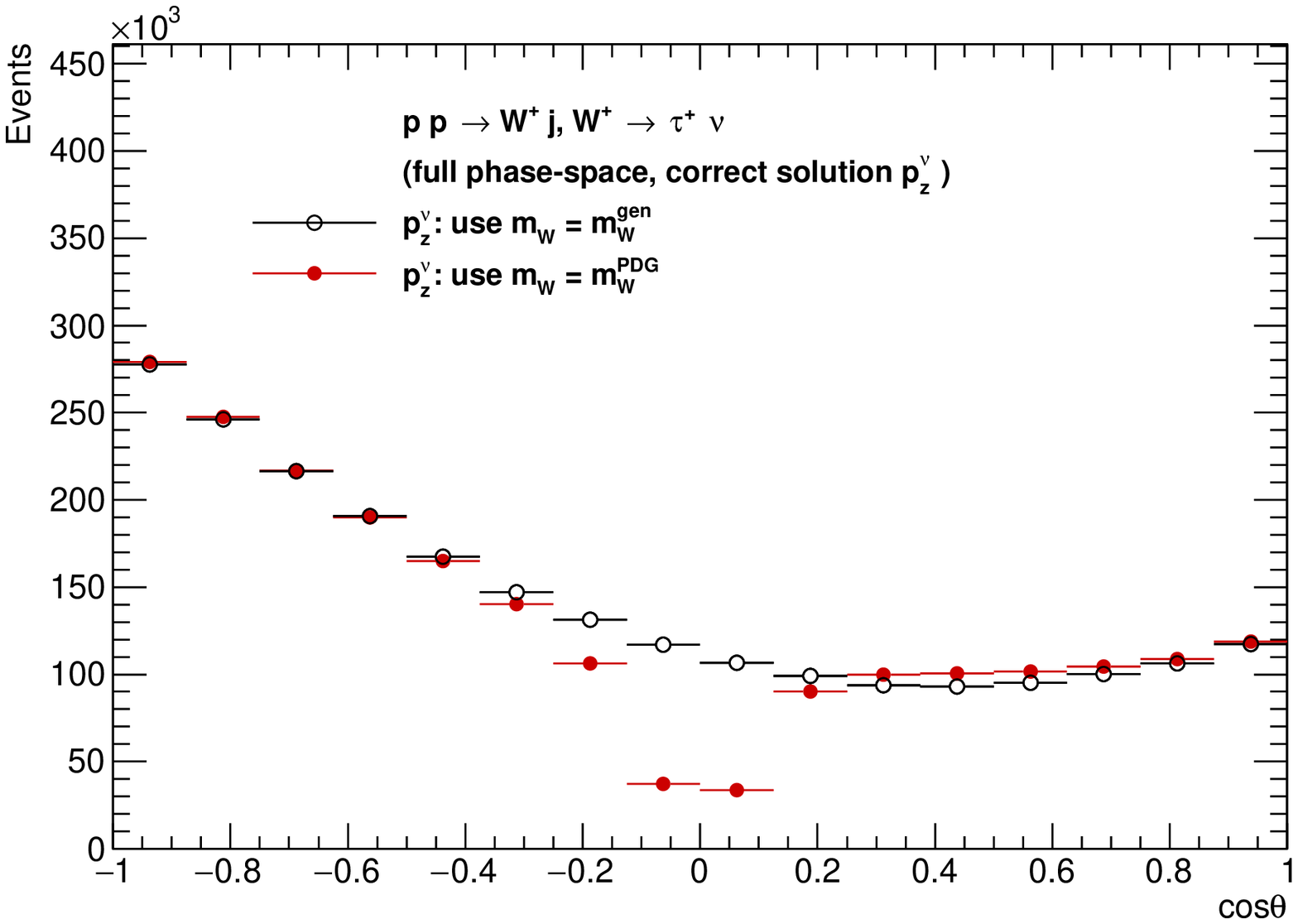}
   \includegraphics[width=6.5cm,angle=0]{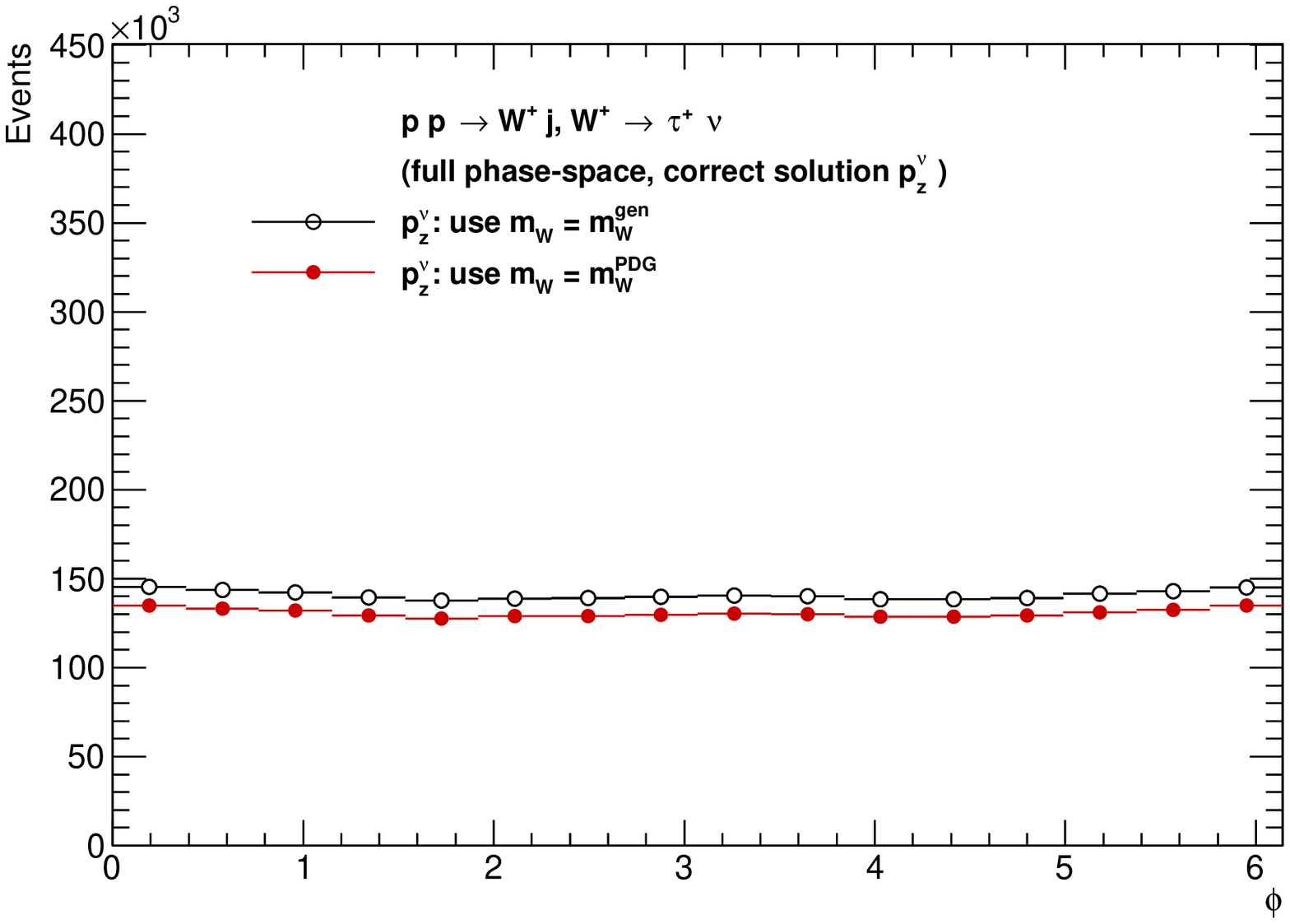}
}
\end{center}
\caption{As Fig.~\ref{Fig:Wmevtloss} but for $W^{+} \to \tau^{+} \nu$. 
The distributions of $\cos \theta$ and $\phi$ of charged lepton in the Collins-Soper rest frame. 
Effect from events loss due to non-existing solution for the neutrino momenta, 
when $m_W = m_W^{PDG}$ is used for  Eq.~(\ref{Eq:pznu}) is concentrated in the 
central bins of the left plot.
\label{FigApp:Wpevtloss} }
\end{figure}

In this Appendix, we collect plots corresponding to the ones shown 
in Section~\ref{sec:Wenu} but for $W^+ \to \ell^+ \nu$.
Figure~\ref{FigApp:Wpevtloss}  shows $\cos \theta$ and $\phi$ distributions of 
charged lepton in the Collins-Soper 
rest frame. 
We use generated $W$ boson mass in a given event $m_W = m_W^{gen}$ 
and compare to the case when
fixed PDG value $m_W = m_W^{PDG}$ for calculating neutrino
momenta $p_z^{\nu}$, taking {\it correct} solution for $p_z^{\nu}$ is used.
The losses due to nonexisting solution of Eq.~(\ref{Eq:pznu}) are concentrated at $\cos \theta = 0$ but are uniformly 
distributed over the whole $\phi$ range.

\begin{figure}
  \begin{center}                               
{
   \includegraphics[width=7.0cm,angle=0]{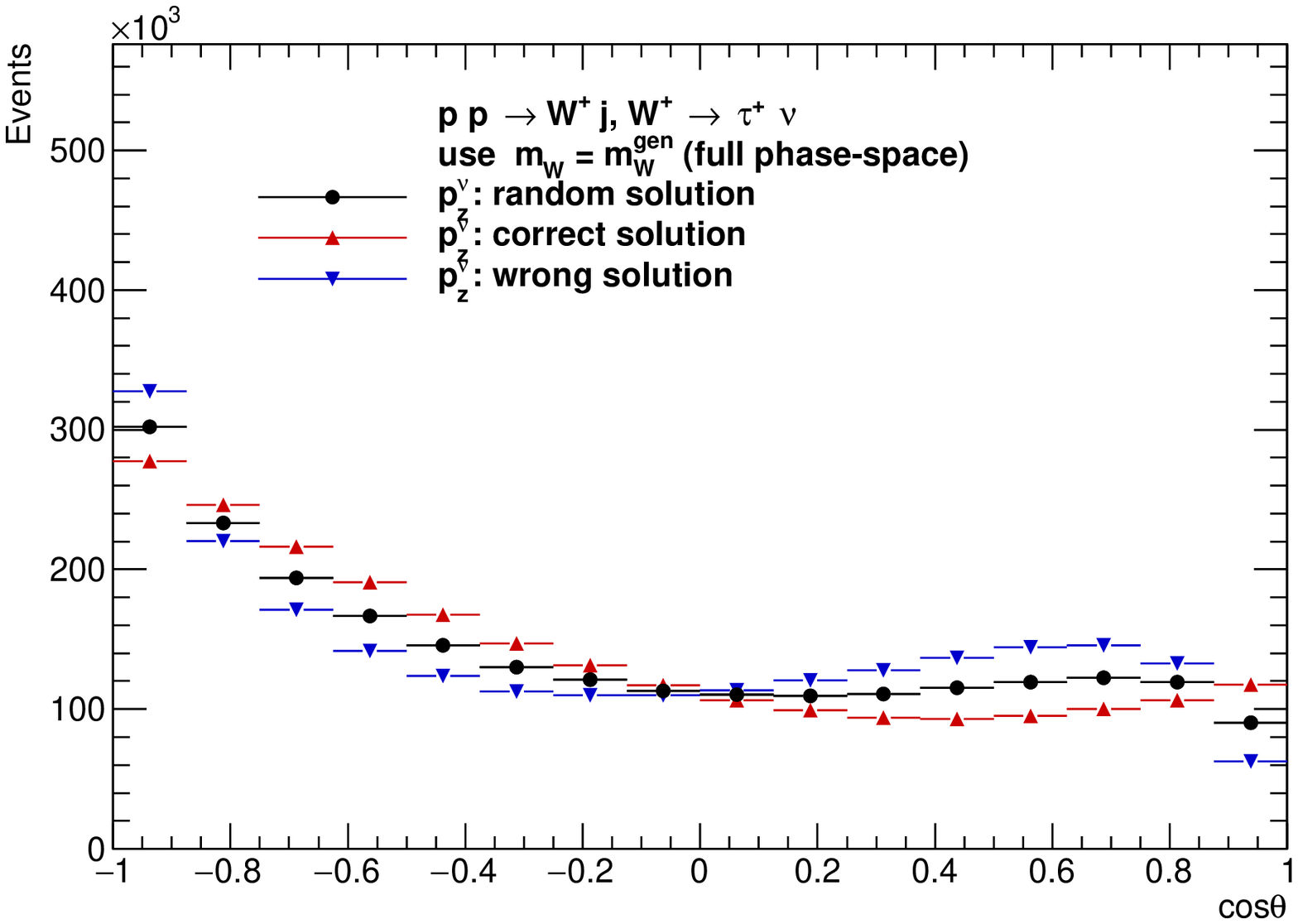}
   \includegraphics[width=7.0cm,angle=0]{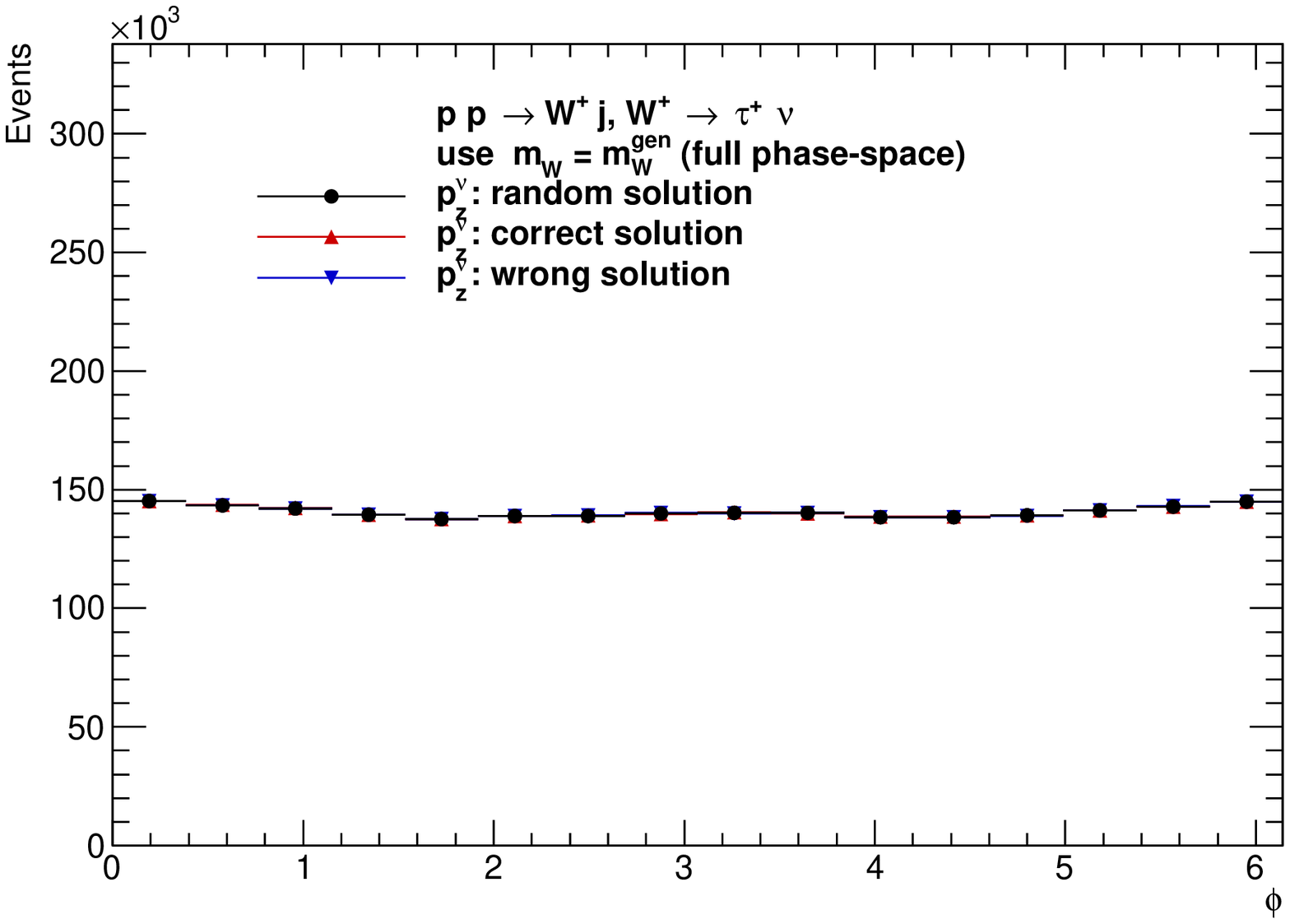}
   \includegraphics[width=7.0cm,angle=0]{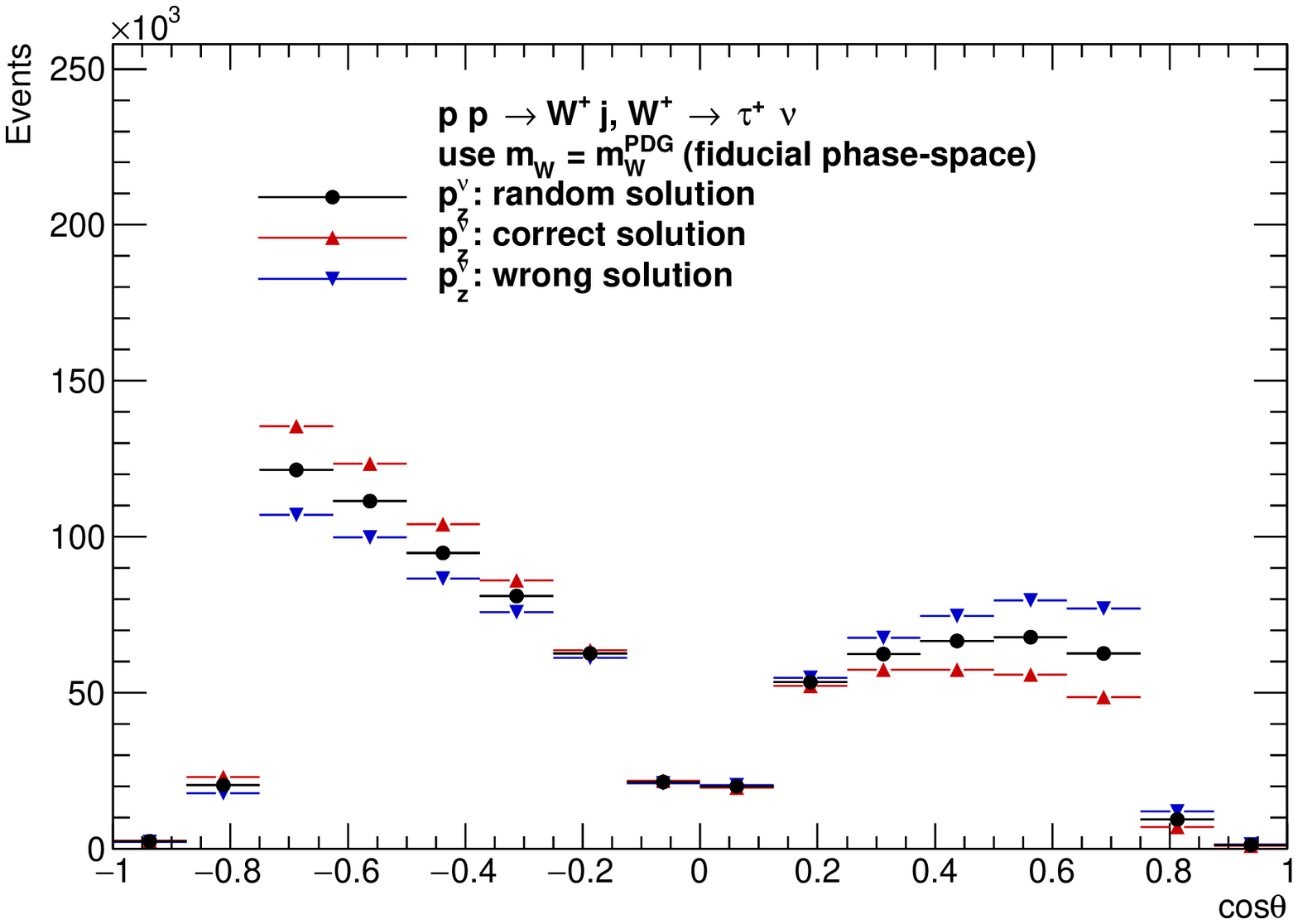}
   \includegraphics[width=7.0cm,angle=0]{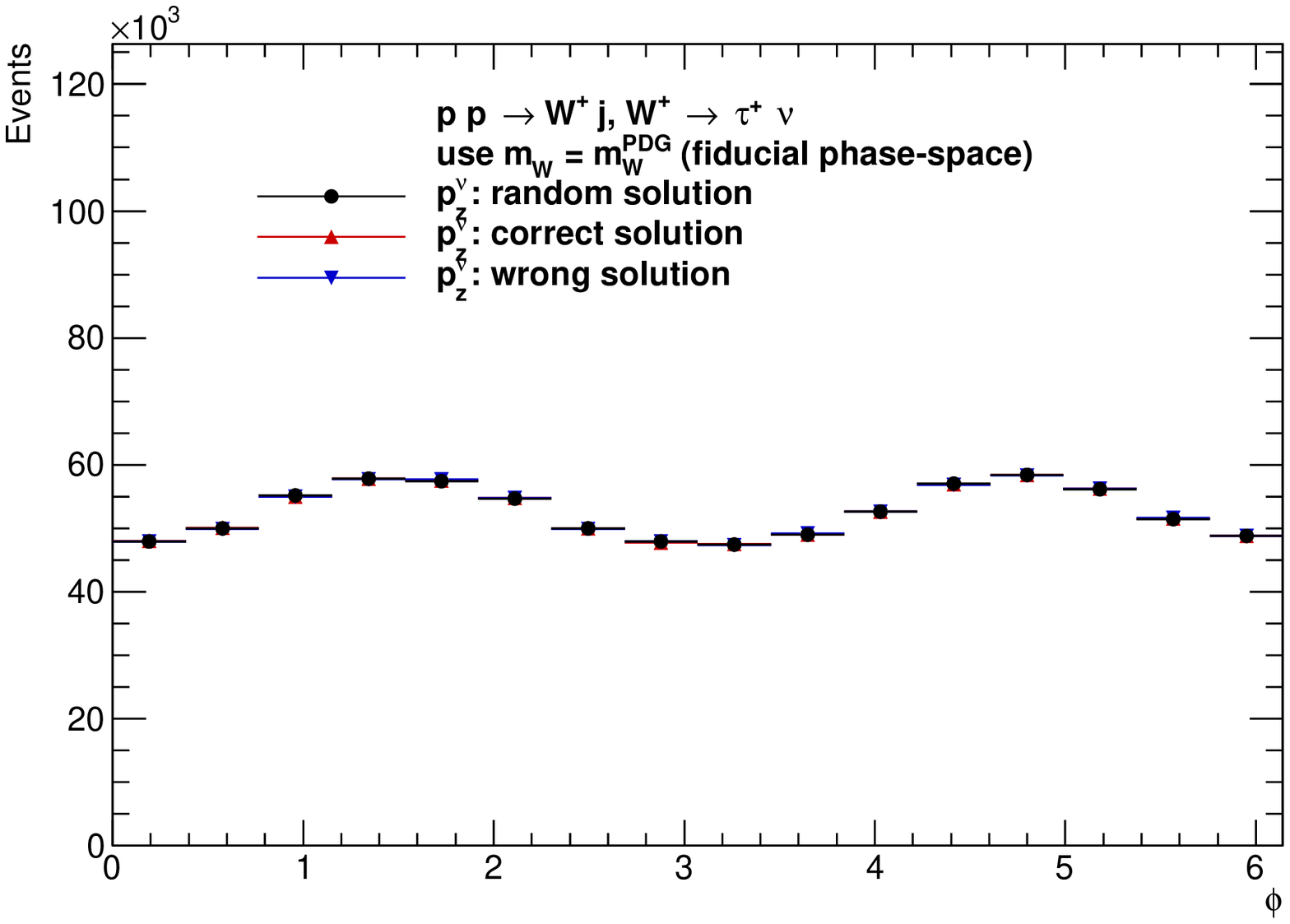}
}
\end{center}
\caption{As Fig.~\ref{Fig:Wm_cosThetaCSphiCS} but for  $W^{+} \to \tau^{+} \nu$ events. 
The distributions of $\cos \theta$ and $\phi$ of charged lepton in the Collins-Soper 
rest frame. Cases of $m_W = m_W^{PDG}$ for solving 
Eq.~(\ref{Eq:pznu}) where {\it correct}, {\it wrong} or  {\it random} solution for $p_z^{\nu}$ are taken.
Top plots are for the distributions in the full phase-space, bottom ones for the fiducial phase-space. 
\label{FigApp:Wp_cosThetaCSphiCS} }
\end{figure}

Figure~\ref{FigApp:Wp_cosThetaCSphiCS} shows  variation of $\cos \theta$ and $\phi$ distributions 
for charged lepton from $W^+$ decays when $m_W = m_W^{gen}$ or $m_W = m_W^{PDG}$ are used for solving  Eq.~(\ref{Eq:pznu}).
In the second case  selection 
of the fiducial region is applied. In each case distributions are shown for {\it correct}, {\it wrong} and  {\it random} solution for $p_z^{\nu}$.
One can notice, comparing with Figure~\ref{Fig:Wm_cosThetaCSphiCS}, that the shapes of the $\cos \theta$ distributions
in case of $W^+$ and $W^-$ are not mirrored when  {\it random} or {\it wrong} solution for the neutrino momenta are used. 

\begin{figure}
  \begin{center}                               
{
   \includegraphics[width=7.0cm,angle=0]{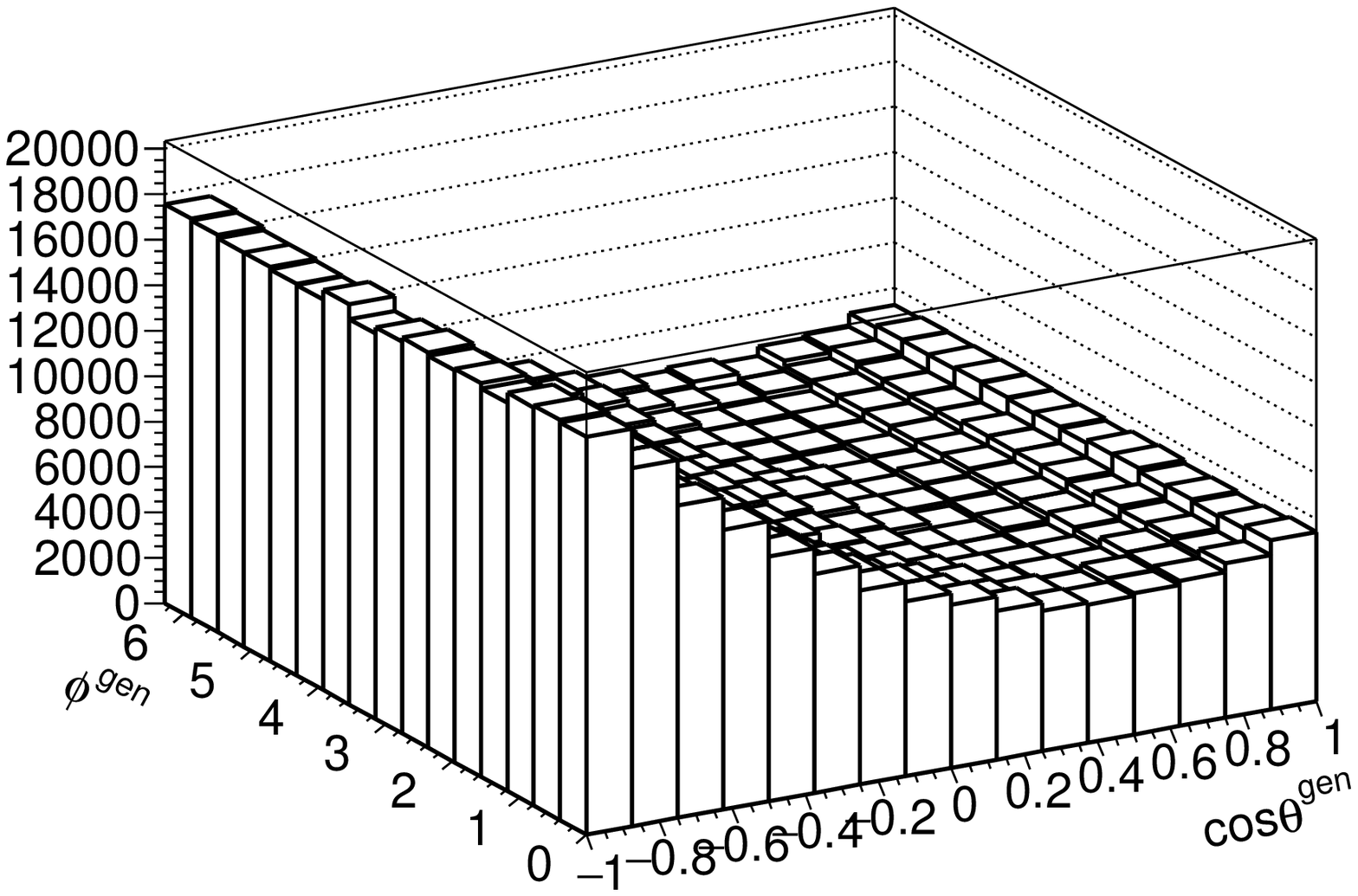}
   \includegraphics[width=7.0cm,angle=0]{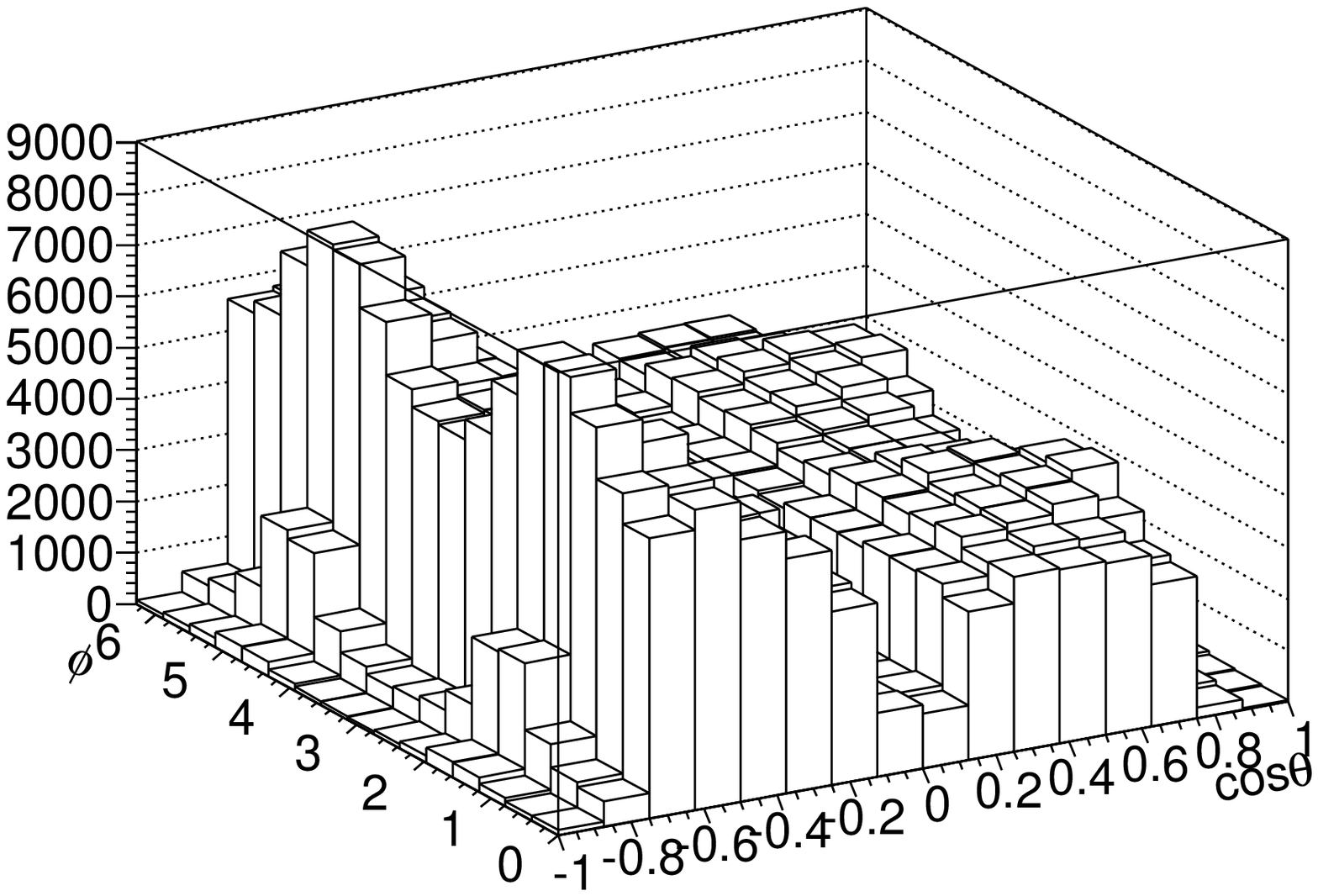}
}
\end{center}
\caption{As Fig.~\ref{Fig:2DWm_cosThetaCSphiCS} but for  $W^{+} \to \tau^{+} \nu$ events.
The 2D distribution of charged lepton $\cos \theta$ and $\phi$.
Case of  the full-phase space with  generated neutrino momentum, left side plot.
Case of the fiducial phase-space and  $m_W = m_W^{PDG}$ used in
Eq.~(\ref{Eq:pznu}) and {\it random} solution for  $p_z^{\nu}$, right side plot. 
\label{FigApp:2DWpcosThetaCSphiCS} }
\end{figure}

Figure~\ref{FigApp:2DWpcosThetaCSphiCS} shows 2D distributions 
in $(\cos \theta, \phi)$  for events in: full phase-space with generated neutrino momenta used and in the 
fiducial phase-space (also $m_W = m_W^{PDG}$ is used for solving Eq.~(\ref{Eq:pznu}) with {\it random} solution of neutrino momenta taken). 


\section{Further plots for templated shapes and extracting $A_i$'s coefficients.}
\label{App:Templates}

\begin{figure}
  \begin{center}                               
{
   \includegraphics[width=6.0cm,angle=0]{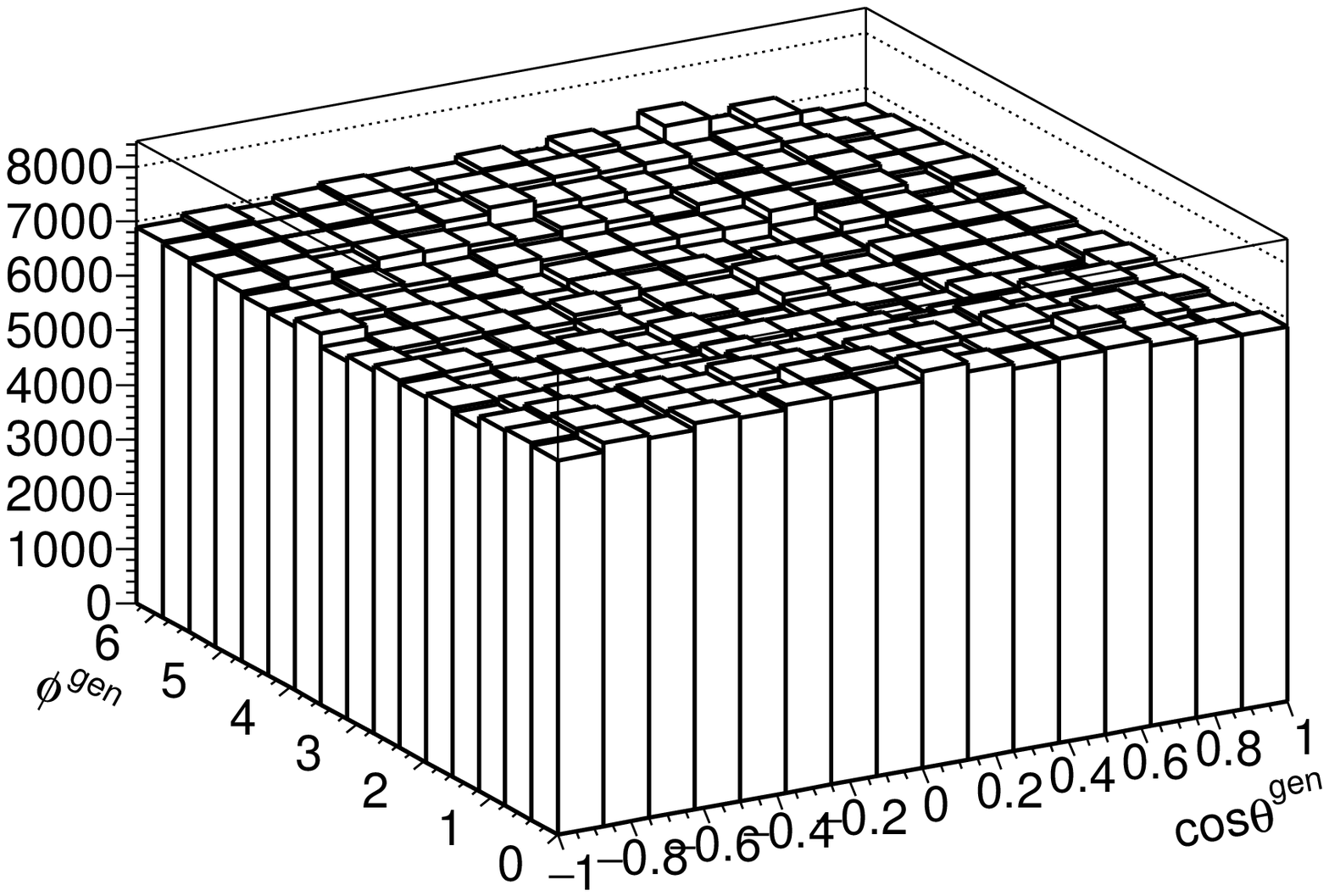}
   \includegraphics[width=6.0cm,angle=0]{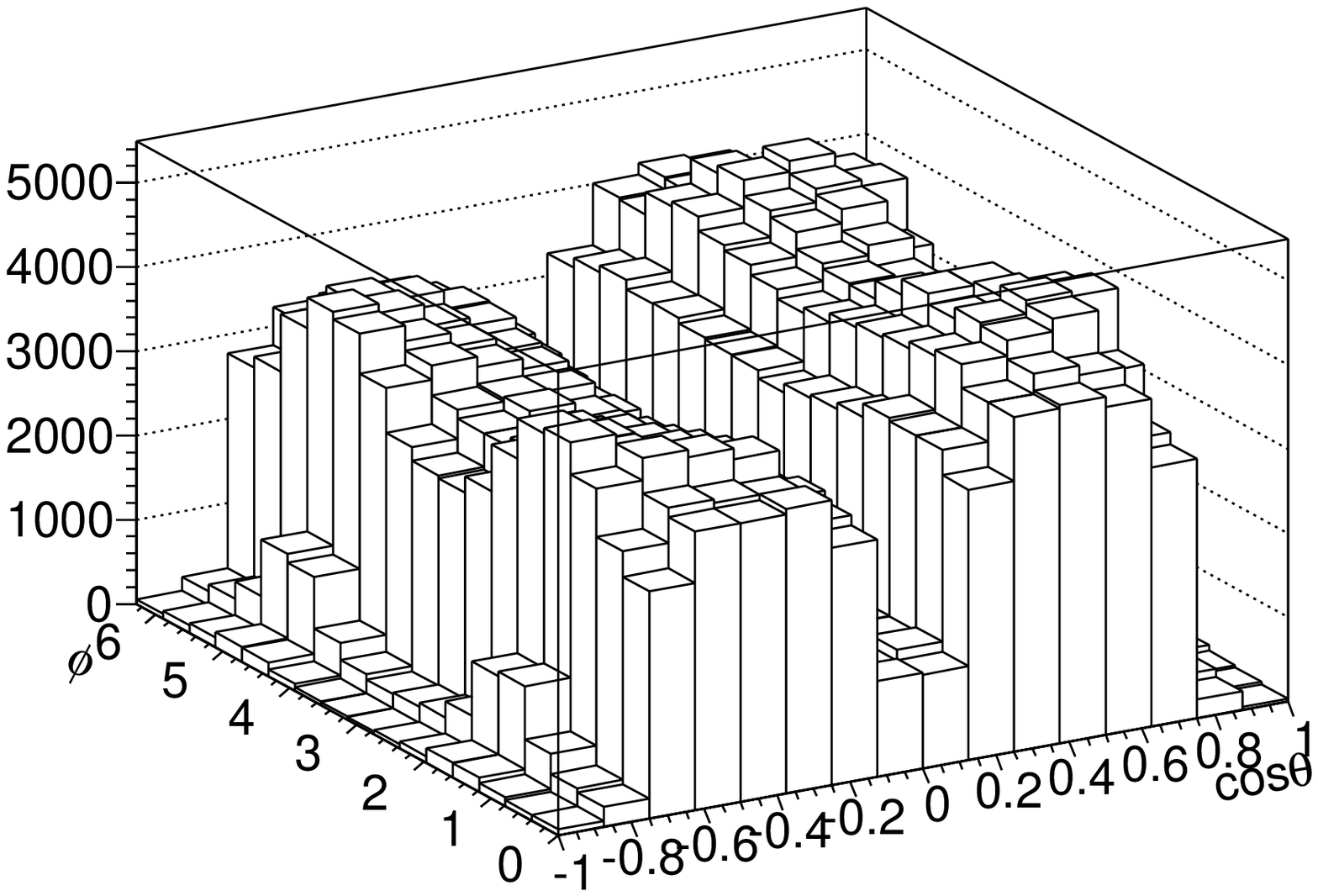}
}
\end{center}
\caption{As Fig.~\ref{Fig:Wm_2Dclo} but for $W^{+} \to \tau^{+} \nu$ events.
The 2D distribution of charged lepton $\cos \theta$ and $\phi$.
On left side distribution of the full phase-space, with generated neutrino momentum used, and  events  weighted $wt_{\Sigma AiPi}$. 
On right, the same distribution is shown, but:  $m_W^{PDG}$ is used for solving Eq.~(\ref{Eq:pznu}), randomly one of the 
solutions for $p_z^{\nu}$ is taken and fiducial selection is applied. The weight $wt_{\Sigma AiPi}$ is still
calculated with generated neutrino momenta.
%
\label{FigApp:Wp_2Dclo} }
\end{figure}

Figure~\ref{FigApp:Wp_2Dclo}  (right) shows  $(\cos \theta, \phi)$ distribution  of $W^+ \to \ell^+ \nu$ events,
where $\theta, \phi $ are calculated using generated neutrino momentum and events are weighted with $wt_{\Sigma AiPi}$.
Right plot of Figure~\ref{FigApp:Wp_2Dclo} show how the initially flat distributions are distorted by this folding 
procedure. Note that the shape is not the same as in case of $W^- \to \ell^- \nu$ shown in Figure~\ref{Fig:Wm_2Dclo}.
This is due to different distributions e.g. the pseudorapidity of charged lepton and thus different acceptance of 
the fiducial selection.

\begin{figure}
  \begin{center}                               
{
   \includegraphics[width=5.0cm,angle=0]{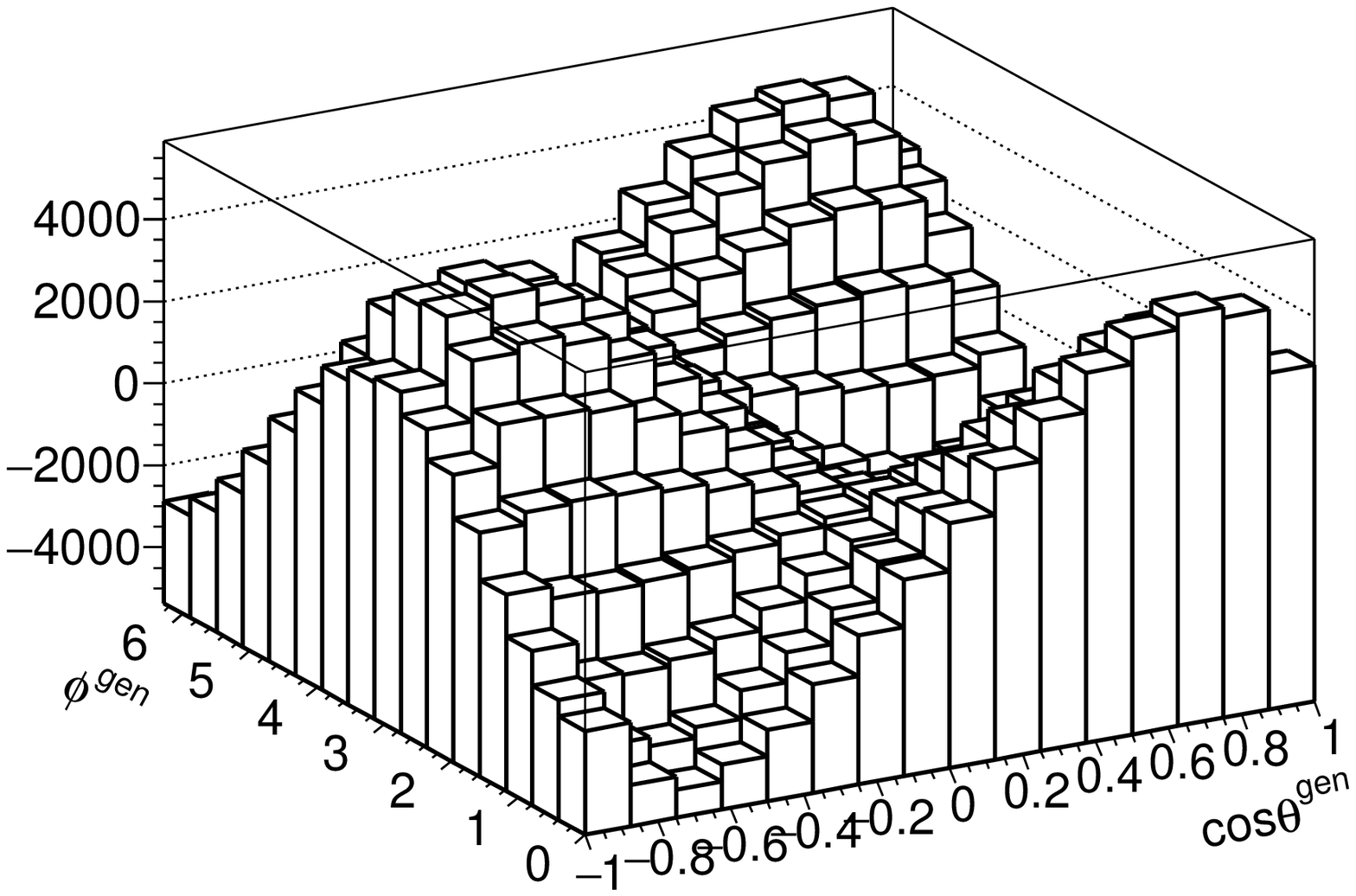}
   \includegraphics[width=5.0cm,angle=0]{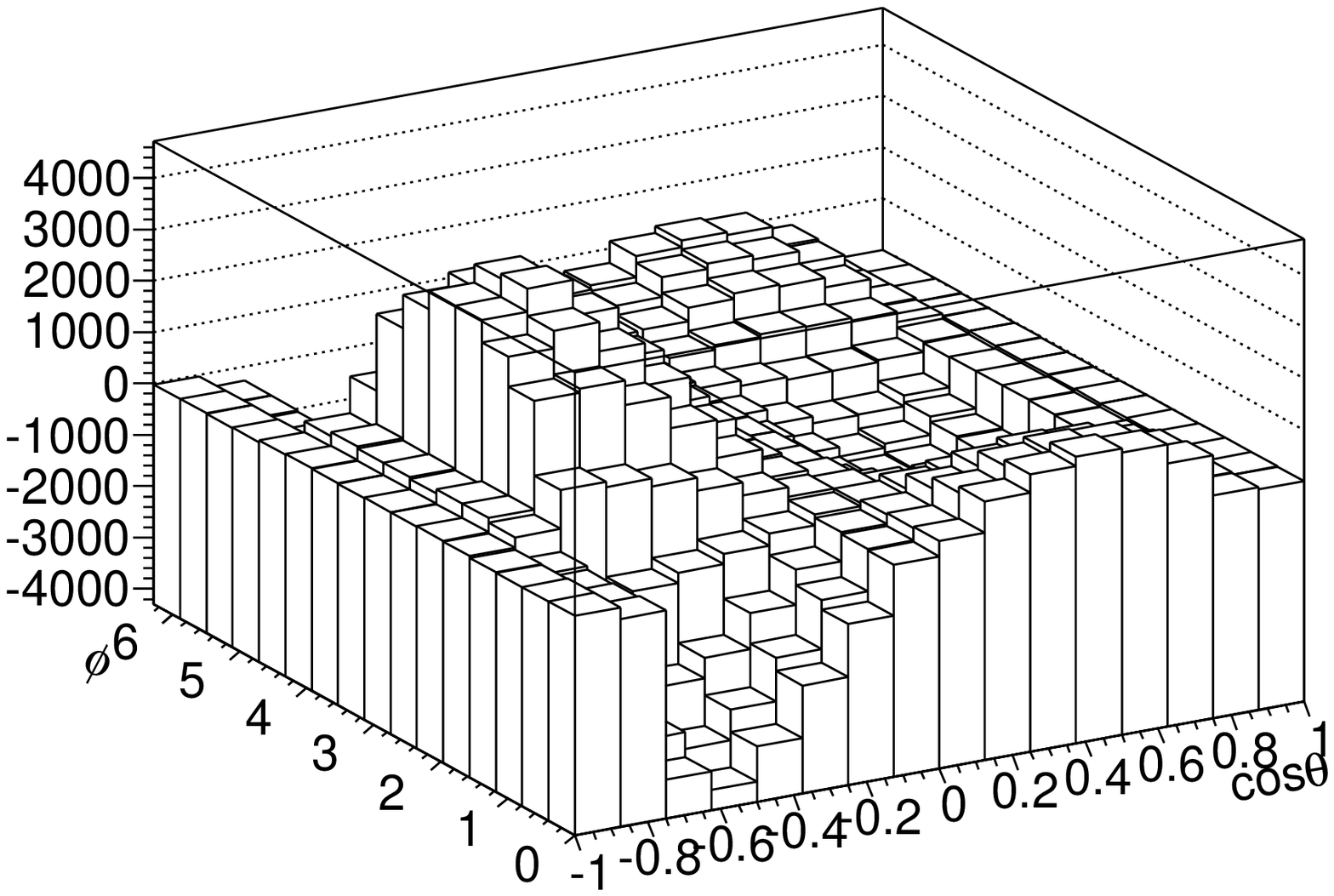}
   \includegraphics[width=5.0cm,angle=0]{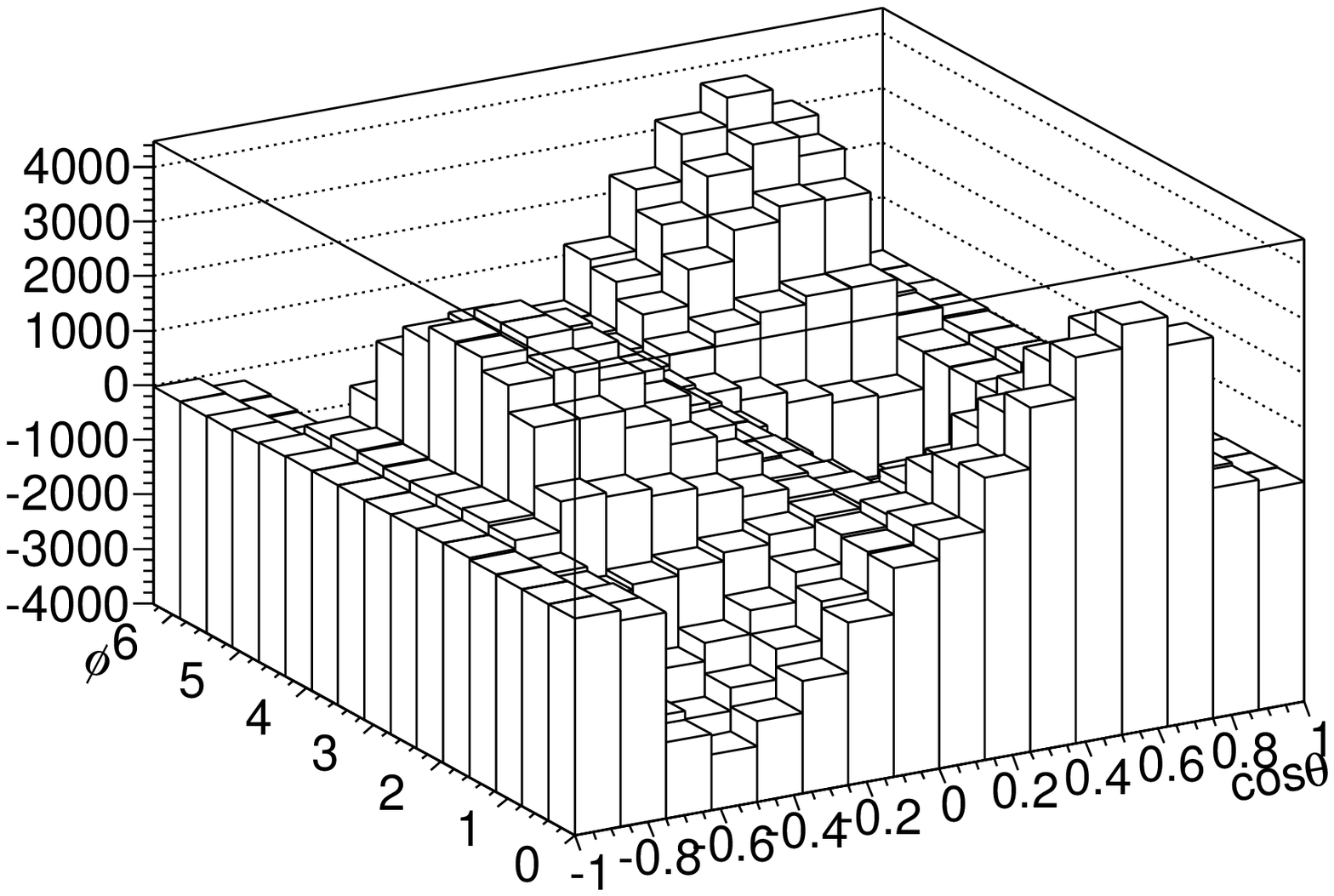}
   \includegraphics[width=5.0cm,angle=0]{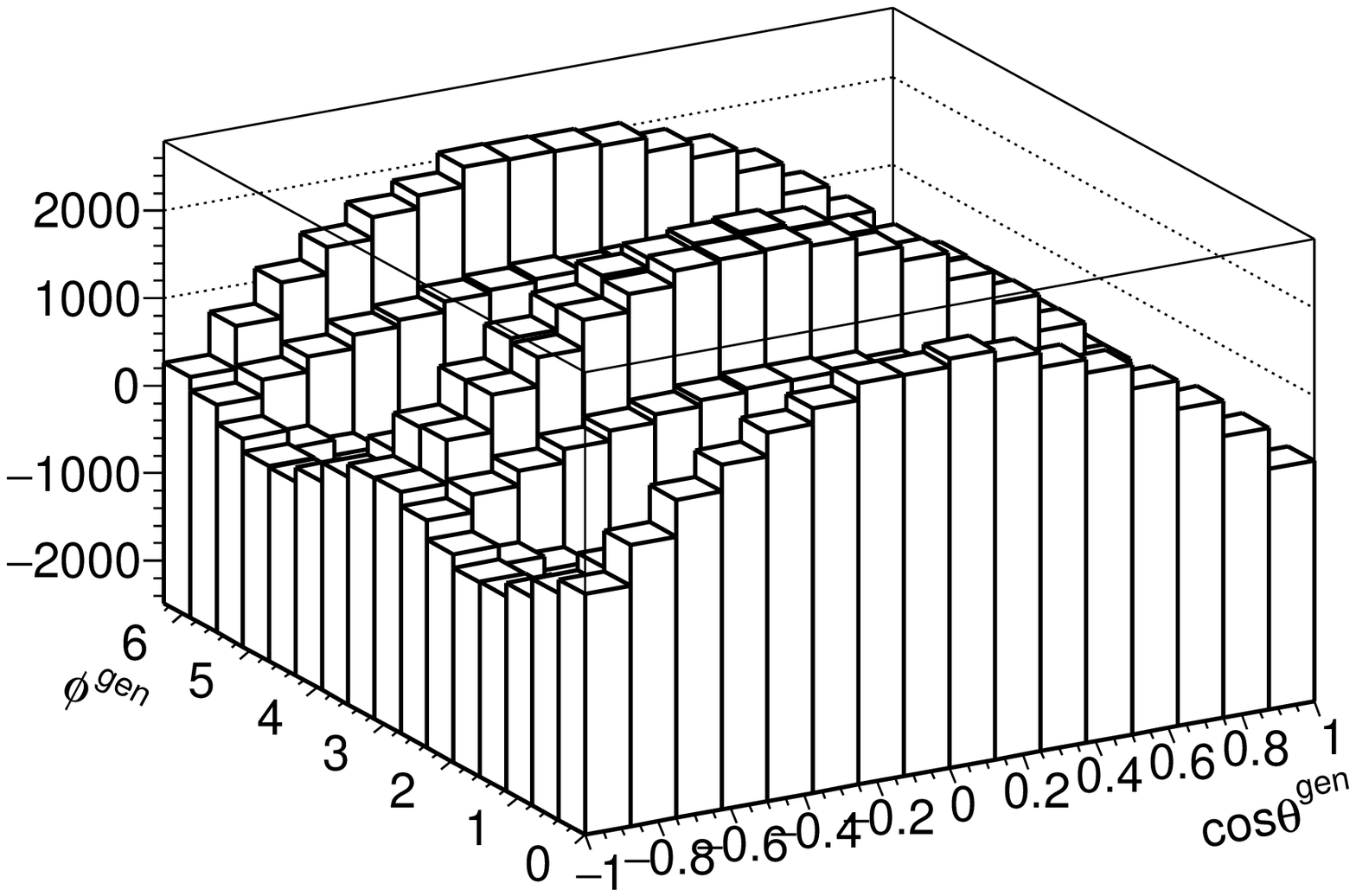}
   \includegraphics[width=5.0cm,angle=0]{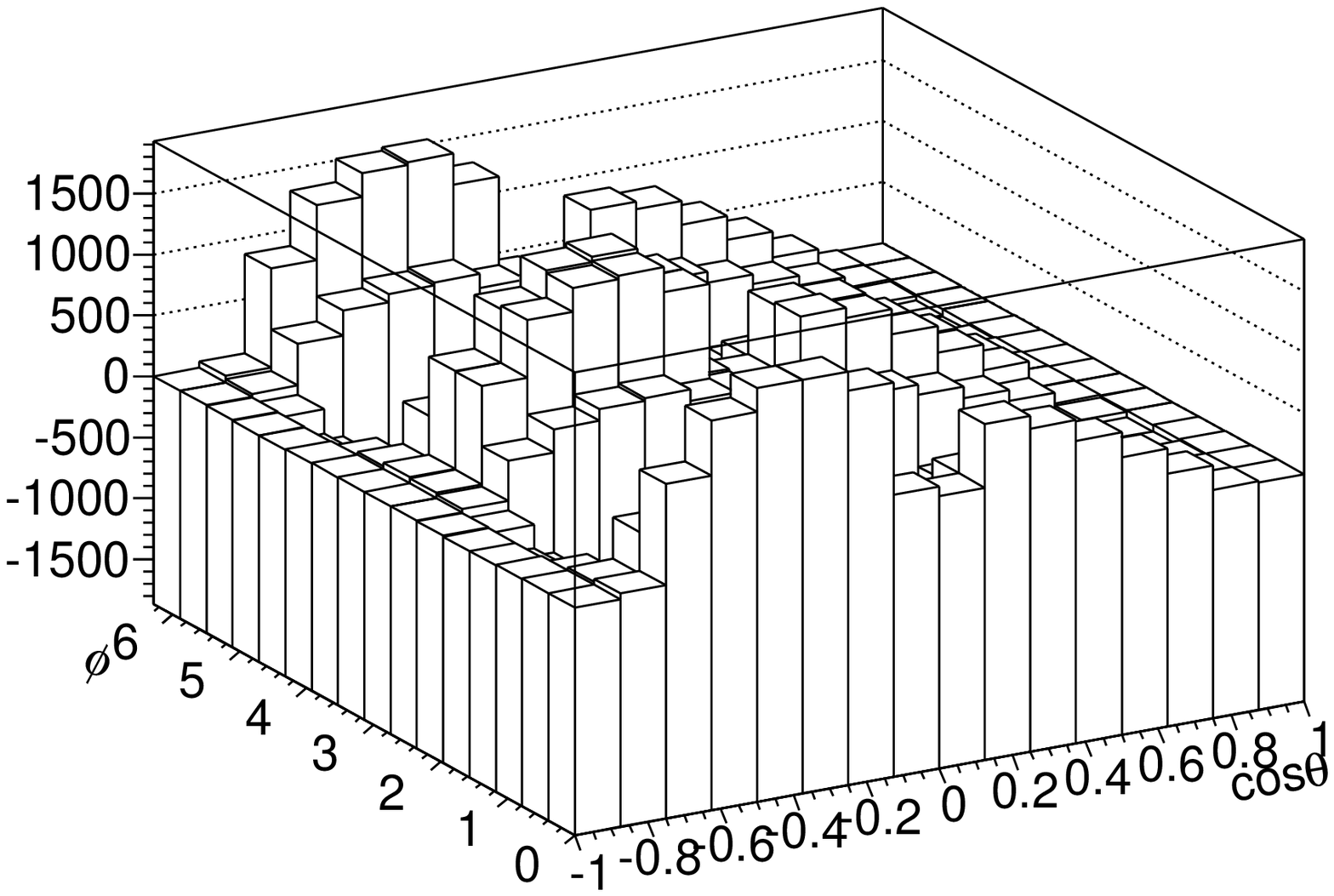}
   \includegraphics[width=5.0cm,angle=0]{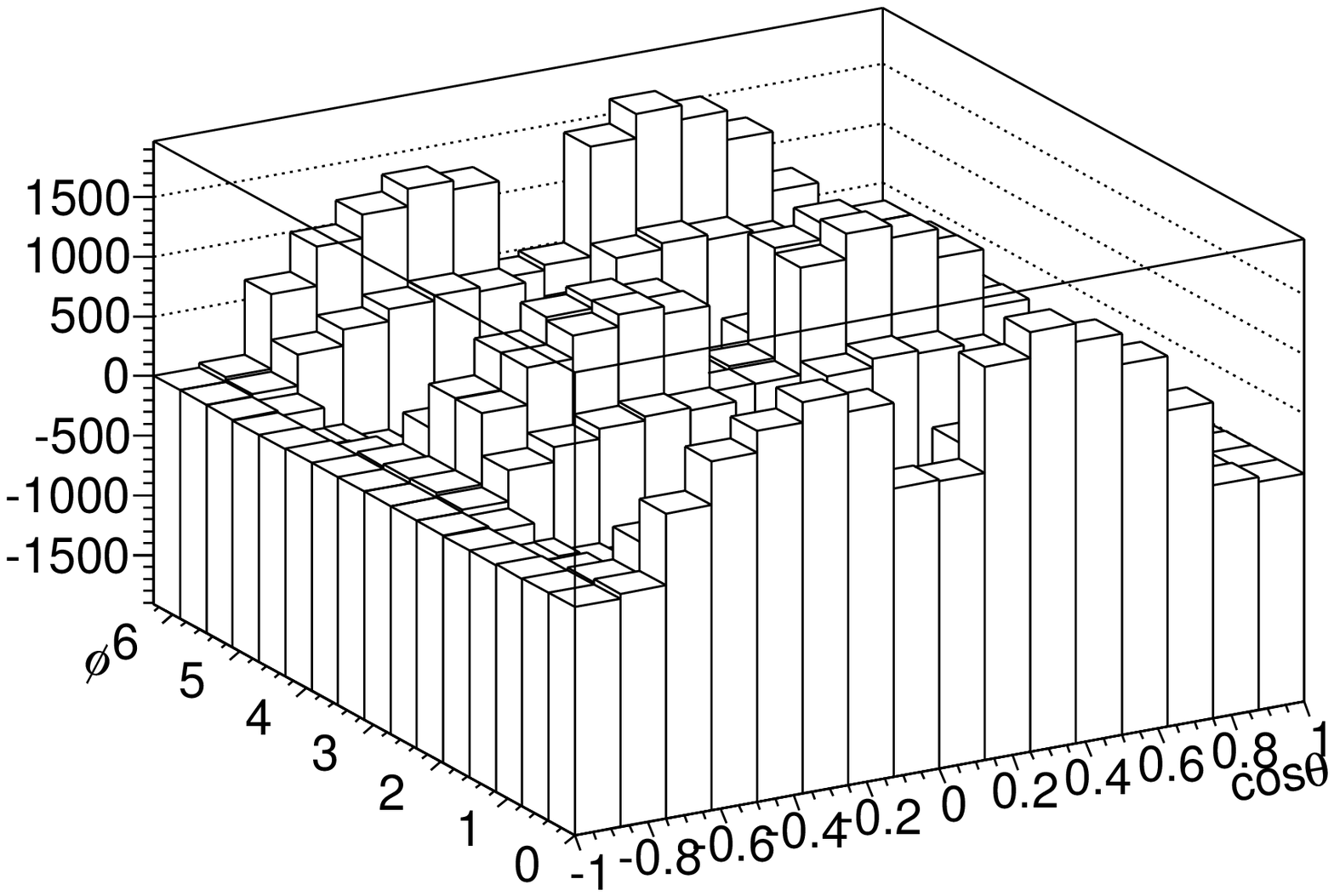}
   \includegraphics[width=5.0cm,angle=0]{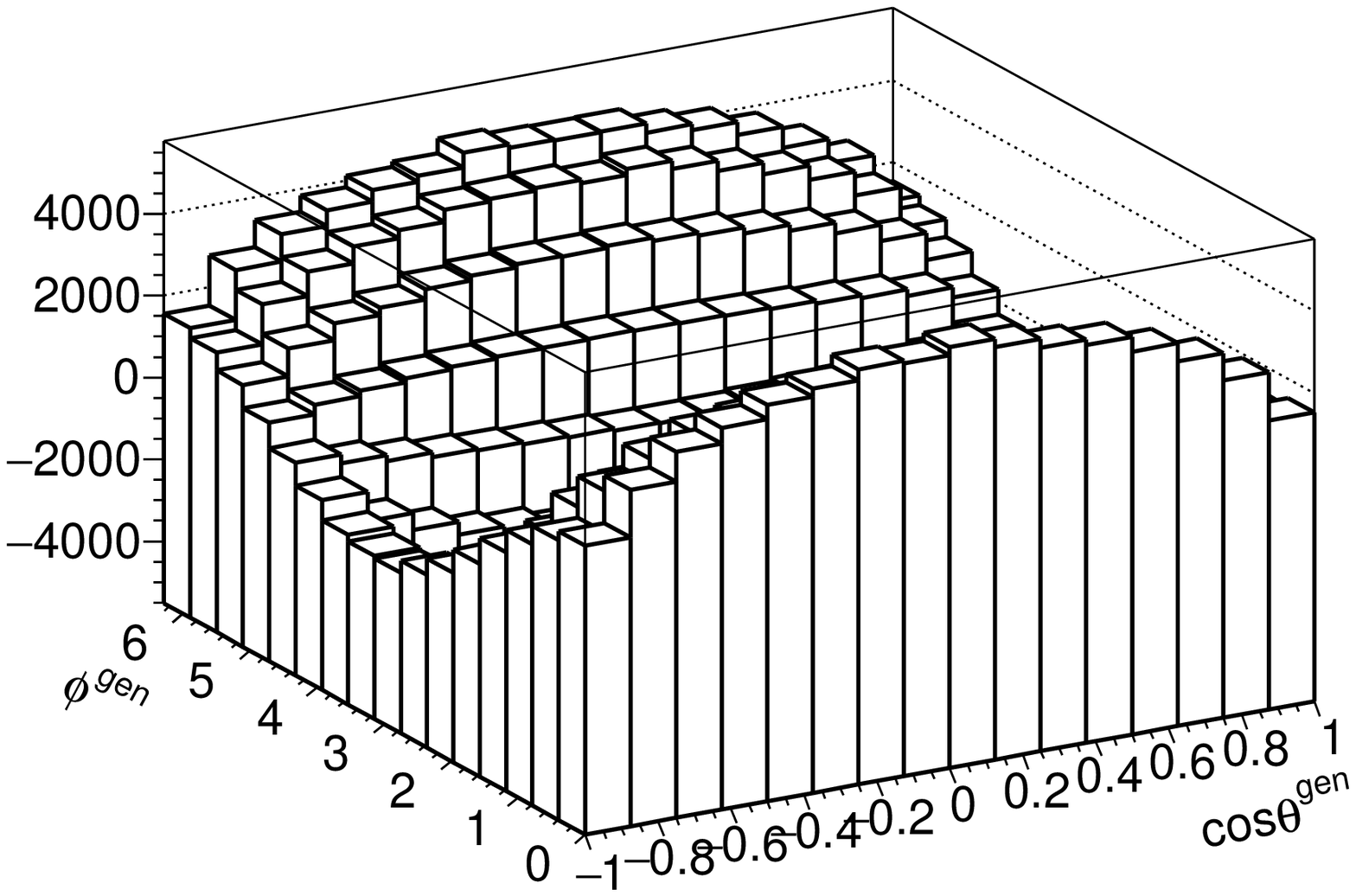}
   \includegraphics[width=5.0cm,angle=0]{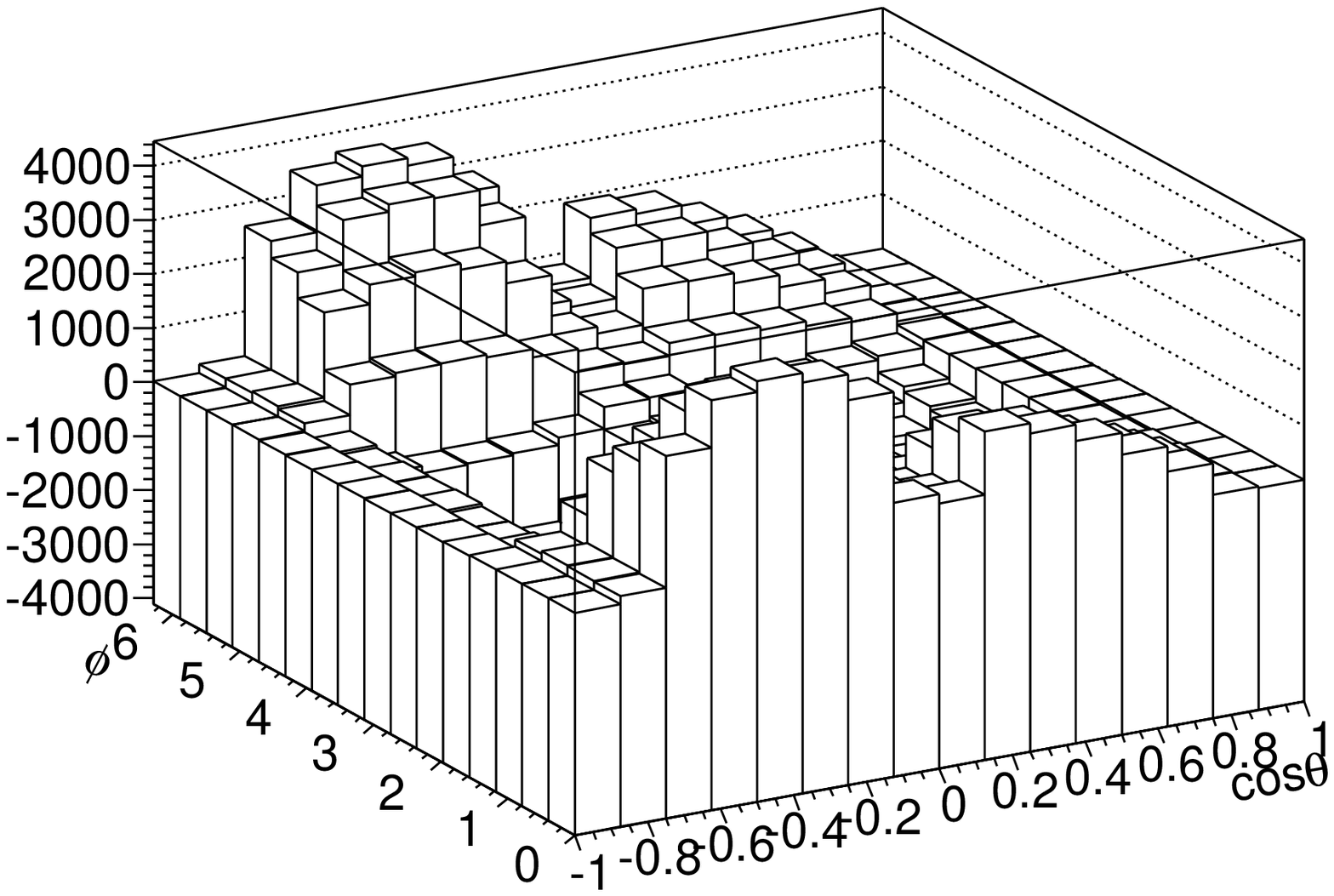}
   \includegraphics[width=5.0cm,angle=0]{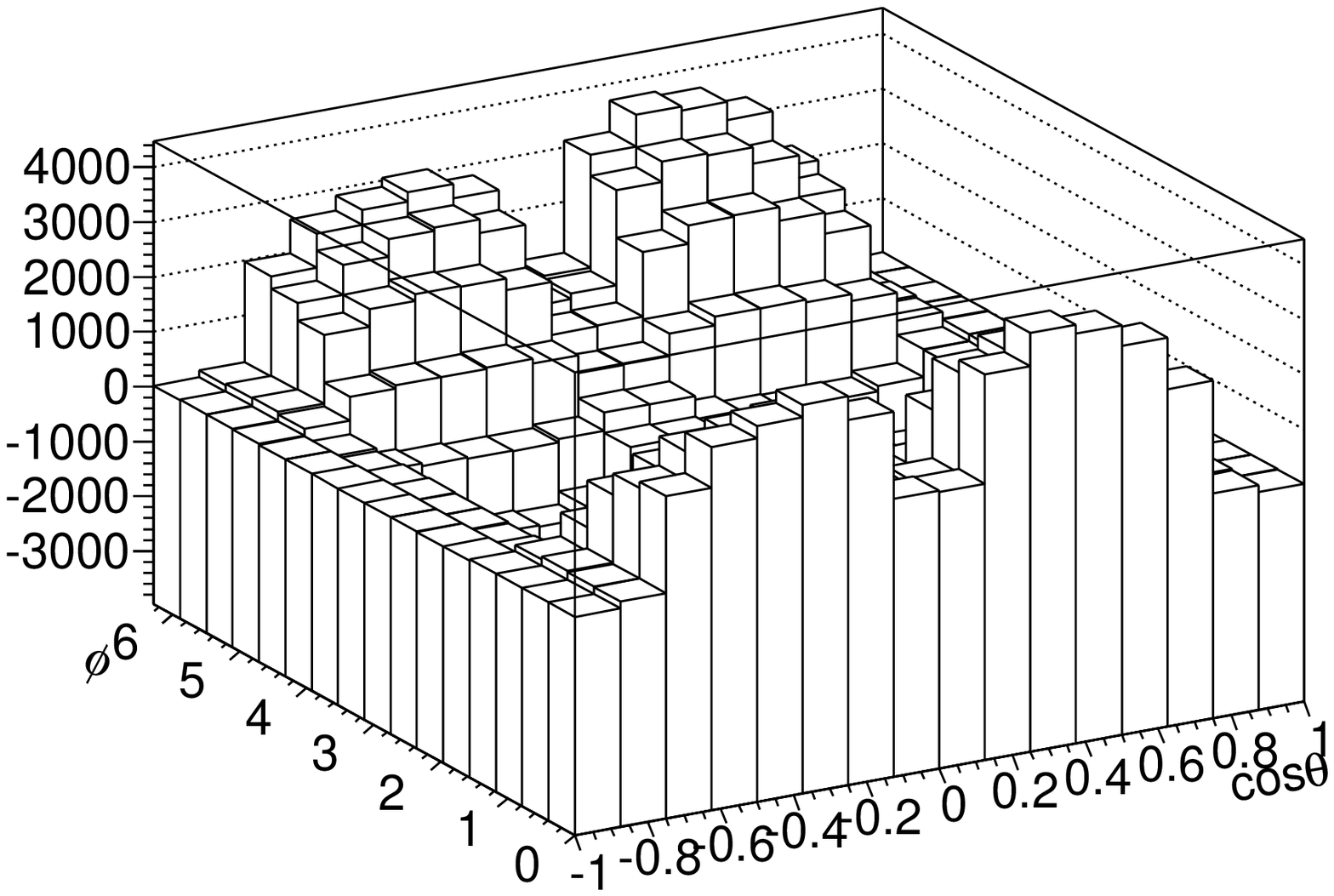}
}
\end{center}
\caption{Analytical shape of the polynomials $P_{i}$ in the full phase-space (left) and templates for polynomials
after reconstructing $p_Z^{\nu}$ with fiducial selection applied: for  $W^{-}$ (middle) and for  $W^{+}$ (right). 
From top  to bottom: $P_{1}, P_{2}, P_{3}$. 
\label{FigApp:Wmp_2Dtempl} }
\end{figure}

Figure~\ref{FigApp:Wmp_2Dtempl} collects examples of 2D distributions 
for polynomials $P_1(\cos\theta,\phi)$, $P_2(\cos\theta,\phi)$ and , $P_3(\cos\theta,\phi)$ 
in the full and fiducial phase-space. Now for each event we 
 reconstruct neutrino momenta using $m_W = m_W^{PDG}$, 
take randomly one of the solutions to 
recalculate $\theta, \phi$ angles from Eq.~(\ref{Eq:pznu}) and to apply kinematical selection of the fiducial phase-space. 

\begin{figure}
  \begin{center}                               
{
   \includegraphics[width=7.5cm,angle=0]{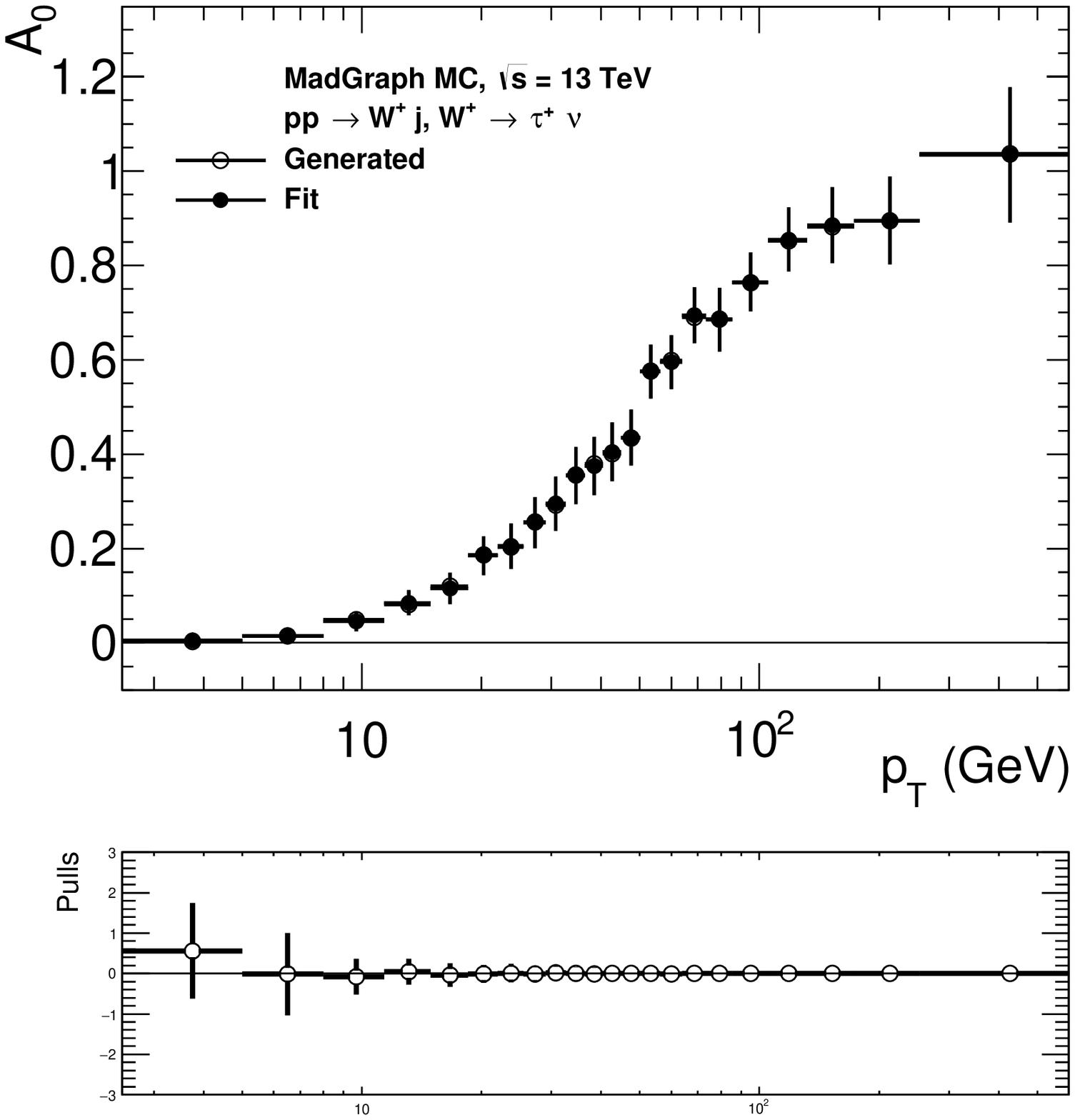}
   \includegraphics[width=7.5cm,angle=0]{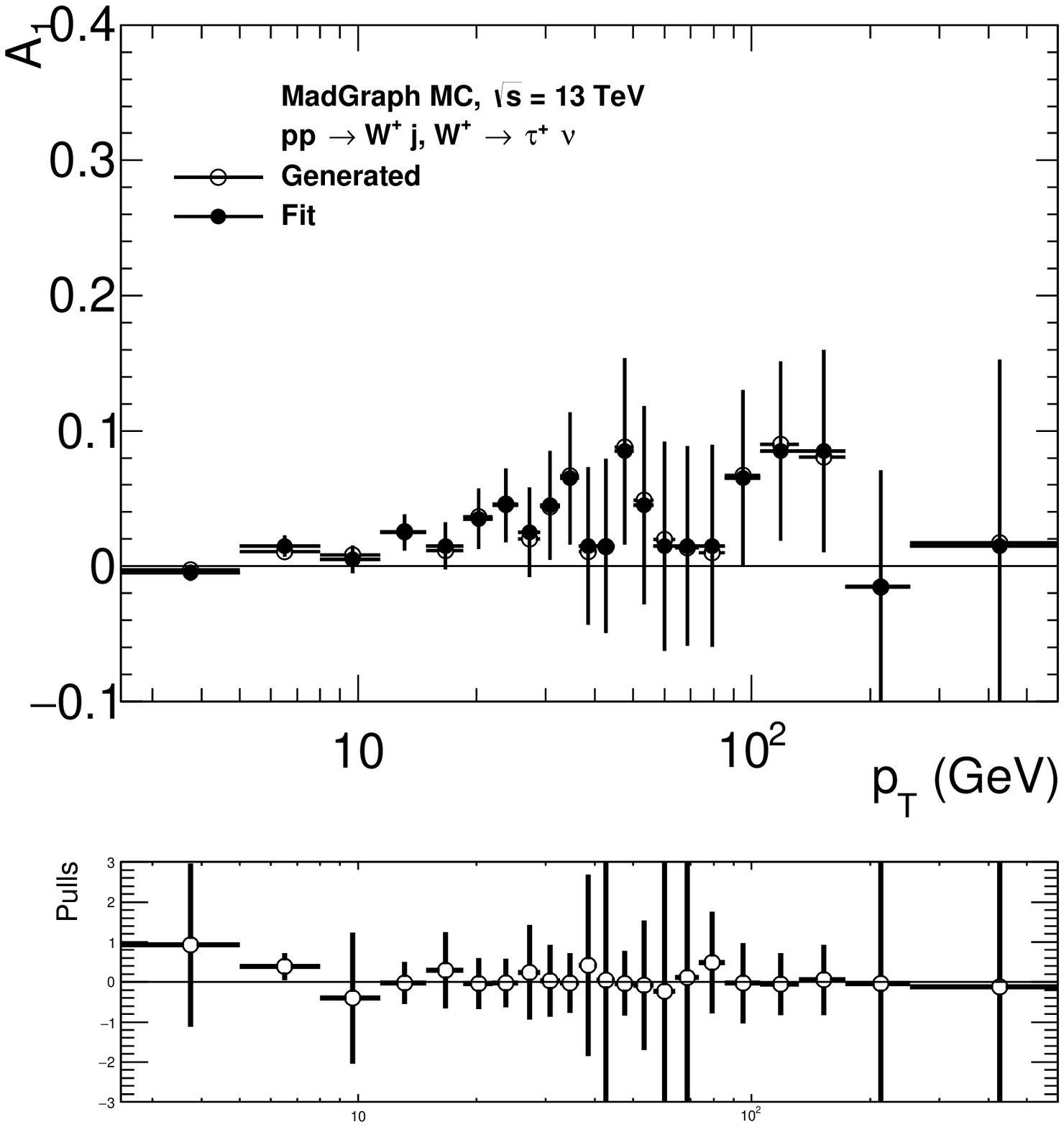}
   \includegraphics[width=7.5cm,angle=0]{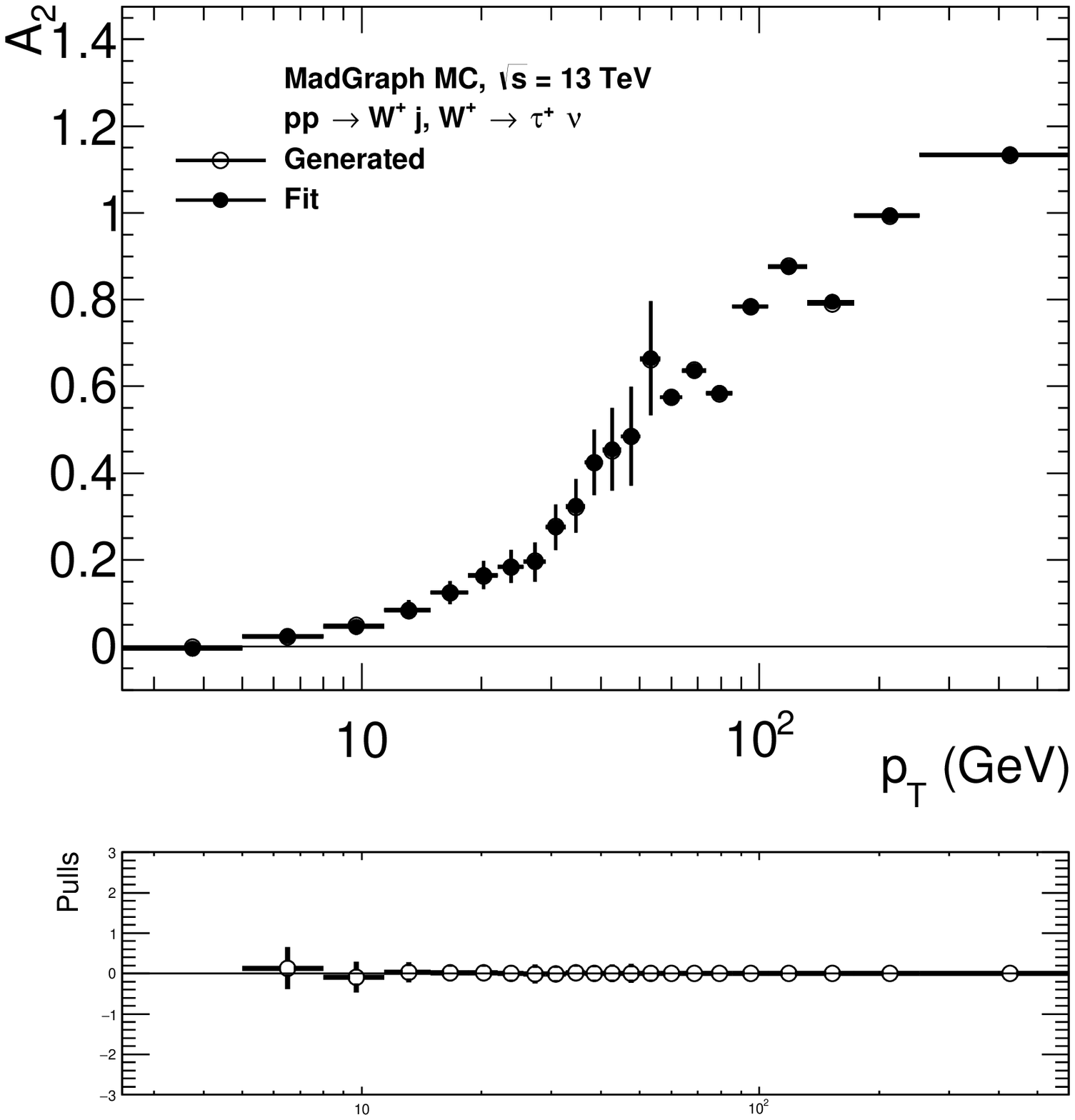}
   \includegraphics[width=7.5cm,angle=0]{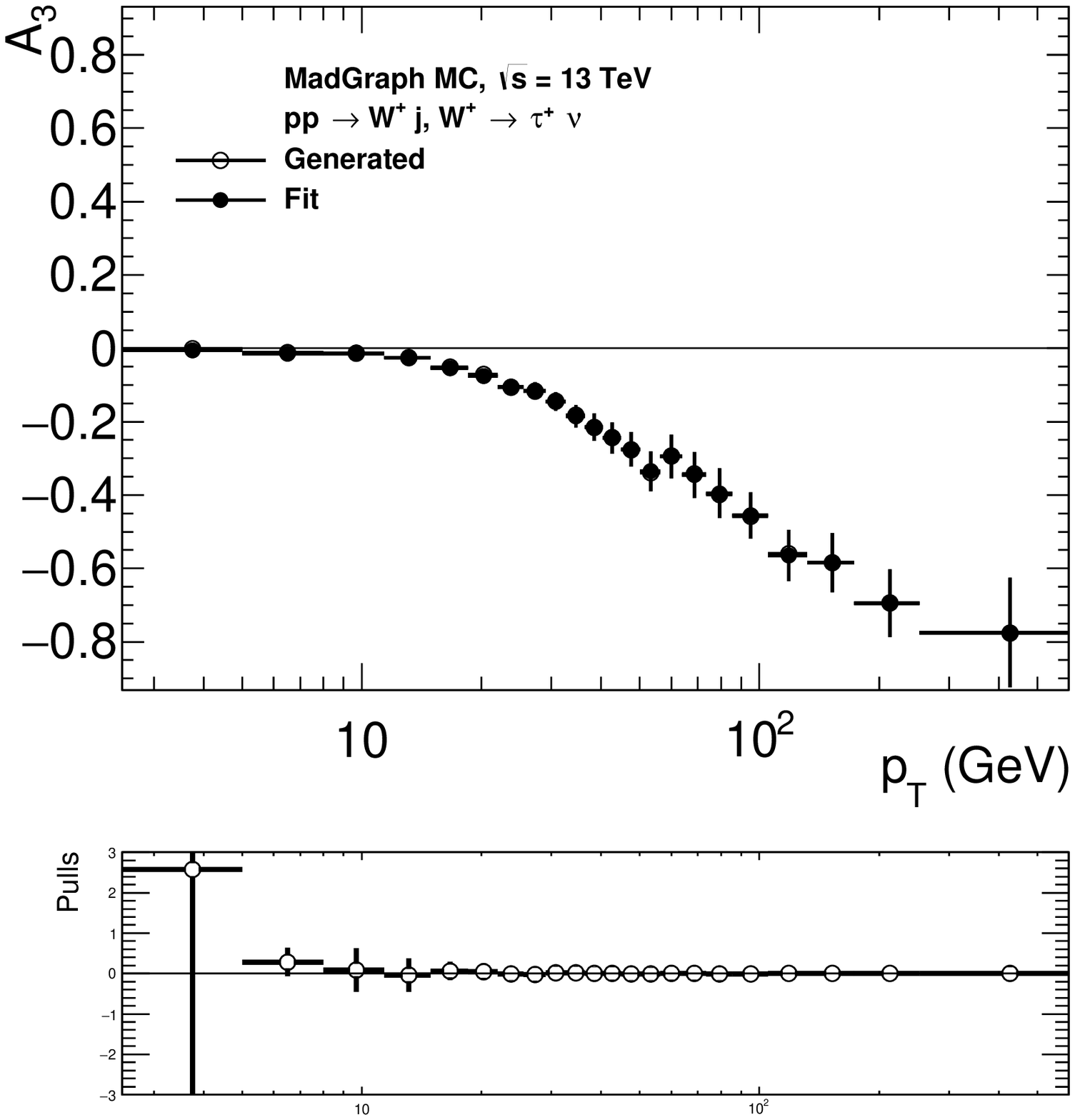}
   \includegraphics[width=7.5cm,angle=0]{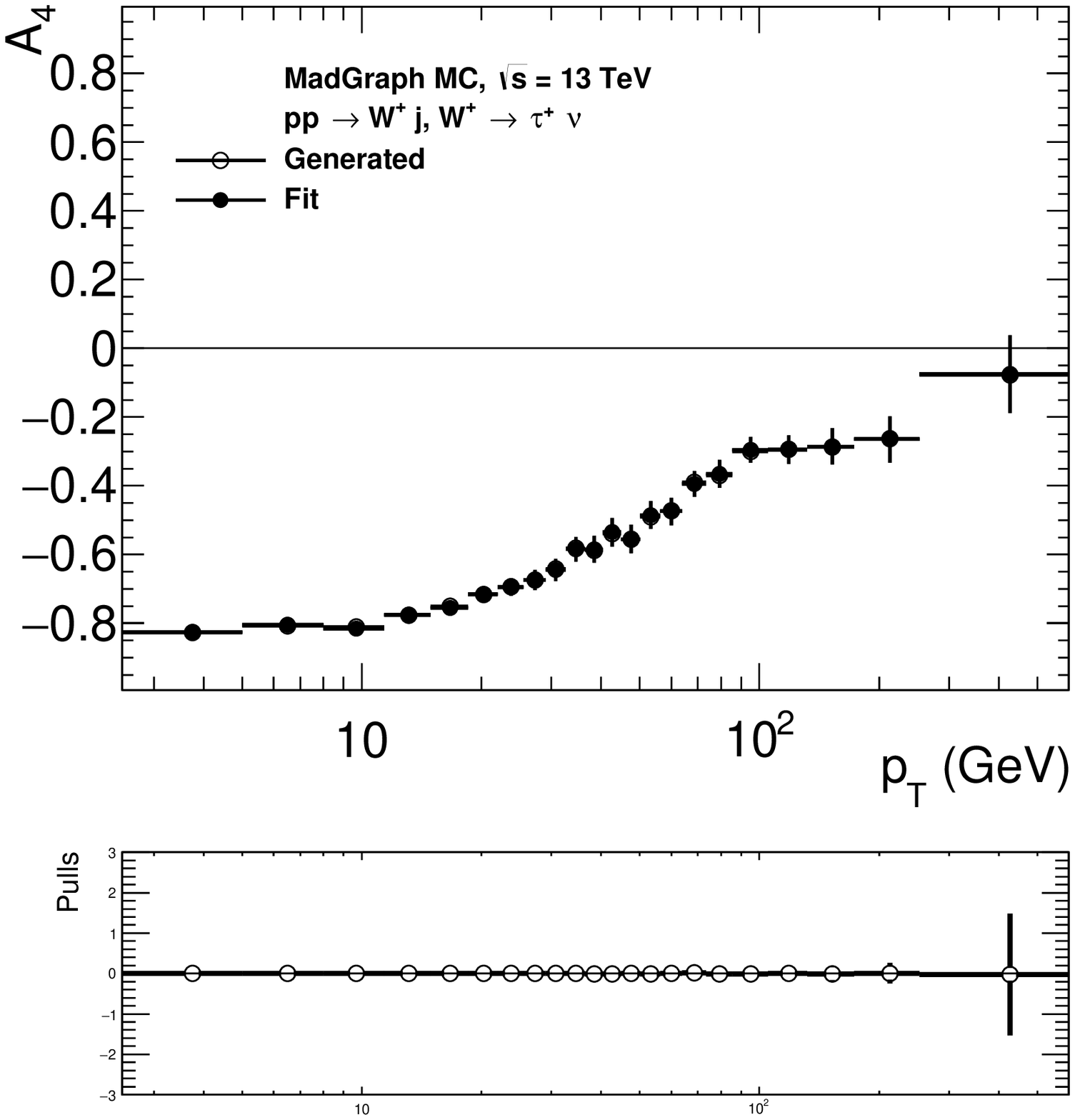}
}
\end{center}
\caption{As Fig.~\ref{Fig:WmLLfits} but for  $W^+ \to \ell^+ \nu$ events. 
Closure test on the fitting of angular coefficients $A_i$'s. Fit is performed in the fiducial phase-space. 
Shown are generated $A_i$'s coefficients (open circles) and their fitted values (black points). In the bottom panels shown are pulls 
(difference between generated and fit value, divided by the statistical error of the fit). 
Pulls are smaller than one could expect. This is because events of pseudo-data and templates are statistically correlated.
\label{FigApp:WpLLfits} }
\end{figure}

Figure~\ref{FigApp:WpLLfits} collects results of the multi-likelihood fit of $W^+ \to \ell^+ \nu$, displayed are $A_i$
coefficients as function of $p_T^W$. 

\section{Additional plots on $A_i$s  coefficients}
\label{App:Ais}

\begin{figure}
  \begin{center}                               
{
   \includegraphics[width=7.5cm,angle=0]{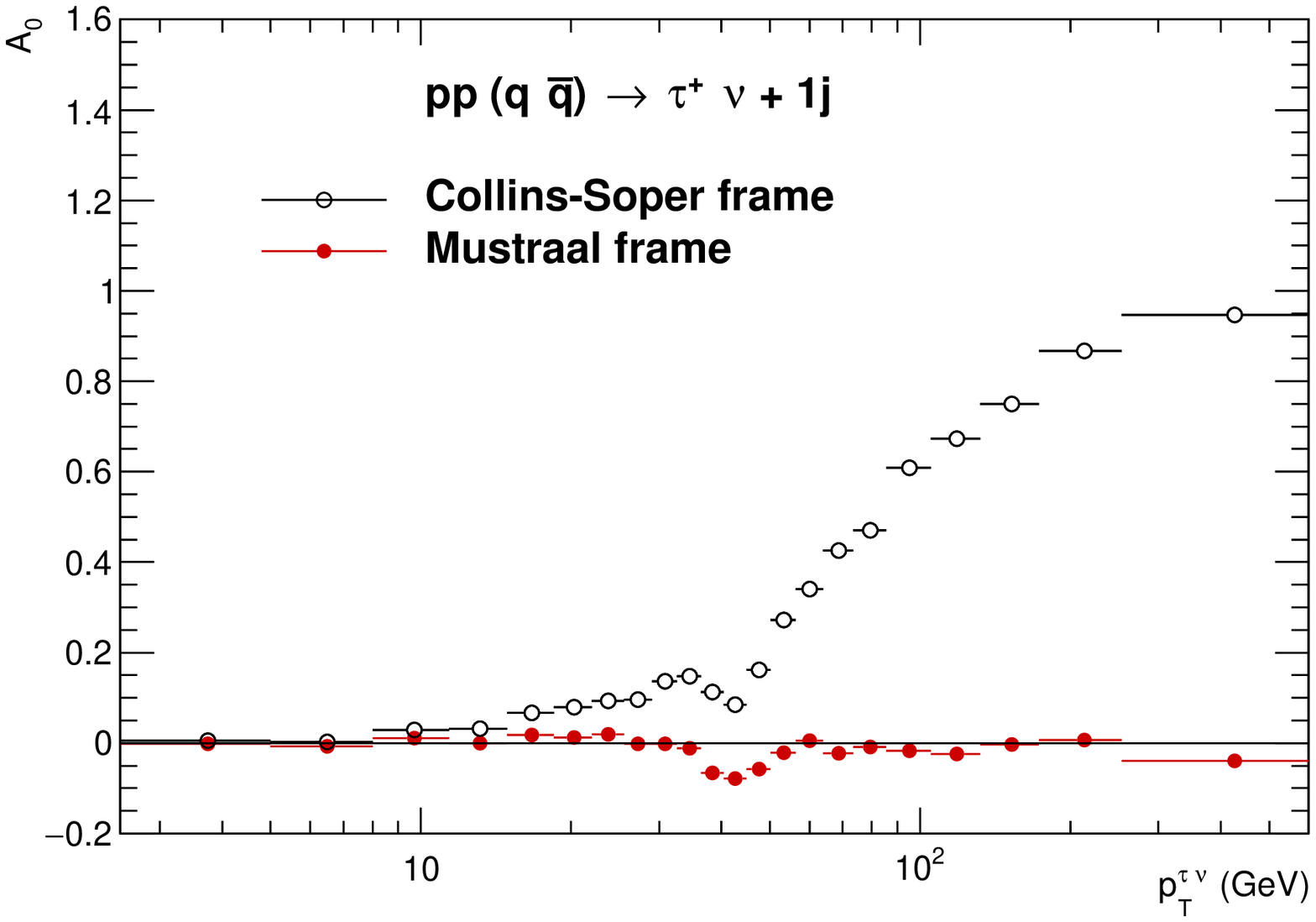}
   \includegraphics[width=7.5cm,angle=0]{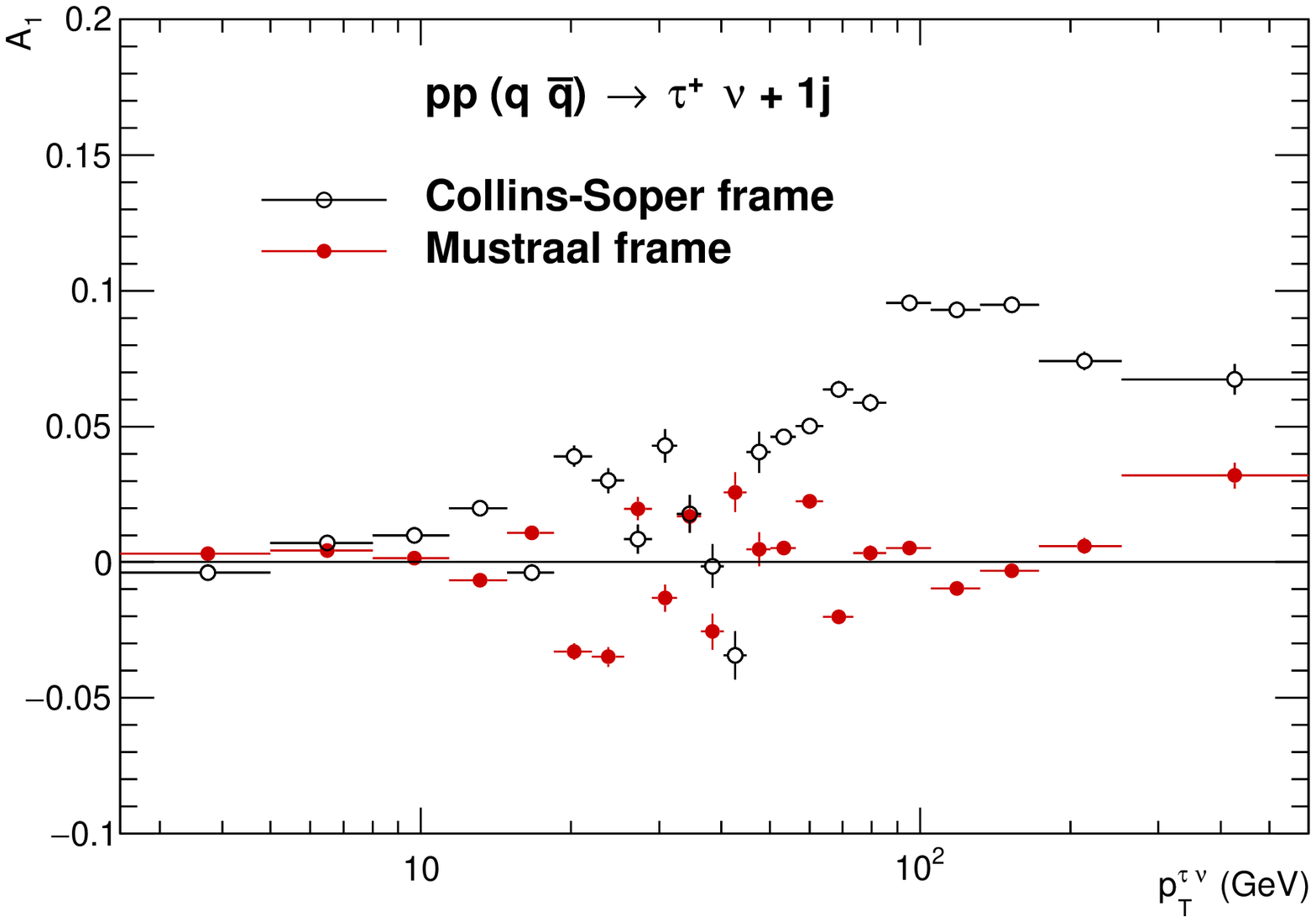}
   \includegraphics[width=7.5cm,angle=0]{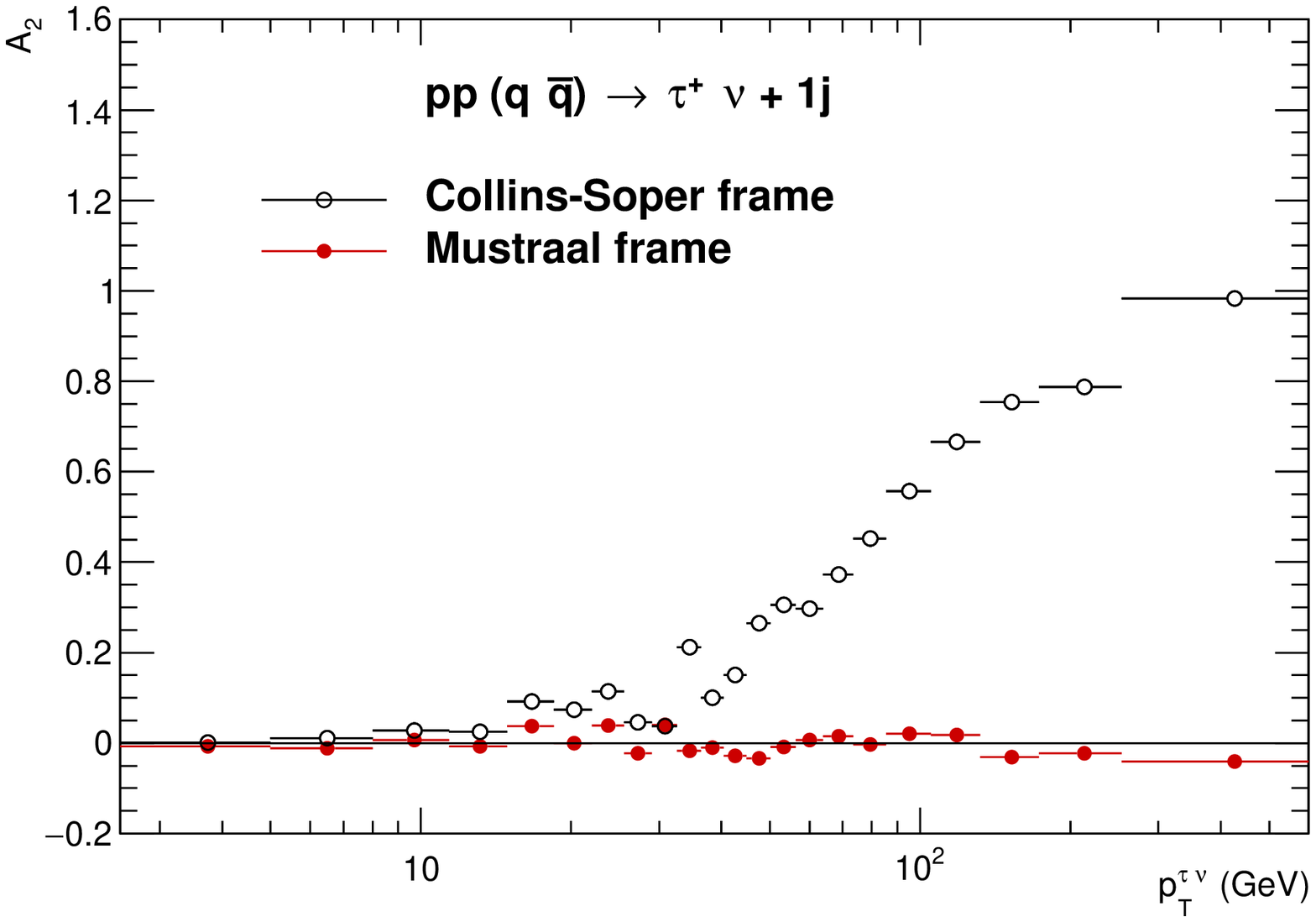}
   \includegraphics[width=7.5cm,angle=0]{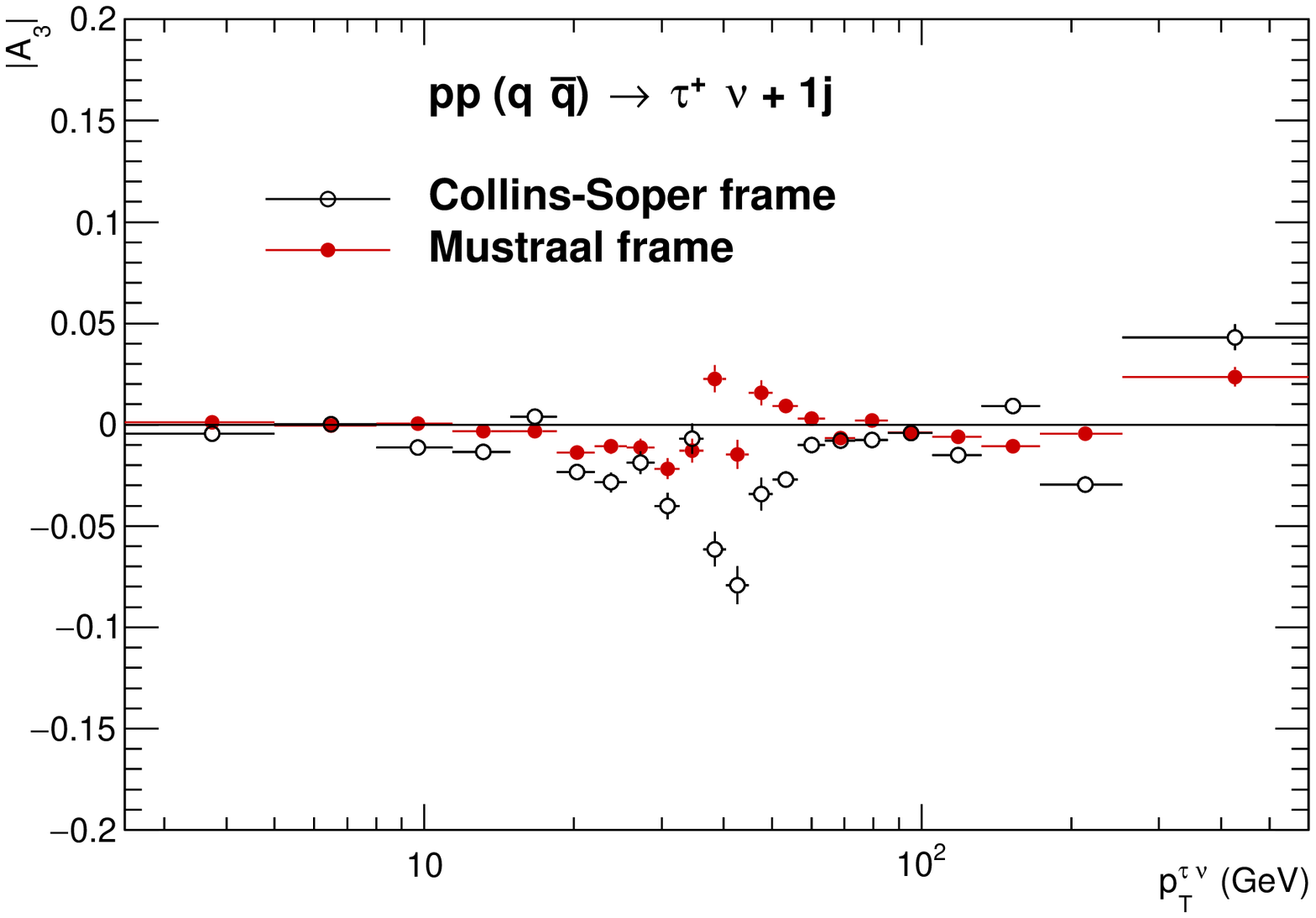}
   \includegraphics[width=7.5cm,angle=0]{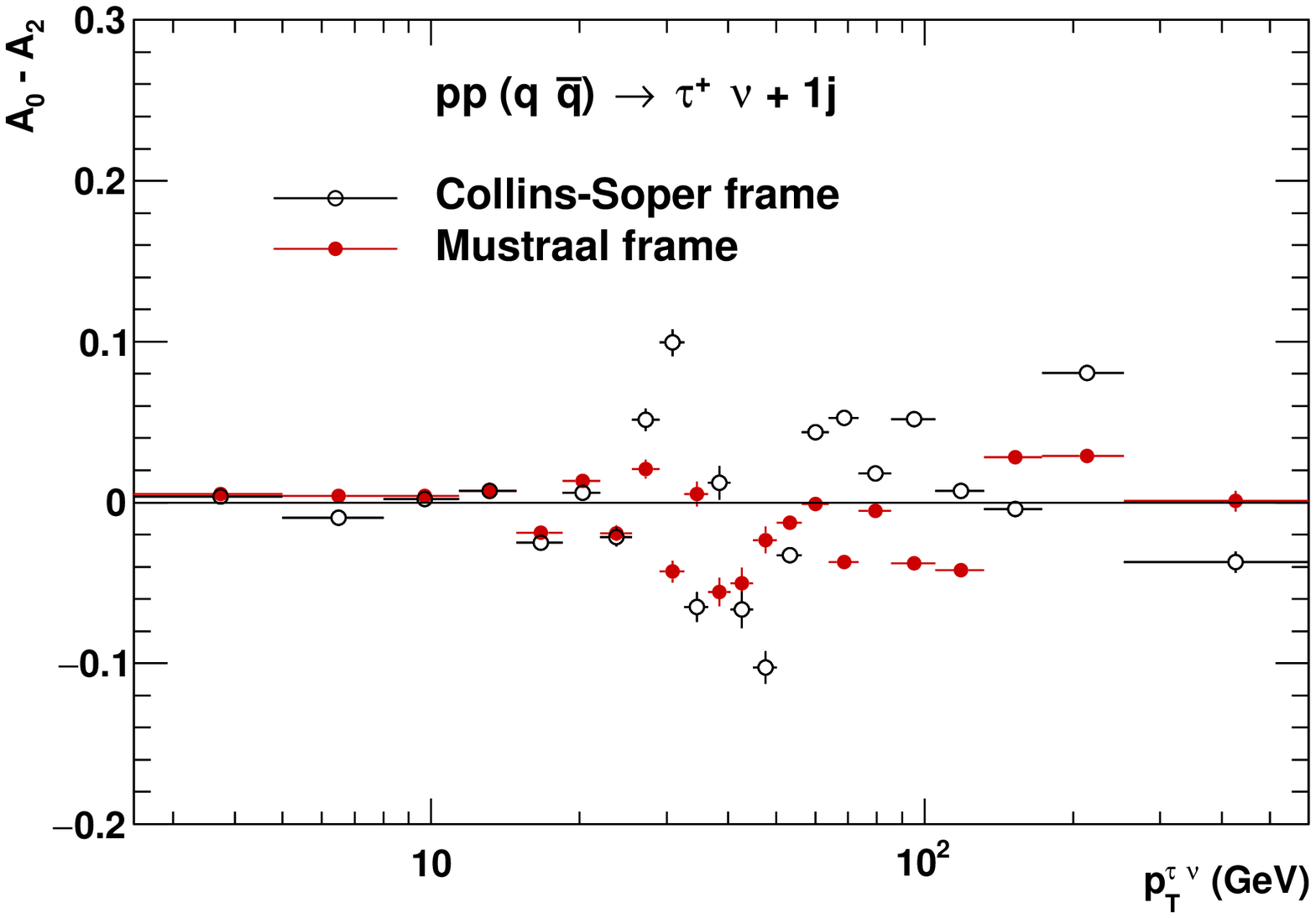}
   \includegraphics[width=7.5cm,angle=0]{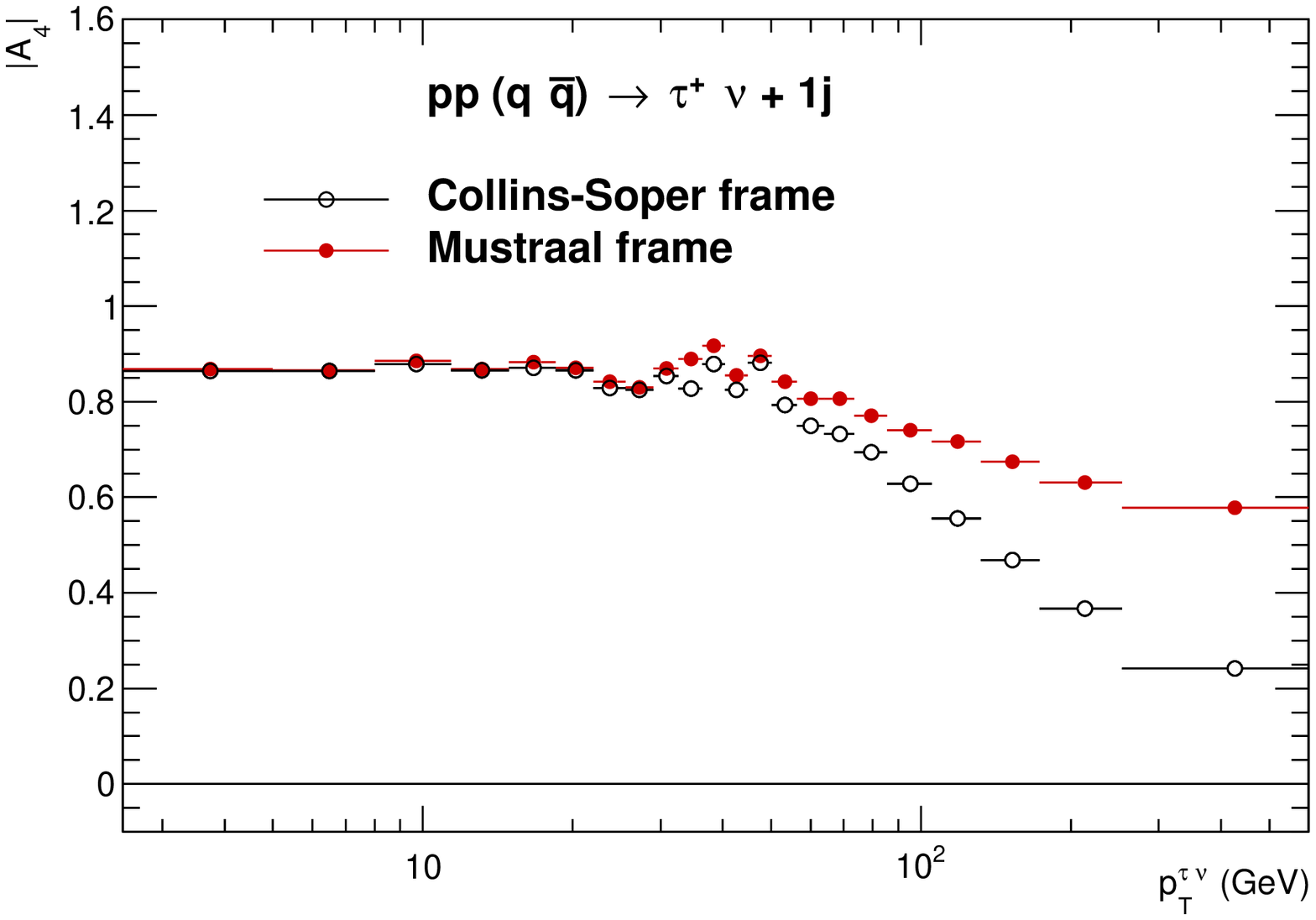}
}
\end{center}
\caption{ 
As Fig.~\ref{Fig:Ai1jet} but for $W^+$ and parton level subprocess only.
The $A_i$ coefficients  calculated in Collins-Soper (black) and in {\tt Mustraal} (red) frames 
for selected parton level process $p p ( q \bar q)  \to \tau^{+} \nu\ + 1j$ generated with {\tt MadGraph}.
\label{FigApp:Ai1quark} }
\end{figure}

\begin{figure}
  \begin{center}                               
{
   \includegraphics[width=7.5cm,angle=0]{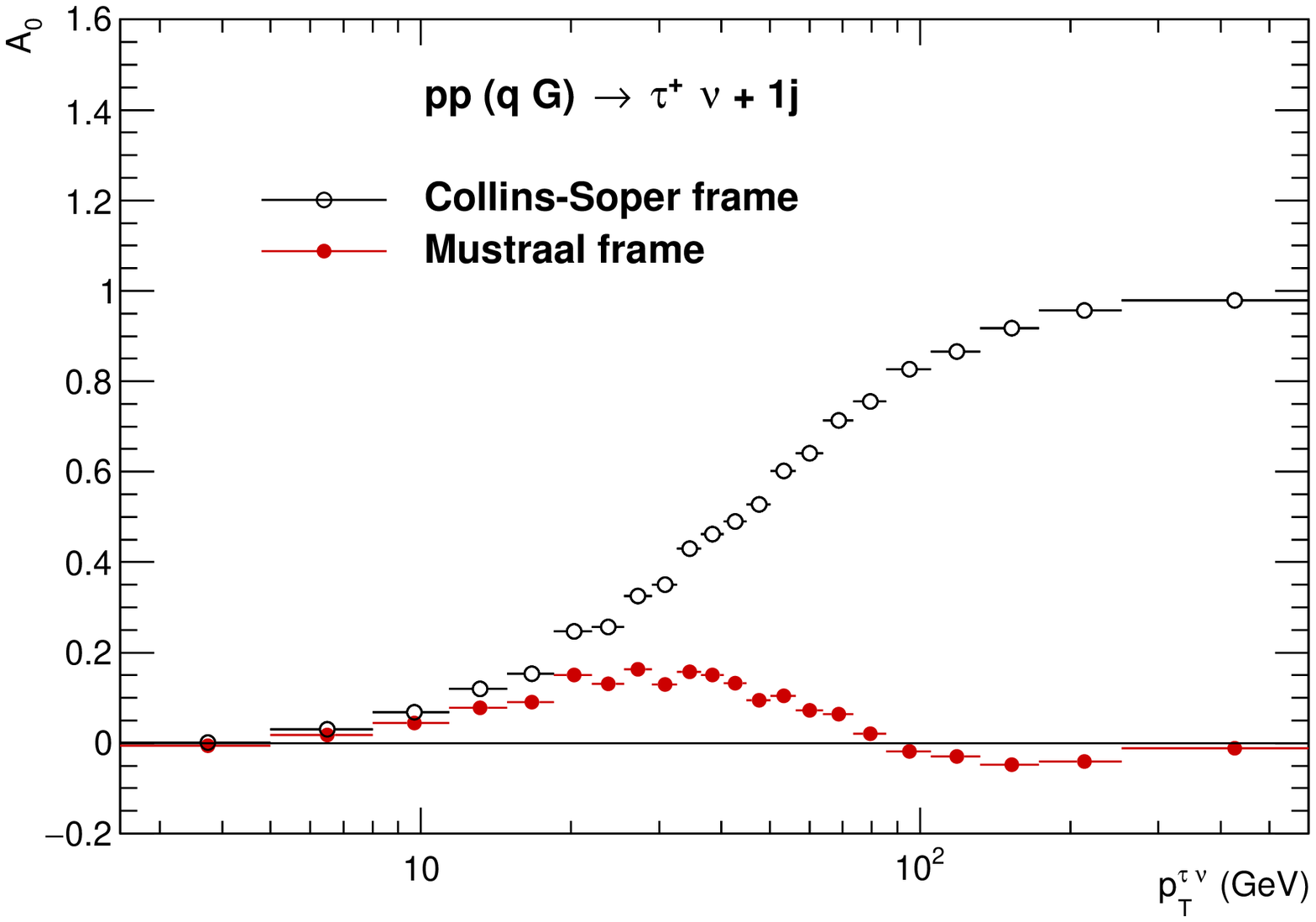}
   \includegraphics[width=7.5cm,angle=0]{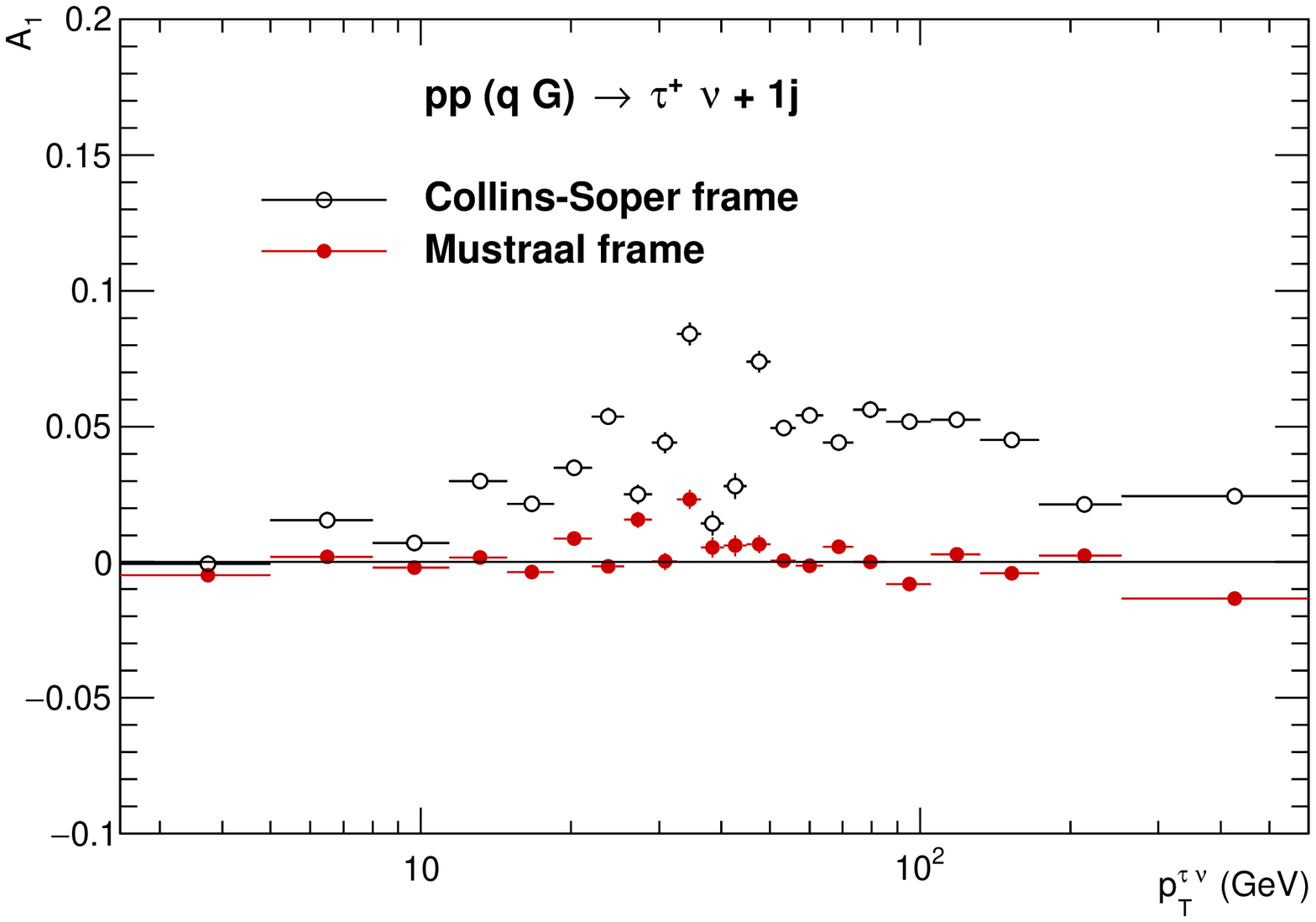}
   \includegraphics[width=7.5cm,angle=0]{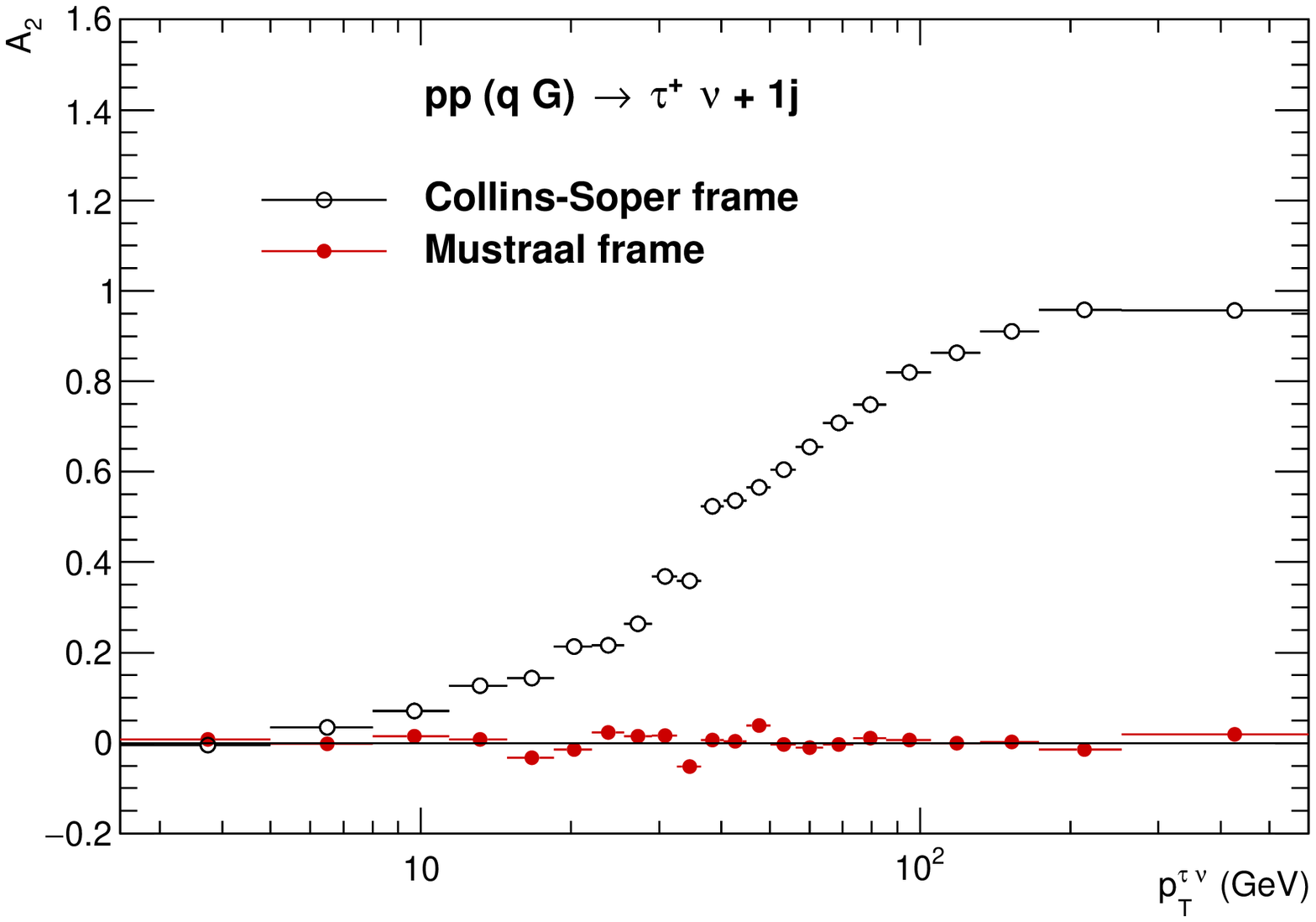}
   \includegraphics[width=7.5cm,angle=0]{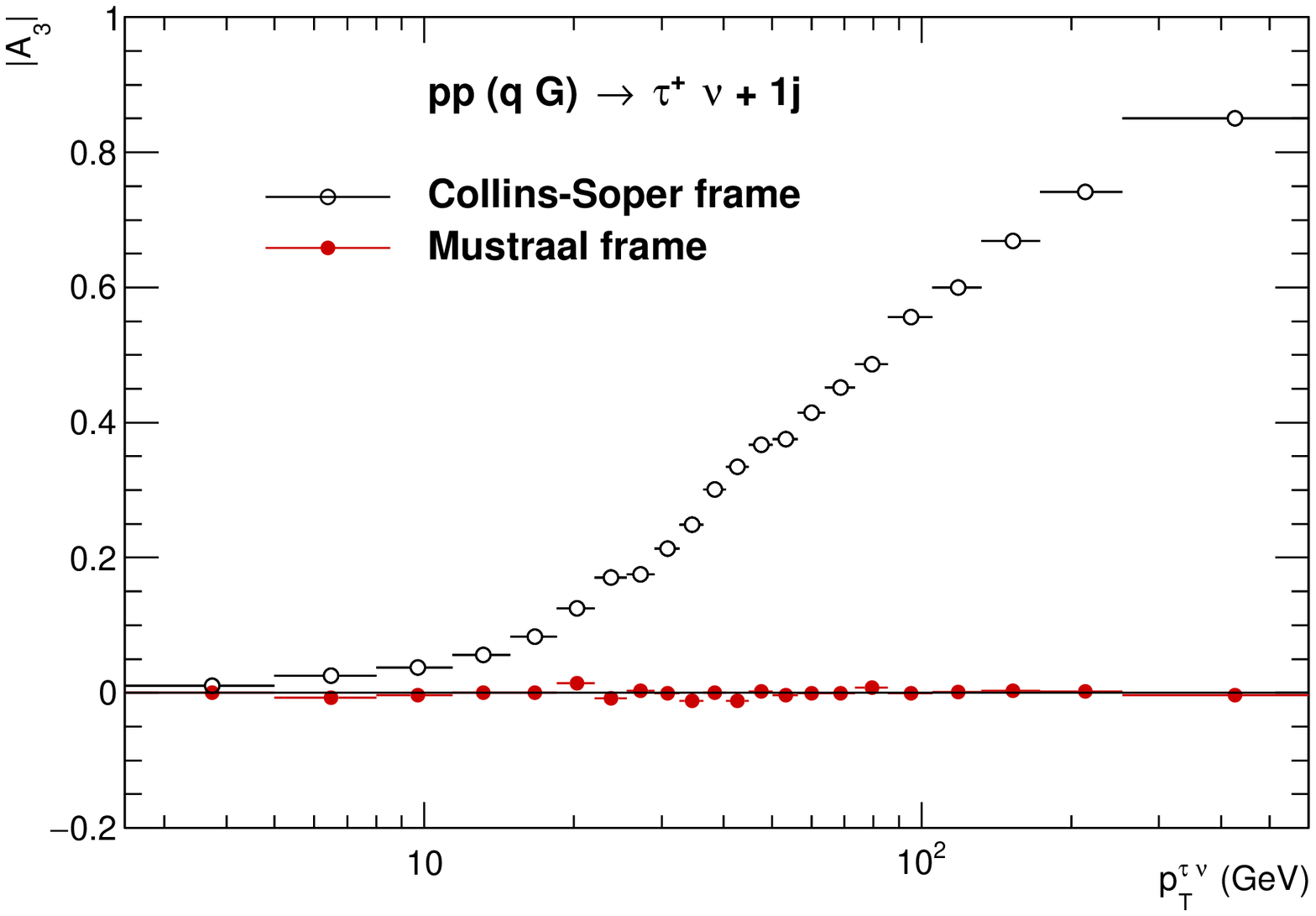}
   \includegraphics[width=7.5cm,angle=0]{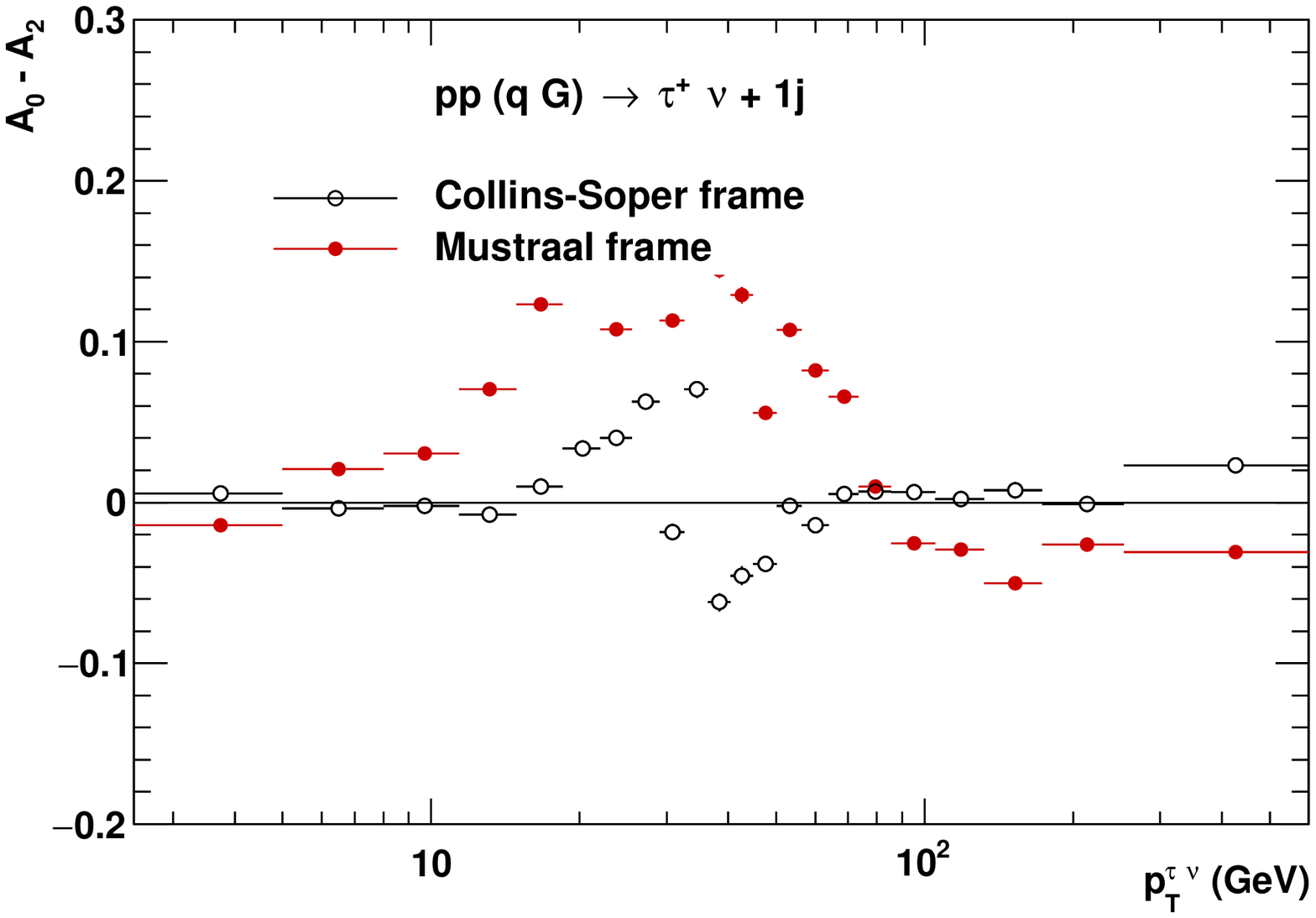}
   \includegraphics[width=7.5cm,angle=0]{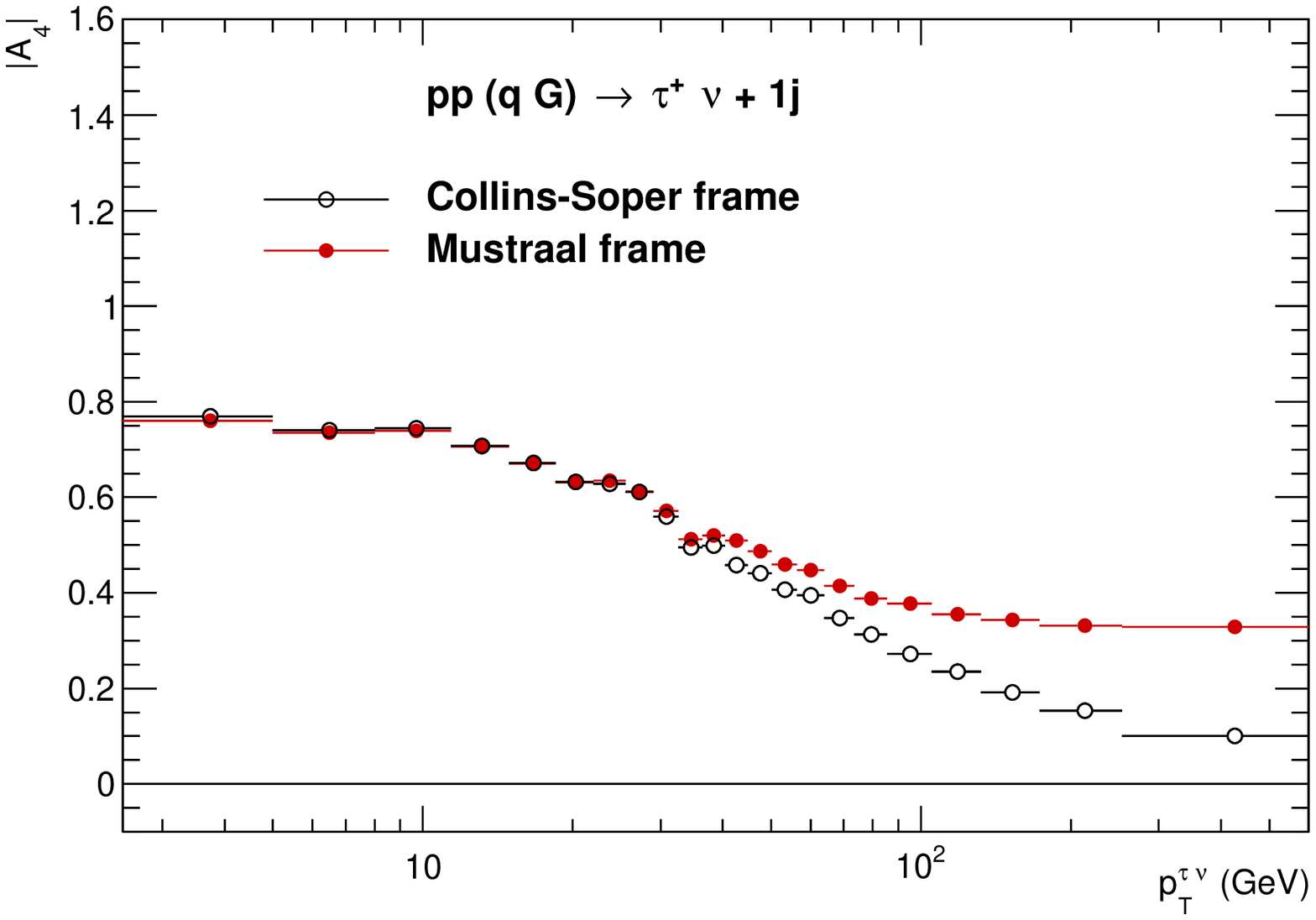}
}
\end{center}
\caption{ 
As Fig.~\ref{Fig:Ai1jet} but for $W^+$ and parton level subprocess only.
The $A_i$ coefficients  calculated in Collins-Soper (black) and in {\tt Mustraal} (red) frames 
for selected parton level process $p p (G \bar q) \to \tau^{+} \nu \ + 1j$ generated with {\tt MadGraph}.
\label{FigApp:Ai1gluon} }
\end{figure}

In the generated sample information on incoming and outgoing partons flavours
is stored. We will use this information for tests, to define sub-samples of $q \bar q$ and $q (\bar q) G$ parton level initial states.
Figures~\ref{FigApp:Ai1quark} and ~\ref{FigApp:Ai1gluon} show predictions for $A_i$'s coefficients for 
$W^+ \to \tau^+ \nu$  and events generated 
with {\tt MadGraph} Monte Carlo for these sub-samples. 

\begin{figure}
  \begin{center}                               
{
   \includegraphics[width=7.5cm,angle=0]{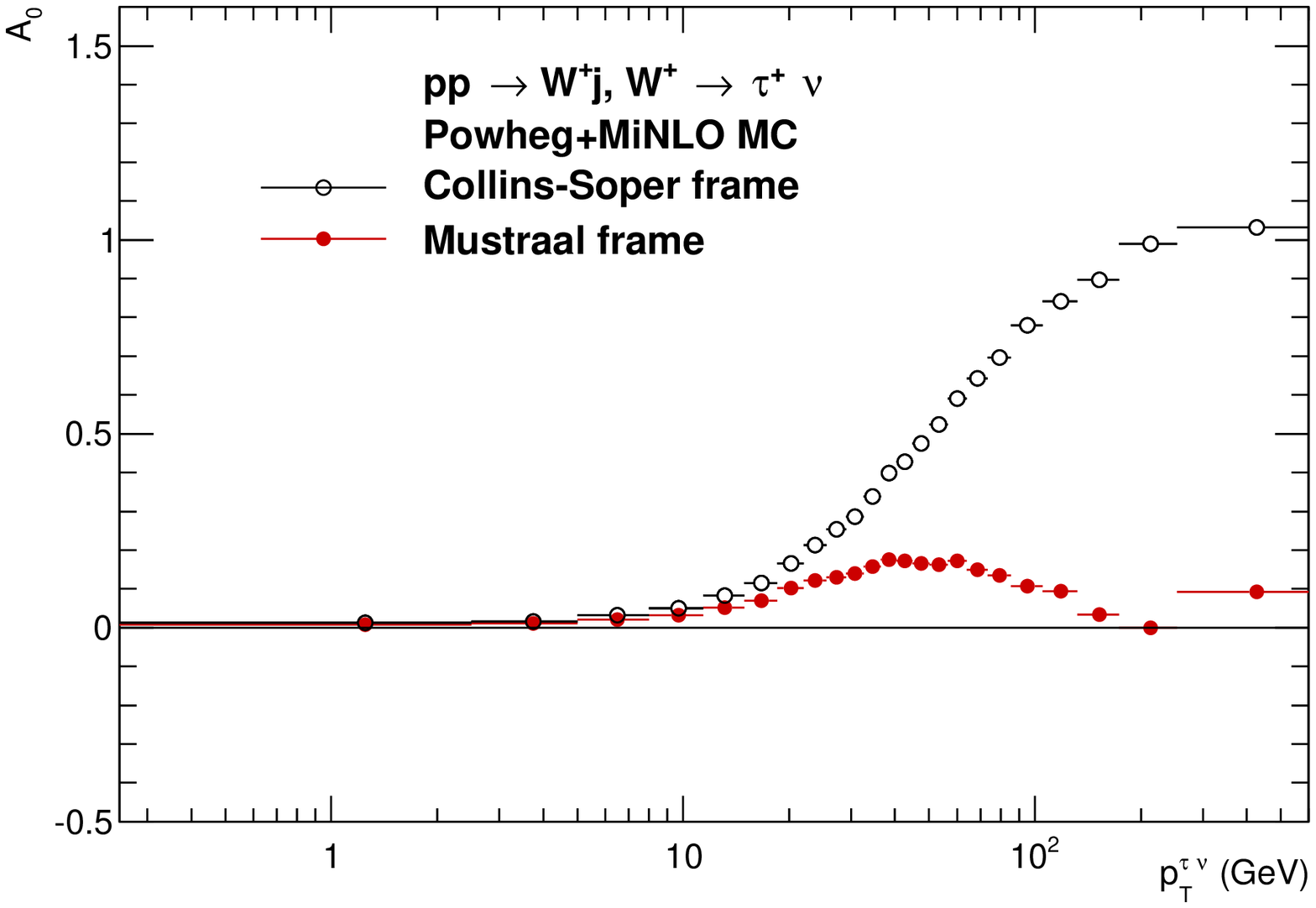}
   \includegraphics[width=7.5cm,angle=0]{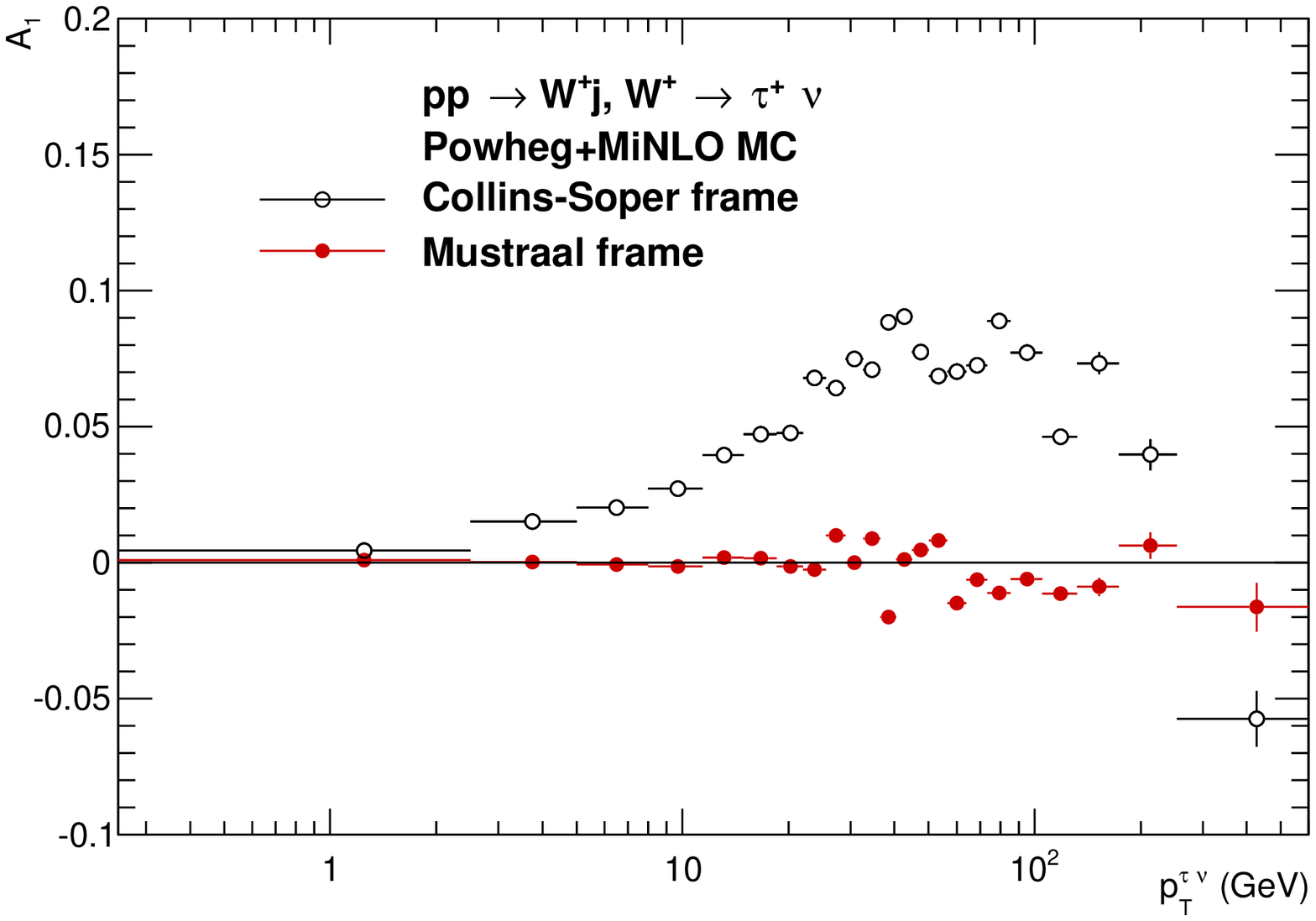}
   \includegraphics[width=7.5cm,angle=0]{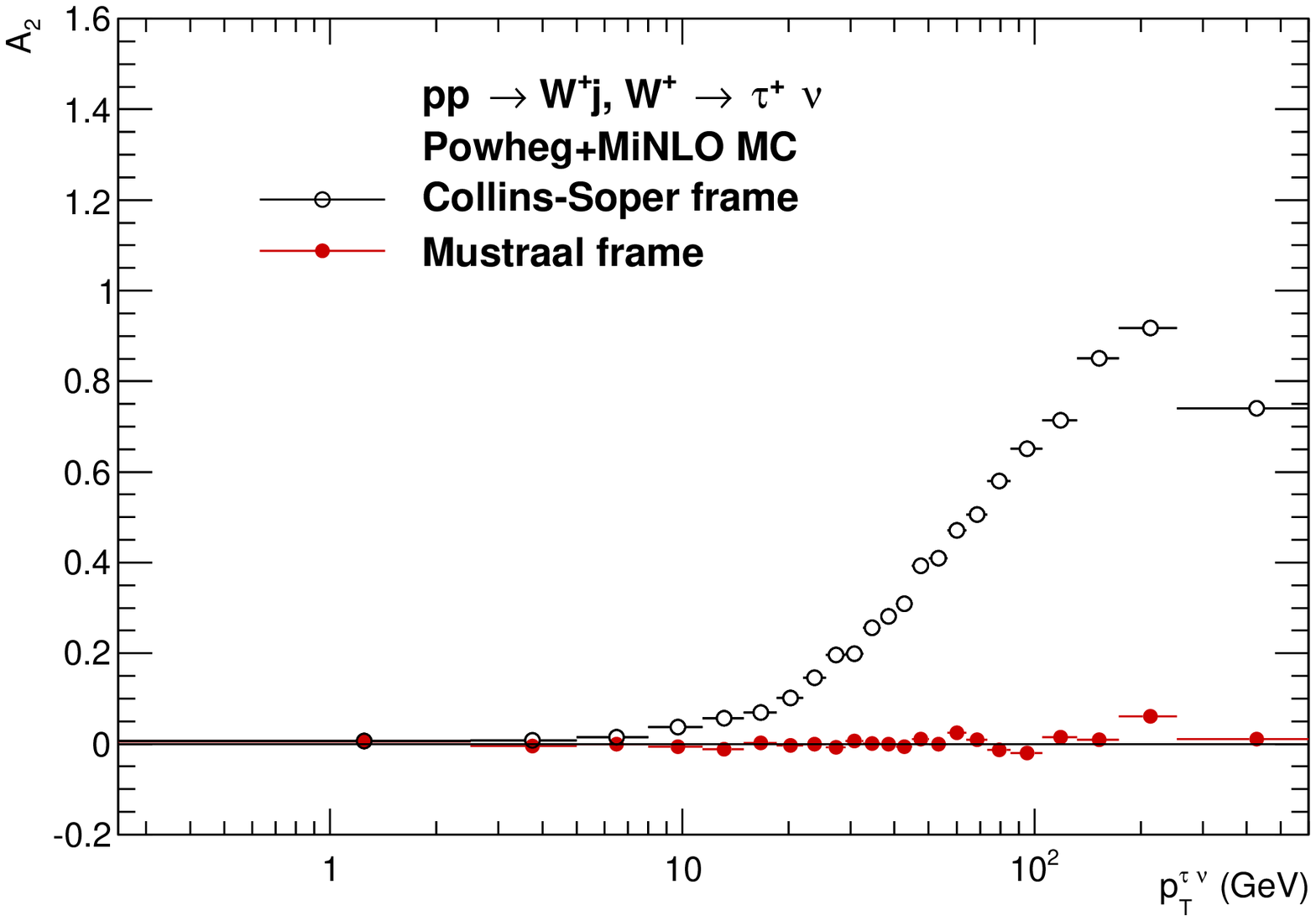}
   \includegraphics[width=7.5cm,angle=0]{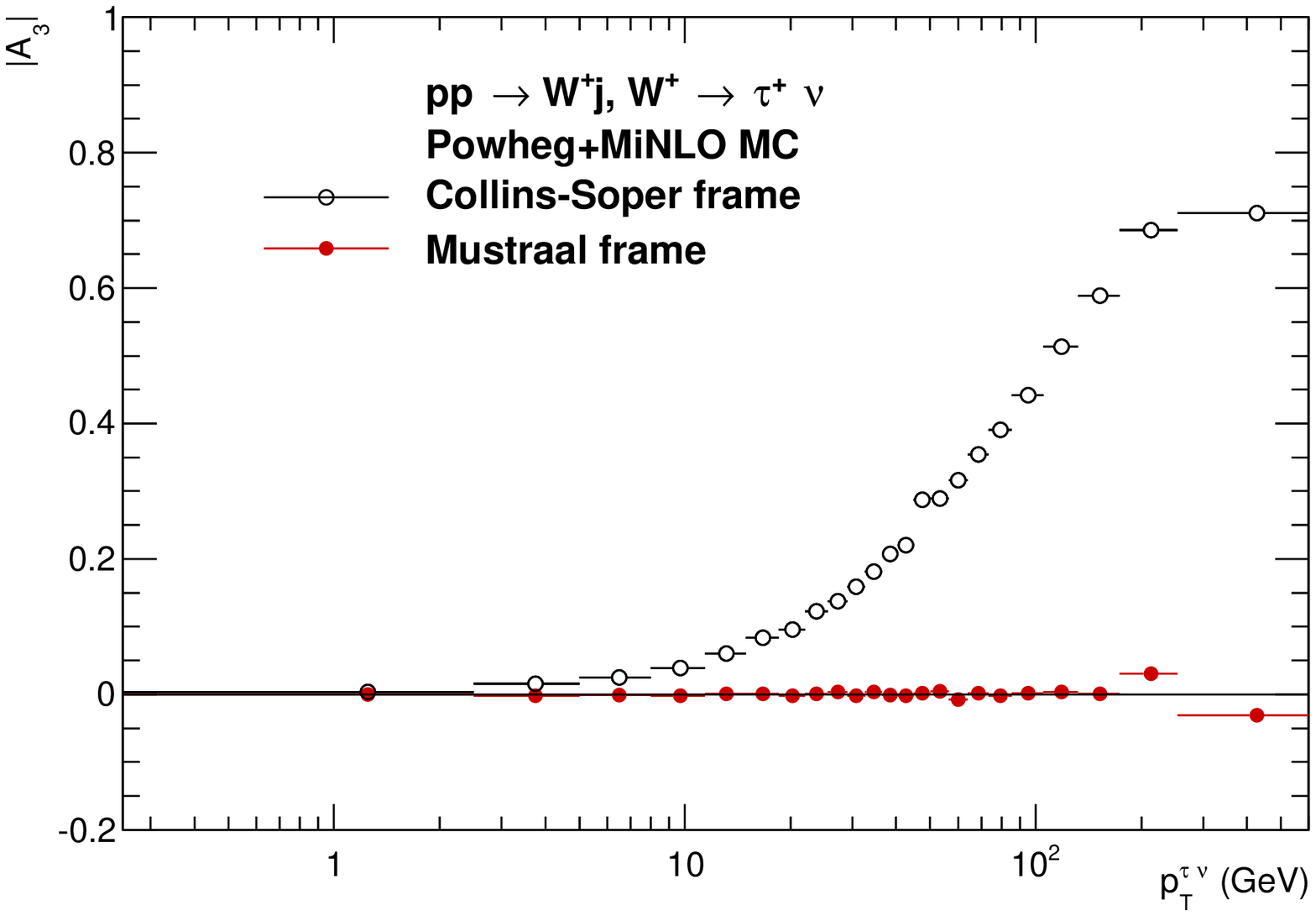}
   \includegraphics[width=7.5cm,angle=0]{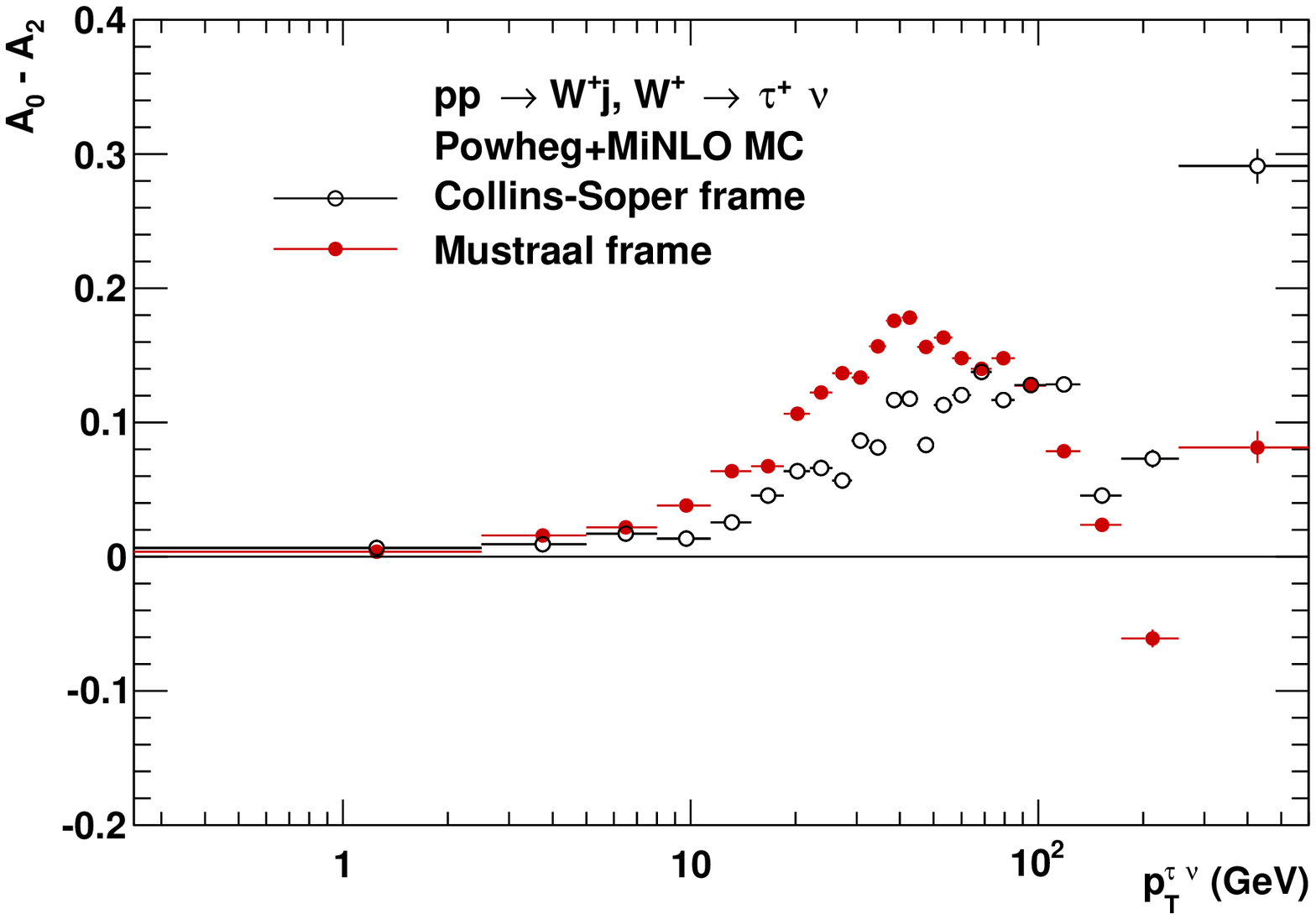}
   \includegraphics[width=7.5cm,angle=0]{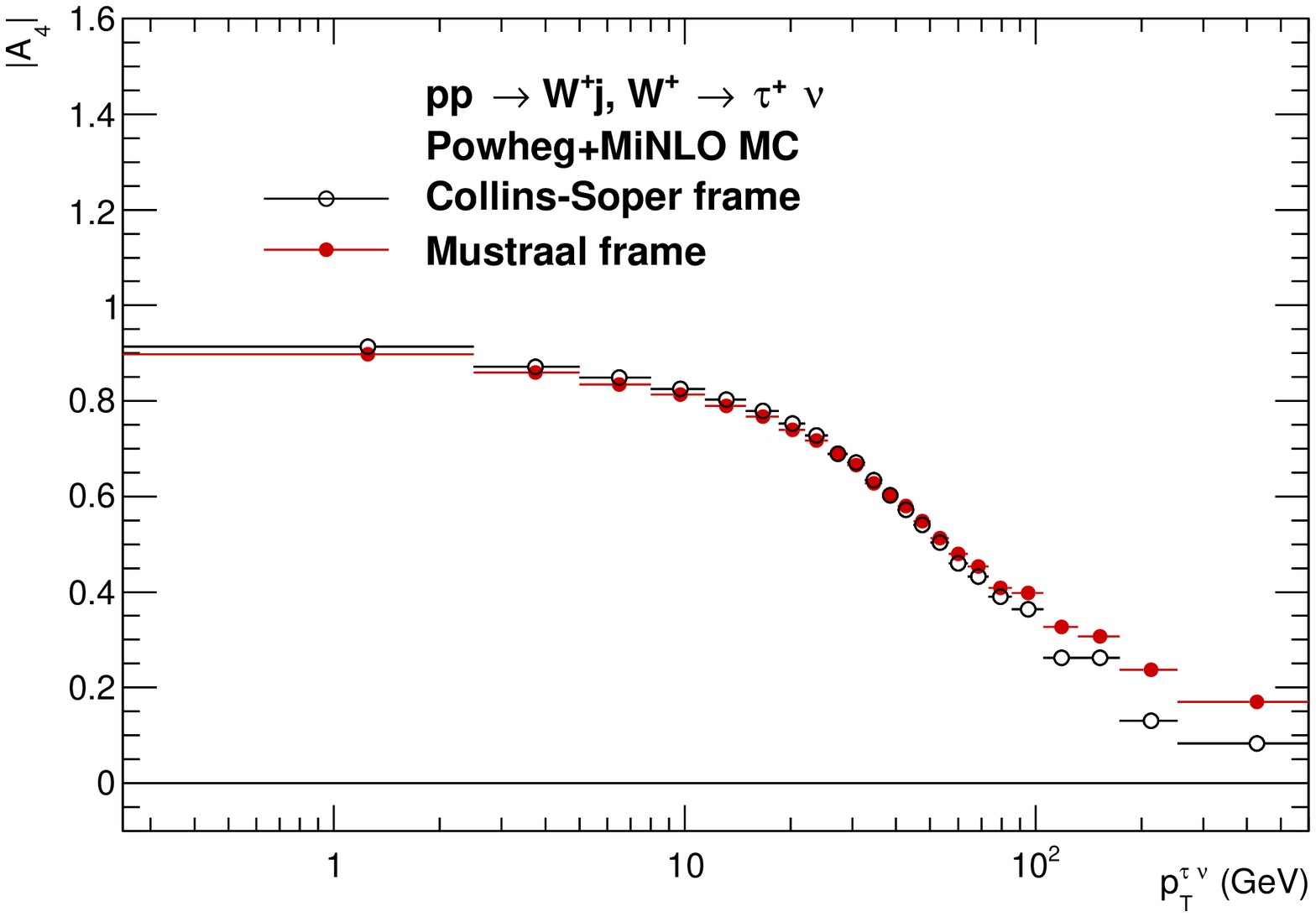}
}
\end{center}
\caption{As Fig.~\ref{Fig:PoWm} but for $W^+ \to \ell^+ \nu$ events.
The $A_i$ coefficients  calculated in Collins-Soper (black) and in {\tt Mustraal} (red) frames 
for $p p \to \tau^{+} \nu\ + 1j$ process generated with {\tt Powheg+MiNLO}.
\label{FigApp:PoWp} }
\end{figure}

Figure~\ref{FigApp:PoWp} shows predictions for  $A_i$'s coefficients of 
$W^+ \to \ell^+ \nu$ and for processes generated with {\tt Powheg+MiNLO}.

\end{document}